\documentclass[a4paper,clock]{article}
\usepackage[dvips]{epsfig}
\usepackage{amssymb}
\usepackage{bm}
\usepackage{dcolumn}

\catcode`\@=11
\def\marginnote#1{}
\newcount\hour
\newcount\minute
\newtoks\amorpm
\hour=\time\divide\hour by60
\minute=\time{\multiply\hour by60 \global\advance\minute
by-\hour}\edef\standardtime{{\ifnum\hour<12
\global\amorpm={am}%
        \else\global\amorpm={pm}\advance\hour by-12 \fi
        \ifnum\hour=0 \hour=12 \fi
        \number\hour:\ifnum\minute<10
0\fi\number\minute\the\amorpm}}
\edef\militarytime{\number\hour:\ifnum\minute<10
0\fi\number\minute}

\def\draftlabel#1{{\@bsphack\if@filesw {\let\thepage\relax
   \xdef\@gtempa{\write\@auxout{\string
      \newlabel{#1}{{\@currentlabel}{\thepage}}}}}\@gtempa
   \if@nobreak \ifvmode\nobreak\fi\fi\fi\@esphack}
        \gdef\@eqnlabel{#1}}
\def\@eqnlabel{}
\def\@vacuum{}
\def\draftmarginnote#1{\marginpar{\raggedright\scriptsize\tt#1}}
\def\draft{\oddsidemargin -.5truein
        \def\@oddfoot{\sl preliminary draft \hfil
        \rm\thepage\hfil\sl\today\quad\militarytime}
        \let\@evenfoot\@oddfoot \overfullrule 3pt
        \let\label=\draftlabel
        \let\marginnote=\draftmarginnote

\def\@eqnnum{(\theequation)\rlap{\kern\marginparsep\tt\@eqnlabel}%
\global\let\@eqnlabel\@vacuum}  }

\def\numberbysection{\@addtoreset{equation}{section}
        \def\theequation{\thesection.\arabic{equation}}}

\def\underline#1{\relax\ifmmode\@@underline#1\else
 $\@@underline{\hbox{#1}}$\relax\fi}

\catcode`@=12
\relax

\numberbysection

\advance \topmargin by -\headheight
\advance \topmargin by -\headsep

\evensidemargin \oddsidemargin
\def\br{\begin{eqnarray}}
\def\er{\end{eqnarray}}
\def\be{\begin{equation}}
\def\ee{\end{equation}}

\def\({\left(}
\def\){\right)}

\relax

\def\D{\Delta}

\def\g{\gamma}

\def\l{\lambda}

\def\pa{\partial}

\def\tp0{\Theta_{+}^{(0)}}
\def\tm0{\Theta_{-}^{(0)}}

\def\f#1#2#3 {f^{#1#2}_{#3}}

\def\win1{{\sf w_{1+\infty}}}

\def\Win1{{\sf W_{1+\infty}}}

\def\rlx{\relax\leavevmode}
\def\inbar{\vrule height1.5ex width.4pt depth0pt}
\def\IZ{\rlx\hbox{\sf Z\kern-.4em Z}}
\def\IR{\rlx\hbox{\rm I\kern-.18em R}}
\def\IC{\rlx\hbox{\,$\inbar\kern-.3em{\rm C}$}}
\def\IN{\rlx\hbox{\rm I\kern-.18em N}}
\def\IO{\rlx\hbox{\,$\inbar\kern-.3em{\rm O}$}}
\def\IP{\rlx\hbox{\rm I\kern-.18em P}}
\def\IQ{\rlx\hbox{\,$\inbar\kern-.3em{\rm Q}$}}
\def\IF{\rlx\hbox{\rm I\kern-.18em F}}
\def\IG{\rlx\hbox{\,$\inbar\kern-.3em{\rm G}$}}
\def\IH{\rlx\hbox{\rm I\kern-.18em H}}
\def\II{\rlx\hbox{\rm I\kern-.18em I}}
\def\IK{\rlx\hbox{\rm I\kern-.18em K}}
\def\IL{\rlx\hbox{\rm I\kern-.18em L}}
\def\one{\hbox{{1}\kern-.25em\hbox{l}}}
\def\0#1{\relax\ifmmode\mathaccent"7017{#1}%
B        \else\accent23#1\relax\fi}

\def\PRL#1#2#3{{\sl Phys. Rev. Lett.} {\bf#1} (#2) #3}
\def\NPB#1#2#3{{\sl Nucl. Phys.} {\bf B#1} (#2) #3}

\def\PLA#1#2#3{{\sl Phys. Lett.} {\bf #1A} (#2) #3}

\def\PLB#1#2#3{{\sl Phys. Lett.} {\bf #1B} (#2) #3}
\def\JMP#1#2#3{{\sl J. Math. Phys.} {\bf #1} (#2) #3}
\def\PTP#1#2#3{{\sl Prog. Theor. Phys.} {\bf #1} (#2) #3}

\def\IJMPA#1#2#3{{\sl Int. J. Mod. Phys.} {\bf A#1} (#2) #3}

\def\TMP#1#2#3{{\sl Theor. Mat. Phys.} {\bf #1} (#2) #3}

\def\PHSD#1#2#3{{\sl Physica} {\bf D#1} (#2) #3}

\def\JHEP#1#2#3{{\sl JHEP} {\bf #1} (#2) #3}

\def\JCP#1#2#3{{\sl Journal of Computational Physics} {\bf #1} (#2) #3}

\def\Nonl#1#2#3{{\sl Nonlinearity} {\bf #1} (#2) #3}

\def\JPAMG#1#2#3{{\sl J. Physics A: Math. Gen.} {\bf A#1} (#2) #3}
\def\JGP#1#2#3{{\sl Journal of Geometry and Physics} {\bf #1} (#2) #3}
\def\EPL#1#2#3{{\sl Europhysics Letters} {\bf #1} (#2) #3}
\def\AdP#1#2#3{{\sl Annalen der Physik} {\bf #1} (#2) #3}

\begin{document}

\begin{titlepage}

\vspace{.2in}
\begin{center}
{\large\bf Riccati-type pseudo-potentials, conservation laws and solitons of deformed sine-Gordon models}
\end{center}

\vspace{.2in}

\begin{center}

H. Blas$^{a,b,c}$, H. F. Callisaya$^{d}$ and J.P.R. Campos$^{a}$

\par \vskip .2in \noindent

$^{a}$ Instituto de F\'{\i}sica\\
Universidade Federal de Mato Grosso\\
Av. Fernando Correa, $N^{0}$ \, 2367\\
Bairro Boa Esperan\c ca, Cep 78060-900, Cuiab\'a - MT - Brazil. \\
              $^{b}$ Facultad de Ciencias\\ Universidad Nacional de Ingenier\'ia, Av. Tupac Amaru, s/n, Lima-Per\'u.\\
               $^{c}$ Facultad de Ciencias\\Universidad Nacional Santiago Ant\'unez de Mayolo, Ciudad Universit\'aria, Huaraz-Per\'u.\\
$^{(d)}$~Departamento de Matem\'atica\\
Universidade Federal  de Mato Grosso  \\ 
Bairro Boa Esperan\c ca, Cep 78060-900, Cuiab\'a - MT - Brazil.\\

\normalsize
\end{center}

\vspace{.3in}

\begin{abstract}
\vspace{.3in}

Deformed sine-Gordon (DSG) models $\pa_\xi \pa_\eta \, w +  \frac{d}{dw}V(w) = 0$, with $V(w)$ being the deformed potential, are considered in the context of the Riccati-type pseudo-potential approach. A compatibility condition of the deformed system of Riccati-type equations reproduces the equation of motion of the DSG models. Then, we provide a pair of linear systems of equations for the DSG model and an associated infinite tower of non-local conservation laws. Through a direct construction and supported by numerical simulations of soliton scatterings, we show that the DSG models, which have recently been defined as quasi-integrable in the anomalous zero-curvature approach [Ferreira-Zakrzewski, JHEP05(2011)130], possess new towers of infinite number of quasi-conservation laws. We compute numerically the first sets of non-trivial and independent charges (beyond energy and momentum) of the DSG model: the two third order conserved charges and the two fifth order asymptotically conserved charges in the pseudo-potential approach, and the first four anomalies of the new towers of charges, respectively. We consider kink-kink, kink-antikink and breather configurations for the Bazeia {\sl et al.} potential $V_{q}(w) = \frac{64}{q^2} \tan^2{\frac{w}{2}} (1-|\sin{\frac{w}{2}}|^q)^2 \, (q \in \IR)$, which contains the usual SG potential $V_2(w) = 2[1- \cos{(2 w)}]$. The numerical simulations are performed using the 4th  order Runge-Kutta method supplied with non-reflecting boundary conditions. 
\end{abstract}

\end{titlepage}

\section{Introduction}

While the soliton type solutions and the presence of infinite number of conserved charges are among the remarkable properties of the integrable models, some non-linear field theory models with important physical applications and solitary wave solutions are not integrable. Recently, it has been performed certain deformations of integrable models such that they exhibit soliton-type solutions with some properties resembling to their counterparts of the truly integrable ones. In this context, it has been introduced the quasi-integrability concept related to the anomalous zero-curvature approach to deformed integrable models \cite{jhep1, jhep2}. For earlier results on some non-linear field theories with solitary waves and the study of their collisions, see e.g. \cite{hietarinta} and references therein. 

Recently, in a series of papers, the quasi-integrability concept has been developed and certain deformations of the  sine-Gordon (SG), Toda, Bullough-Dodd, KdV and non-linear Schr\"odinger (NLS) models \cite{jhep1, jhep2, jhep6, jhep3, toda, arxiv1} have been studied using their relevant anomalous zero-curvature representations. The main developments have been focused on the construction of an infinite number of  their quasi-conservation laws through the so called abelianization procedure and the numerical simulations of two-soliton collisions, in order to examine the behavior of the so-called `anomalies' present in the in-homogeneous quasi-conservation laws for the relevant currents. In this way, it has been shown that the quasi-integrable models possess an infinite number of charges that are asymptotically conserved, i.e. conserved charges, such that their values vary during the scattering of the two-solitons only. As a strong support for the quasi-integrability concept it has recently been considered the relevant charges associated to three-soliton collisions of the various deformations of the KdV model, and they  have been shown to be also asymptotically conserved \cite{arxiv1}; i.e. the quasi-conservation laws exhibit certain anomaly terms which vanish when integrated on the space-time plane.

In  \cite{cnsns}, by two of us, strengthening the results of \cite{jhep1, jhep2},  it has been shown the existence of several towers of exactly conserved charges. In fact,  it has been shown that the deformed SG models indeed possess a subset of infinite number of exactly conserved charges for two-soliton field configurations being eigenstates of the space-reflection parity symmetry \cite{cnsns}. Similar results were found for the deformed defocusing (focusing ) NLS model with dark (bright) solitons \cite{jhep4, jhep5} for a variety of two-soliton configurations. These results have been obtained by combining analytical and numerical methods. 

Here, we provide, by direct construction, new types of two sets of dual towers of asymptotically conserved charges with true anomalies. Through numerical simulations we verify the vanishing of the lowest order anomalies with space-reflection odd parity. These novel charges turn out to be anomalous even for the integrable sine-Gordon model. The only analytical explanation we have found, so far, for the unexpected appearance of these anomalous charges are the space-time symmetry properties which the 2-soliton solutions of the standard sine-Gordon model exhibit. It is expected that those types of charges will play an important role in the study of soliton gases and formation of certain structures in (quasi-)integrable systems, such as soliton turbulence, soliton gas dynamics and rogue waves \cite{pla1, new}. In addition, these new kind of  charges are expected to appear in the other quasi-integrable theories considered in the literature.

Moreover, this work presents the first steps toward deformations of the sine-Gordon model following the pseudo-potential approach. Our main goal is to uncover more specific integrability structures associated to the deformed integrable models mentioned above. It starts by introducing a particular deformation of the Riccati-type pseudo-potential equations related to the ordinary SG model \cite{nucci, prl1}. We introduce a deformed sine-Gordon potential $V$ into the Riccati-type system of equations and a new system of equations for a set of auxiliary fields, such that the compatibility condition applied to the extended system gives rise to the deformed sine-Gordon model (DSG) equation of motion. Then, we construct a first type of infinite number of dual conservation laws, such that an infinite set of quasi-conserved charges, in the laboratory coordinates, emerges as linear combinations of the relevant charges in the both dual formulations.

Then, by combining the analytical pseudo-potential and numerical methods, we provide explicitly the first six conservation laws and the relevant four exactly and two asymptotically conserved charges, including the energy and momentum charges, organized in powers of the spectral parameter, respectively. Remarkably, we have shown that the so-called `anomaly' terms, defined in the quasi-integrability approach, are indeed immersed in the relevant higher order exact conservation laws, beyond energy and momentum conservation laws. We show that the same holds for the third and fifth order quasi-conservation laws presented in \cite{jhep1}. In fact, their `anomalies' can be removed such that the inhomogeneous quasi-conservation laws become truly conservation laws for the conveniently redefined new currents.    
 
New pseudo-potential representations are introduced for the deformed sine-Gordon model. This is achieved by performing certain transformations of the system of Riccati-type equations and writing them in terms of convenient  pseudo-potentials which carry the information of the deformed sine-Gordon potential.  In addition, in the framework  of the pseudo-potential approach \cite{nucci}, we propose two sets of  linear system of equations whose compatibility condition gives rise to the DSG equation of motion. As an application of the proposed linear system of equations and its pair of linear operators, we have obtained the energy and momentum conservation laws of the DSG model, and an infinite set of non-local conservation laws. One of the lowest order non-local conservation law hides a related quasi-conservation law obtained by a direct construction.  
  
In order to simulate the soliton collisions we used the 4th order Runge-Kutta method provided that the non-reflecting boundary conditions, allowing the radiation to cross the boundary points $x= \pm L$ freely  \cite{nonreflec}, are assumed.  Our simulations show that some radiation is produced by the soliton configurations  and the rate of loss of the energy depends on the initial conditions of the system.   
     
The paper is organized as follows. In the next section we introduce the deformed sine-Gordon model (DSG) and briefly discuss the anomalous conserved charges. In subsection \ref{sec:newtowers} new towers of quasi-conservation laws are obtained by direct construction. The section \ref{sec:numerical} presents our results on numerical simulations. In subsection \ref{sec:symmetry} we discuss the space-reflection symmetries of the associated charge densities. We numerically simulate soliton collisions for kink-kink, kink-antikink and breather configurations of the DSG model and compute the energy, momentum and the two sets of third and fifth order conserved charges in subsections \ref{sec:third} and  \ref{sec:fifth}, respectively. In \ref{sec:secth} we numerical simulate the lowest order anomalies of the second and third type of towers. Next,  in section \ref{sec:riccati},  in the framework of the deformed Riccati-type pseudo-potential equations, we construct a dual set of infinite towers of conservation laws. In section \ref{newdiff} new pseudo-potential representations are introduced. In subsection \ref{sec:linear} the Riccati-type  pseudo-potential framework  is used to construct a linear system of equations associated to the DSG model. In \ref{ss:nonlo} the non-local conservation laws are constructed. In section \ref{sec:conclu} we present some conclusions and point out the future prospects of our formalism. The appendices A, B,..., and F present the relevant quantities which have been used to construct the series of charges.

\section{The model and quasi-conservation laws}
\label{sec:themodel}

Let us consider Lorentz invariant field theories in $(1 + 1)$-dimensions with equation of motion, in light-cone coordinates $(\eta,\xi)$, given by\footnote{In the $x$ and $t$ laboratory coordinates: $\eta=\frac{t+x}{2},\,\xi=\frac{t-x}{2},\, \pa_{\eta} = \pa_{t}+\pa_{x},\, \pa_{\xi} = \pa_{t}-\pa_{x},\, \pa_{\eta}\pa_{\xi}= \pa_{t}^2-\pa_{x}^2$}
\br
\label{sg1}
\pa_\xi \pa_\eta \, w +  V^{(1)}(w) &= &0, 
\er
where $w$ is a real scalar field, $V(w)$ is the scalar potential, $\pa_{\xi}\,$ and $\pa_{\eta}$ are partial derivatives and $V^{(1)}(w) \equiv \frac{d}{dw} V(w)$. The family of potentials $V(w)$ will represent certain deformations of the usual SG model, and the eq. (\ref{sg1}) will be defined as a deformed sine-Gordon (DSG)  model equation of motion. We would like to study the properties 
of the theory using some modifications of the techniques of integrable field theories, such as the deformations of the Riccati-type equations \cite{nucci}. 

In \cite{jhep1} the authors have considered the so-called anomalous zero-curvature formulation  and discussed the quasi-integrability  properties of the model (\ref{sg1}), such as the asymptotically conserved charges associated to certain anomalous conservation laws. Following this formalism, in a previous paper by two of us \cite{cnsns}, it has been introduced an infinite  subset of exactly conserved charges associated to space-reflection symmetric kink-antikink, antisymmetric  kink-kink and symmetric breather configurations, respectively, of the model (\ref{sg1}). In this way, for this subset of anomalous conservation laws, the so-called 'anomaly' terms of the quasi-integrability formulation vanish. On the other hand, in recent papers by one of us and collaborators \cite{jhep4, jhep5}, it has been shown that the quasi-integrable modified (focusing and defocusing) non-linear Schr\"odinger models supports a tower of infinite number of exactly conserved charges for two-soliton configurations (bright-bright  or dark-dark) possessing definite parity under space-reflection symmetry. Moreover, in the both type of deformed NLS models it has been reported that for various two-soliton configurations without parity symmetry, the first  nontrivial fourth order charge which presents an `anomalous' term in the quasi-integrability formulation, is exactly conserved, within numerical accuracy; i.e. its associated anomaly vanishes.

In the context of the anomalous zero-curvature formulation of deformed sine-Gordon models in \cite{jhep1, jhep2, cnsns} the authors have introduced the set of quasi-conservation laws by defining the so-called anomalies $\beta^{(2n+1)}$, such that 
\br
\label{asymp}
\frac{d  }{dt} q_{a}^{(2n+1)} = \int dx\, \beta^{(2n+1)}, \,\,\,\,n=1,2,3,...
\er
where the quantities  $q_{a}^{(2n+1)}$ define the so-called asymptotically conserved charges, provided that the time-integrated `anomalies' $ \int dt  \int dx\, \beta^{(2n+1)}$ vanish for some two-soliton configurations. This condition, when combined with eq. (\ref{asymp}), implies that the relationship  $q_{a}^{(2n+1)}(t\rightarrow +\infty) = q_{a}^{(2n+1)}(t\rightarrow -\infty) $ must hold, realizing in this way the concept of asymptotically conserved charges. 

It is a well known fact  in $(1 + 1)$-dimensional Lorentz invariant integrable field 
theories to have dual integrability conditions or Lax equations. So, as we will show below, there exist a dual formulation for each equation  as in (\ref{asymp}) by interchanging $\xi \leftrightarrow \eta$ in the procedure to obtain the relevant conservation laws. So, one can get
\br
\label{asympd}
\frac{d  }{dt} \widetilde{q}_{a}^{(2n+1)} = \int dx\, \widetilde{\beta}^{(2n+1)}, \,\,\,\,n=1,2,3,...
\er
where the quantities  $\widetilde{q}_{a}^{(2n+1)}$ define the dual asymptotically conserved charges, provided that the time-integrated `anomalies' $ \int dt  \int dx\, \widetilde{\beta}^{(2n+1)}$ vanish. This result implies $\widetilde{q}_{a}^{(2n+1)}(t\rightarrow +\infty) = \widetilde{q}_{a}^{(2n+1)}(t\rightarrow -\infty)$. 

The importance and the relevance of such a dual construction will become clear below when  linear combinations of relevant charges of the new dual towers of asymptotically conserved charges are in fact exactly conserved for special two-soliton solutions, a result first obtained by two of us in \cite{cnsns}, for a couple of towers involving charges bearing  the same form as the standard sine-Gordon model. These types of quasi-conservation laws which reproduce the same polynomial form as the usual sine-Gordon charges will be called as the first types of towers.

\subsection{New towers of quasi-conservation laws}
\label{sec:newtowers}

The above first types of dual towers of quasi-conservation laws (\ref{asymp}) and (\ref{asympd}) are characterized by the fact that their r.h.s. terms provide the relevant anomalies; whereas the l.h.s. terms bear the same polynomial form as the usual sine-Gordon charges. In  \cite{jhep1, jhep2} the relevant anomalies were shown to vanish upon space-time integration, then giving rise to asymptotically conserved charges, provided that the field $w$ and the potential $V$ satisfy the symmetry  
\br
\label{pxt}
{\cal P}:  w \rightarrow - w  + \mbox{const.};\,\,\,\,\,V(w) \rightarrow  V(w),
\er
under the space-time reflection around a given point $(x_{\D},t_{\D})$ 
\br
\label{pxt1}
{\cal P}: (\widetilde{x},\widetilde{t}) \rightarrow   (-\widetilde{x},-\widetilde{t}),\,\,\,\,\, \widetilde{x}\equiv x-x_{\D},\,\,\,\widetilde{t}=t-t_{\D}.
\er
In the formulation of  \cite{jhep1, jhep2} the relevant anomalies possess odd parities under (\ref{pxt})-(\ref{pxt1}), so that they must  vanish upon space-time integration. Below we will construct new towers of quasi-conservation laws, such that their anomaly terms also possess odd parities under (\ref{pxt})-(\ref{pxt1}).
 
\subsubsection{Second type of towers}

Multiplying by $(\pa_{\xi} w)^{N-1}$ on the both sides of the eq. (\ref{sg1})  one can rewrite it as
\br
\label{qc}
\partial_{\eta}[\frac{1}{N} (\partial_{\xi} w)^{N}]+\partial_{\xi}[V(\partial_{\xi}w)^{N-2}]=(N-2)(\partial_{\xi}w)^{N-3}\partial_{\xi}^{2}w V,\,\,\,\,\,\,N=3,4,5,...
\er
This tower of infinite number of equations defines a family of anomalous conservation laws in the so-called quasi-integrability approach to deformed integrable field theories. In fact, one can define the quasi-conservation laws
\br
\label{asy1}
\frac{d}{dt} Q^{(N)}_a &=& \mbox{a}^{(N)},\\
Q^{(N)}_a &\equiv & \int dx \,  [\frac{1}{N} (\partial_{\xi} w)^{N} + V(\partial_{\xi}w)^{N-2}],\,\,\,\,\,\,\, \mbox{a}^{(N)} \equiv \int dx \, (N-2)(\partial_{\xi}w)^{N-3}\partial_{\xi}^{2}w V,\,\,\,\,\,\,N\geq 3, \label{asy11}
\er
where we have introduced the  asymptotically conserved charges $Q^{(N)}_a$  and the corresponding  anomalies $\mbox{a}^{(N)}$.

The construction of the dual quasi-conservation laws is performed by multiplying by $(\pa_{\eta} w)^{N-1}$ on the both sides of the eq. (\ref{sg1}). So one has
\br
\label{qcd}
\partial_{\xi}[\frac{1}{N} (\partial_{\eta} w)^{N}]+\partial_{\eta}[V(\partial_{\eta}w)^{N-2}]=(N-2)(\partial_{\eta}w)^{N-3}\partial_{\eta}^{2}w V,\,\,\,\,\,\,N=3,4,5,...
\er
Similarly, this is another tower of infinite number of equations defining a set of anomalous conservation laws. These eqs. allow us to define the quasi-conservation laws
\br
\label{asy2}
\frac{d}{dt} \widetilde{Q}^{(N)}_a &=& \widetilde{\mbox{a}}^{(N)},\\
\widetilde{Q}^{(N)}_a &\equiv & \int dx \,  [\frac{1}{N} (\partial_{\eta} w)^{N} + V(\partial_{\eta}w)^{N-2}],\,\,\,\,\,\,\, \widetilde{\mbox{a}}^{(N)} \equiv \int dx \, (N-2)(\partial_{\eta}w)^{N-3}\partial_{\eta}^{2}w V,\,\,\,\,\,\,N\geq 3,\label{asy21}
\er
where we have introduced the dual asymptotically conserved charges $\widetilde{Q}^{(N)}_a$  and the relevant anomalies $\widetilde{\mbox{a}}^{(N)}$. 

The densities of the anomalies $\mbox{a}^{(N)}$ and $\widetilde{\mbox{a}}^{(N)}$ in (\ref{asy11}) and (\ref{asy21}), respectively, possess odd parities under (\ref{pxt})-(\ref{pxt1}), so the quasi-conservation laws (\ref{qc}) and (\ref{qcd}), respectively,  allow the construction of asymptotically conserved charges according to \cite{jhep1, jhep2}. 

\subsubsection{Third type of towers}

Multiplying by $V^{N-1}$ on the both sides of the eq. (\ref{sg1})  one can rewrite it as
\br
\label{qc1}
\partial_{\eta}[\frac{1}{2} V^{N-1} (\partial_{\xi} w)^2]+\frac{1}{N}\partial_{\xi}V^N=\frac{1}{2}(\partial_{\xi}w)^2\partial_{\eta} V^{N-1},\,\,\,\,\,\,N=2,3,4,5,...
\er
This tower of infinite number of equations define a new family of anomalous conservations laws. In fact, one can define the quasi-conservation laws
\br
\label{asy3}
\frac{d}{dt} {\cal Q}^{(N)}_a &=& {\gamma}^{(N)},\\
{\cal Q}^{(N)}_a &\equiv & \int dx \,  [\frac{1}{2} V^{N-1} (\partial_{\xi} w)^2+\frac{1}{N} V^N],\,\,\,\,\,\,\, \g^{(N)} \equiv \int dx \,\frac{1}{2}(\partial_{\xi}w)^2\partial_{\eta} V^{N-1},\,\,\,\,\,\,N\geq 2, \label{asy331}
\er
where we have introduced the  asymptotically conserved charges $\hat{Q}^{(N)}_a$  and the corresponding  anomalies $\g^{(N)}$.

The interchange $\eta \leftrightarrow \xi$ allows us to reproduce the dual quasi-conservation laws. So, one has
\br
\label{emd1}
\partial_{\xi}[\frac{1}{2} V^{N-1} (\partial_{\eta} w)^2]+\frac{1}{N}\partial_{\eta}V^N=\frac{1}{2}(\partial_{\eta}w)^2\partial_{\xi} V^{N-1},\,\,\,\,\,\,N=2,3,4,5,...
\er
These eqs. allow us to define the quasi-conservation laws
\br
\label{asy4}
\frac{d}{dt} \widetilde{\cal Q}^{(N)}_a &=& \widetilde{\g}^{(N)},\\
\widetilde{\cal Q}^{(N)}_a &\equiv & \int dx \,  [\frac{1}{2} V^{N-1} (\partial_{\xi} w)^2+\frac{1}{N} V^N],\,\,\,\,\,\,\, \widetilde{\g}^{(N)} \equiv \int dx \,\frac{1}{2}(\partial_{\eta}w)^2\partial_{\xi} V^{N-1},\,\,\,\,\,\,N\geq 2,\label{asy41}
\er
where we have defined the dual asymptotically conserved charges $\widetilde{\cal Q}^{(N)}_a$  and the anomalies $\widetilde{\g}^{(N)}$. 

Similarly, the densities of the anomalies $\g^{(N)}$ and $\widetilde{\g}^{(N)}$ in (\ref{asy331}) and (\ref{asy41}), respectively, possess odd parities under (\ref{pxt})-(\ref{pxt1}), so the quasi-conservation laws (\ref{qc1}) and (\ref{emd1}), respectively,  allow the construction of  asymptotically conserved charges as in the previous cases. 

The relevant anomalies of the  lowest order asymptotically conserved charges of the above towers of quasi-conservation laws will be simulated below for 2-soliton interactions.

The above charges turn out to be anomalous even for the integrable sine-Gordon model. In fact, the relevant 2-soliton solutions have been constructed analytically \cite{jhep1, jhep2, cnsns} which possess a definite parity under (\ref{pxt})-(\ref{pxt1}), such that the relevant anomaly densities, with odd parities, vanish upon space-time integration. The only explanation we have found, so far, for the appearance of new towers of quasi-conserved charges in the standard sine-Gordon model is the symmetry argument. The appearance of new towers of anomalous charges in the (quasi-)integrable models such as KdV \cite{new} is currently under investigation.

Let us briefly comment on some consequences and importance of that behavior for integrable systems. The above type of integrals have been computed for two-soliton interactions which are thought to play an important role in the study of soliton gases and formation of certain structures in integrable systems, such as integrable turbulence and rogue waves. In the context of the integrable KdV model it has been analyzed the behavior of the so-called statistical moments defined by the integrals of type (see e.g. \cite{pla1}) $M_n(t) = \int_{-\infty}^{+\infty} v^n \, dx,\,\,n \geq 1$; where $v$ is the KdV field. The cases $M_{1,2}$ are exact conserved charges of the model. It is remarkable that  the 3rd and 4th
moments, $M_{3,4}$, respectively, in the interaction region of two-solitons, exhibit a qualitatively similar behavior to the asymptotically conserved charges present in quasi-integrable KdV models \cite{arxiv1}. In fact, in the quasi-integrable KdV models the moments $M_{2,3}$ are actually asymptotically conserved charges \cite{arxiv1, new}. So, since the two-soliton interaction behavior is thought to play an important role in the formation of soliton turbulence and the dynamics of soliton gases, we may expect they will also play an important role in the quasi-integrable counterparts. Certainly, in the present case of  the SG model and its related soliton ensemble, to our knowledge, it is needed a further theoretical development.

\section{Numerical simulations}
\label{sec:numerical}

In order to check our results we have performed several numerical simulations of the linear combinations of the conserved charges: $q^{(3)}_{a}\pm \widetilde{q}^{(3)}_{a}$ and  $q^{(5)}_{a}\pm \widetilde{q}^{(5)}_{a}$, respectively, in (\ref{asymp})  and (\ref{asympd}), of the first type of towers of quasi-conserved charges. In addition, we numerically simulate the linear combinations of the lowest order anomalies of the second and third type of towers (\ref{asy1}) and  (\ref{asy2}), and (\ref{asy3}) and (\ref{asy4}), respectively, of quasi-conserved charges. We consider the Bazeia 
at. al. model, studying kink-antikink, kink-kink and a system involving a kink and an antikink bound state (breather). We used various grid sizes and number of points. The two-soliton (kink-antikink and kink-kink) simulations were performed on a lattice of $1200$ points with spacing of $\D x = 0 . 025$ (in the interval $[ - L, L ] = [ - 15 , 15] $). The time step of our simulations was $\D t = 0 . 0005$, and sometimes $\D t = 0 . 00025$. The breather-like simulations were performed on a lattice of $2000$ lattice points with lattice spacing of $\D x = 0 . 025$ (in the interval $[ −L, L ] = [ - 25 , 25] $). The time evolution was simulated by the fourth order Runge-Kuta method with non-reflecting (transparent) boundary conditions  at the both ends of the lattice grids \cite{nonreflec, cnsns}.  

The  third order conserved charges $q^{(3)}_{a\, \pm}$ 
\br
\label{asy3pm}
q^{(3)}_{a\, \pm} \equiv  \widetilde{q}_{a}^{(3)} \pm  q_{a}^{(3)},
\er
where the dual charges were defined in (\ref{asy32}) and (\ref{asy32d}), respectively. Likewise, we will compute numerically the fifth order conserved charges $q^{(5)}_{a\, \pm}$
\br
\label{fifthpm}
q^{(5)}_{a\,\pm} \equiv \widetilde{q}^{(5)}_{a} \pm q^{(5)}_{a},
\er
where the dual charges were defined in (\ref{charge5}) and (\ref{charge5d}), respectively.  Moreover,  we will compute the lowest order anomalies of the second and third types of towers of charges  (\ref{asy1})-(\ref{asy21}) and (\ref{asy3})-(\ref{asy41}), respectively,
\br
\label{apm}
\mbox{a}_{\pm} &\equiv & \mbox{a}^{(3)} \pm \widetilde{\mbox{a}}^{(3)}\\
\label{gpm}
\g_{\pm} &\equiv & \g^{(2)} \pm \widetilde{\g}^{(2)}.
\er

We will compute numerically the first six charges: the exactly conserved energy and momentum charges, and the two sets of third order (exactly conserved) and fifth order (asymptotically conserved) charges, respectively, for kink-kink, kink-antikink and breather configurations for the Bazeia {\sl et al.} \cite{bazeia, jhep1} potential  
\br
\label{bazeiapot}
V = \frac{64}{q^2} \tan^2{\frac{w}{2}} (1-|\sin{\frac{w}{2}}|^q)^2.
\er
This potential is a one-parameter family of deformations of the sine-Gordon model. Notice that for $q=2$ it reduces to the usual SG potential (\ref{usg}). We have the following deformed SG  equation of motion
\br
\label{sg1lc}
\pa_{t}^2 w - \pa_{x}^2 w + V'(w) & = &0,\\
\nonumber
      V'(w) &=& \frac{64 \tan{\frac{w}{2}}}{q^2(1+ \cos{w})} \{ 1- [(\sin{\frac{w}{2}})^2]^{q/2}\} \{ 2 - [2+q+q\cos{w}][(\sin{\frac{w}{2}})^2]^{q/2}\}.
\er

The kink and anti-kink  solutions of the deformed sine-Gordon model (\ref{sg1lc})   are the following \cite{jhep1,bazeia}
\br
\label{kink1}
w(x,t) = 2\, \eta_{2}\, \arcsin{\{\Big[ \frac{e^{2 \sqrt{2}\, \eta_1\, \g (x - v t)}}{1+ e^{2 \sqrt{2} \eta_1 \g (x - v t)}}\Big]^{\frac{1}{q}}\}} + 2 \pi\, l,\,\,\,\,\,\, \g \equiv \frac{1}{\sqrt{1-v^2}}
\er
where $\eta_i= \pm 1,\, i=1,2$; $l$ is any integer and $v$ is the kink velocity given in units of the speed of light. The topological charge of each solution is provided by the product $\eta_1 \eta_2$.

The simulations of the kink-kink and kink-antikink system of the deformed SG model will consider, as the initial condition, two analytical solitary wave solutions of type (\ref{kink1}). In fact, in order to have a kink-antikink system for $t = 0$ we consider a kink ($ \eta_1 = 1 , \eta_2 = 1 , l = 0 $) 
and an antikink ( $\eta_1 = −1 , \eta_2 = 1 , l = 0 $), according to the solution in (\ref{kink1}), located some distance apart and stitched together at the middle point $x = 0$. Similarly, in order to have a kink-kink system for $t = 0$ we consider two kinks ( $\eta_1 = 1 , \eta_2 = 1 , l = 0 $) from (\ref{kink1}), located some distance apart and stitched together.

{\bf Breather solution of SG} 

Let us write the equation (\ref{sg1}) with potential (\ref{usg}) such that one has the usual SG equation of motion
\br
\label{usg1}
\pa_{t}^2 w - \pa_{x}^2 w + 4 \sin{\( 2 w\)} & = &0.  
\er
Since the general analytic breather solutions of the model (\ref{sg1lc}) are not known, we will consider  the breather type solution of (\ref{usg1}) as an initial condition for our simulations of the conserved charges. The SG breather becomes
\br
\label{breather}
w_{br}(x,t) = 2 \arctan{\Big[ \frac{\sqrt{1-v^2}}{v} \frac{\sin{\(2 \sqrt{2}\, v \,t\)}}{\cosh{\(2 \sqrt{2}\sqrt{1-v^2} \, x \)}}\Big]}.
\er 

Some properties of the breather-like configurations of the deformed sine-Gordon models have been studied in the quasi-integrability formulation \cite{jhep1, jhep2, cnsns} through numerical simulations. Here, we follow the approach of \cite{cnsns} in order to generate a long-lived breather, so the initial condition will be an analytical function of the SG breather at rest (\ref{breather}).  

As we will present below, the numerically simulated energy and momentum charges are very well conserved in all soliton-soliton scatterings and similarly so for the evolution of the breather-like structures. However, in order to simulate the higher order charges, such as $q_{a\, \pm}^{(3)}$ and $q_{a\, \pm}^{(5)}$, and numerically approximate the higher order space derivatives which appear in their relevant charge densities, one must discretize space in some way, and this inevitably introduces fictitious discretization effects into the charge densities, which one should seek to minimize. In order to minimize these effects we will consider the decomposition of the relevant charges as presented in (\ref{asy32}) and (\ref{asy32d}), and in  (\ref{charge5}) and (\ref{charge5d}), respectively. We will write below the relevant components in laboratory coordinates and present the charges (\ref{asy3pm}) and (\ref{fifthpm}), respectively, as summations of certain charge density terms which will exhibit special space-time symmetries and in a manner that their densities present lower order space-time derivatives. This last idea was achieved in the  decomposition above by removing partial time-derivative terms out of the space integrals, by converting the relevant terms in total time derivatives of the $x-$integrated densities. So, in the next sections we will implement this program and make them to be more amenable to numerical simulations.

\subsection{Space-reflection symmetry of the densities of charges and anomalies}

\label{sec:symmetry}

The behaviour of the charges $q^{(3)}_{a\, \pm} $, $ q^{(5)}_{a\,\pm}$ and anomalies $\mbox{a}_{\pm}$ and $\g_{\pm}$ defined above, for soliton collisions, would depend upon the symmetry properties of the relevant field configurations. In order to check the results of the numerical simulations of the densities and their $x-$integrations we will resort to the behavior of them under the space-reflection symmetry. As we will discuss below, some field configurations, such as the kink-kink, kink-antikink and breather solutions of the usual SG model possess definite parity eigenvalues under the  space-reflection symmetry transformation. Moreover, the numerically simulated analogous field configurations of the deformed model will present qualitatively the same properties.     

Let us consider the space-reflection transformation 
\br
\label{px}
{\cal P}_x: x \leftrightarrow - x.
\er
The scalar field $w$ for some soliton configurations is an eigenstate of the operator ${\cal P}_x$, so one has 
\br
\label{pxvp}
{\cal P}_x: w \rightarrow \varrho\,\, w,\,\,\,\,\,\varrho = \pm 1.
\er
In addition, in our discussions below we will consider an even potential $V$ under ${\cal P}_x$
\br 
\label{evenpot}
{\cal P}_x (V) = V.
\er
   
Let us recall the space-reflection symmetries of special two-soliton configurations of the ordinary SG model in the center of mass reference frame \cite{cnsns}. The kink-antikink solution possesses an even parity ($ \rho = +1 $) under the space-reflection transformation (\ref{px})-(\ref{pxvp}). Whereas,  the kink-kink solution possesses an odd parity ($ \rho = -1 $) under the space-reflection transformation (\ref{px})- (\ref{pxvp}). These  two-soliton configurations can be regarded as the zeroth order solution of DSG model in the expansion parameter $\epsilon = q-2$. For those two-soliton (kink-antikink and kink-kink) solutions in laboratory coordinates  without space-reflection  parity symmetry one can  recover the space-reflection symmetries by performing convenient Lorentz transformations to the center of mass frame \cite{cnsns}. In addition, the breather solution (\ref{breather}) satisfies (\ref{pxvp}) with $\rho = 1$, and it could also be used as the zeroth order solution of the deformed SG model in perturbation theory.    

In our numerical simulations of soliton collisions and breather oscillation for the potential (\ref{bazeiapot})  one notices the qualitative realization of the symmetry (\ref{pxvp}) for equal and opposite velocity kink-kink solution (odd parity $\rho = -1$) in the Fig. 1 and kink-antikink solution (even parity $\rho = +1$) in the  Fig. 4. Likewise, for  breather oscillation in Fig. 13 one notices the even parity behavior ($\rho = +1$) of its oscillation around a symmetric vertical axis. On the other hand, the Figs. 7 and 10 show, respectively,  kink-kink and kink-antikink collisions with different velocities and asymmetric behavior.   
   
\subsection{ Third  order conserved charges}
\label{sec:third}

In our numerical computations we will consider directly the expressions of the asymptotically conserved charges $q_{a\, \pm}^{(3)}$, instead of their anomalies as in the previous literature. The explicit form of the charges  (\ref{asy3pm}), written in laboratory coordinates,  become
\br
\label{qq31}
q_{a\, +}^{(3)} &=& \frac{1}{4} \frac{d^2 E}{dt^2}  + \int \, dx \frac{1}{4} \Big\{ (w^{(0,1)})^4 + 4 [V-2] [(w^{(1,0)})^2+ (w^{(0,1)})^2 ] + (w^{(1,0)})^4 + 6 (w^{(0,1)})^2 (w^{(1,0)})^2  - \nonumber\\
&& 4 (w^{(1,1)})^2 -(w^{(0,2)}+w^{(2,0)})^2 \Big\}\\
 &\equiv&  \frac{1}{4} \frac{d^2 E}{dt^2}  + Q_{+}^{(3)} \label{qq311}\\
 \label{qq32}
q_{a\, -}^{(3)} &=&\frac{1}{4} \frac{d^2 P}{dt^2}  + \int \, dx  \Big\{ w^{(0,1)}  w^{(1,0)} [2V -4 + ( w^{(0,1)})^2 + ( w^{(1,0)})^2] -  w^{(1,1)}[ w^{(0,2)}+w^{(2,0)}] \Big\},\\
 &\equiv &  \frac{1}{4} \frac{d^2 P}{dt^2}  + Q_{-}^{(3)} \label{qq322}
\er
where the notation [$w^{(q , p )} \equiv \frac{\pa^{q + p}}{\pa x^q \pa t^p} w(x,t)$] has been used. 

Notice that the charges $q_{a\, \pm}^{(3)}$ have been decomposed as a sum of second order time derivatives of the energy and momentum, respectively, plus the components $Q_{\pm}^{(3)}$, which we will  numerically simulate below. The decomposition of the charges will minimize the fictitious discretization effects into the soliton dynamics, as mentioned above. In fact, in previous computations \cite{cnsns, jhep1}, the asymptotic behavior of the charges $q_{a\, \pm}^{(3)}$ have been examined indirectly by performing the time integration of the anomalous conservation laws of type  (\ref{asymp}). The second order time derivatives of $E$ or $P$ in the r.h.s. of (\ref{qq31})-(\ref{qq322}) can be introduced as partial  time derivatives into their relevant $x-$integral expressions, then it  can be traded by space derivatives using the eqs. of motion. Then, they would be added to the charge densities of the full $q_{a\, \pm}^{(3)}$ expressions; so, this process would introduce higher order space derivatives into them, which amount to introduce more fictitious discretization effects in the simulations of $q_{a\, \pm}^{(3)}$. So, our decomposition above have sought to reduce these effects.       

Another aspect which must be considered in order to check the results of our numerical simulations will be the symmetries of the relevant charge densities above. In particular, the space reflection-symmetries of soliton configurations will reflect on the space-reflection symmetry of each charge density in the integrals of $Q_{\pm}^{(3)}$ in (\ref{qq31})-(\ref{qq311}) and (\ref{qq32})-(\ref{qq322}), respectively, as we will verify numerically.

In fact, the densities of the charge components $Q_{\pm}^{(3)}$ in (\ref{qq31}) and (\ref{qq32}) will be even and odd functions, respectively, under the space-reflection transformation (\ref{px}) for soliton configurations with definite parity $\varrho$ (\ref{pxvp}) and for potentials satisfying (\ref{evenpot}). In particular, this observation holds for kink-kink, kink-antikink and breather configurations. 

Then, by symmetry arguments only, one expects the vanishing of the charge  $q_{-}^{(3)}$ for all soliton configurations possessing definite parity under space-reflection transformations. The numerical results are in accordance with this observation as we will see below.  

The Fig. 1 shows the numerical simulation of kink-kink collision and the energy (E) and momentum (P) charges for equal and opposite velocities $v_2=-v_1=0.15$ and $q=2.01$.  The Figs. 2 and 3 show, respectively,  the charge densities and the conserved charge components  $Q_{\mp}^{(3)}$ for the kink-kink  collision for $v_2=-v_1=0.7$ and $q=2.1$. Whereas, the Figs. 7, 8 and  9  show the same quantities for kink-kink collision with different velocities  $v_2=0.75, v_1=-0.5$ and $q=1.9$. 

The Fig. 4 shows the numerical simulation of kink-antikink collision and the energy (E) and momentum (P) charges for equal and opposite velocities $v_2=-v_1=0.15$ and $q=2.01$.  The Figs. 5 and 6 show, respectively,  the charge densities and the conserved charge components  $Q_{\mp}^{(3)}$ for the kink-antikink  collision for $v_2=-v_1=0.7$ and $q=1.9$. Whereas, the Figs. 10, 11 and  12  show the same quantities for kink-antikink collision with different velocities  $v_2=0.4, v_1=-0.8$ and $q=2.01$. 

The Fig. 13 shows the breather oscillation with period $T = 3.87$ for three successive times and $q=1.97$, and  the plot of the energy versus time. Notice that the energy takes thousands of  units of time to stabilize; in fact, it achieves a constant value after $t\approx 5\times 10^{4}$ of the time interval $t=[0\,,\, 6\times 10^{4}]$. The Figs. 14 and 15 show, respectively,  the charge densities for three successive times within a period and the conserved charge components  $Q_{\pm}^{(3)}(t)$ for the breather configuration.

\begin{figure}
\centering
\label{fig1}
\includegraphics[width=4cm,scale=7, angle=0, height=8cm]{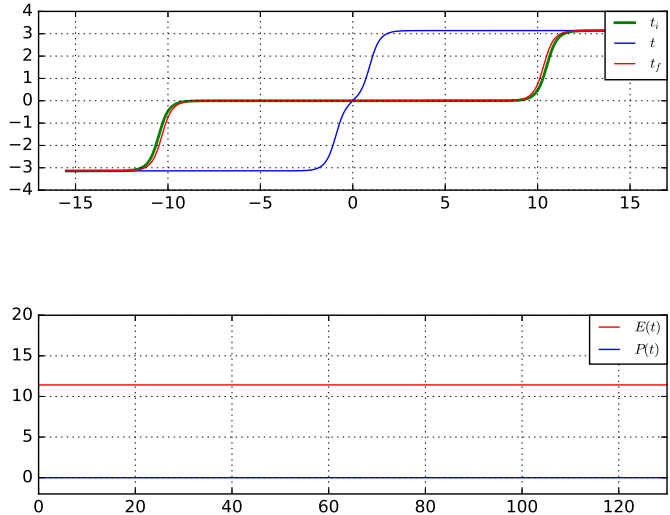}
\parbox{6in}{\caption{(color online) Top Fig. shows kink-kink collision with velocities $v_2=-v_1= 0.15$ for $q = 2.01$, with initial (green), collision (blue) and final (red) configurations. Bottom Fig. shows the conserved energy ($E$) and momentum ($P$) charges of the kink-kink configuration,respectively.}}
\end{figure}

\begin{figure}
\centering
\label{fig2}
\includegraphics[width=2cm,scale=6, angle=0,height=6cm]{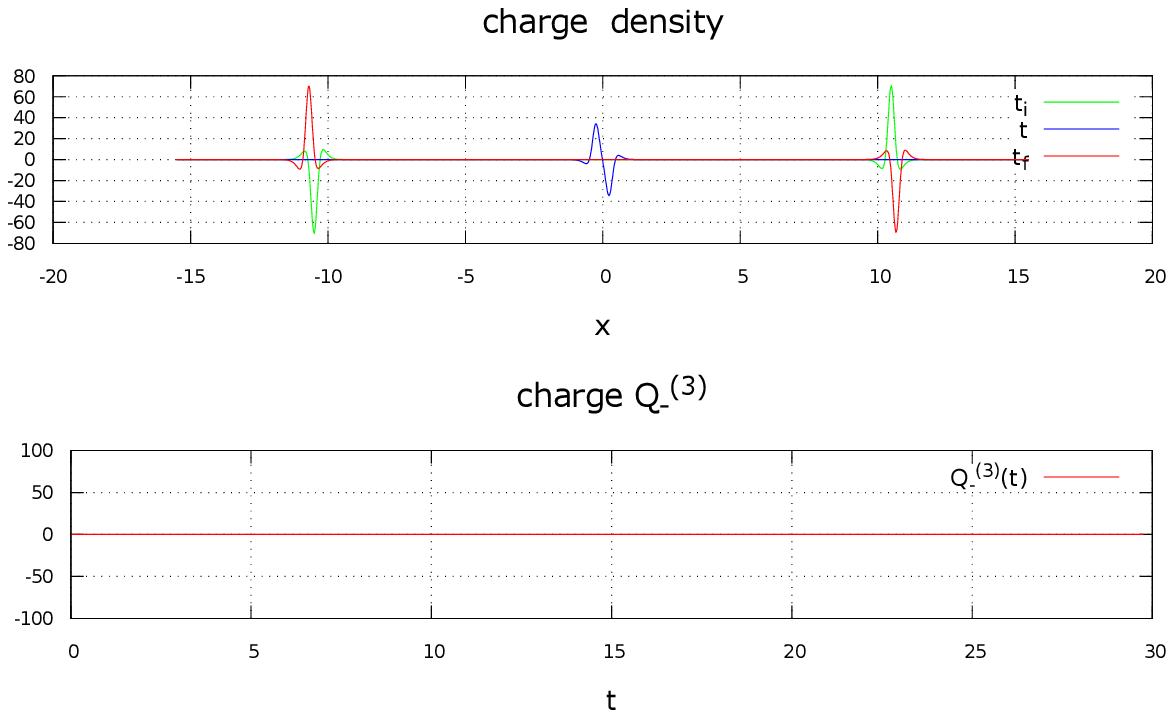} 
\parbox{6in}{\caption{(color online)   Top Fig. shows the charge density of $Q_{-}^{(3)}(t)$ vs $x$ for the kink-kink collision with equal and opposite velocities $v_2=-v_1=0.7$ for $q = 2.1$, with initial (green), collision (blue) and final (red) densities of the kink-kink scattering. Bottom Fig. shows the conserved charge $Q_{-}^{(3)}(t)$ of the kink-kink solution.}}
\end{figure}

\begin{figure}
\centering
\label{fig3} 
\includegraphics[width=2cm,scale=6, angle=0,height=6cm]{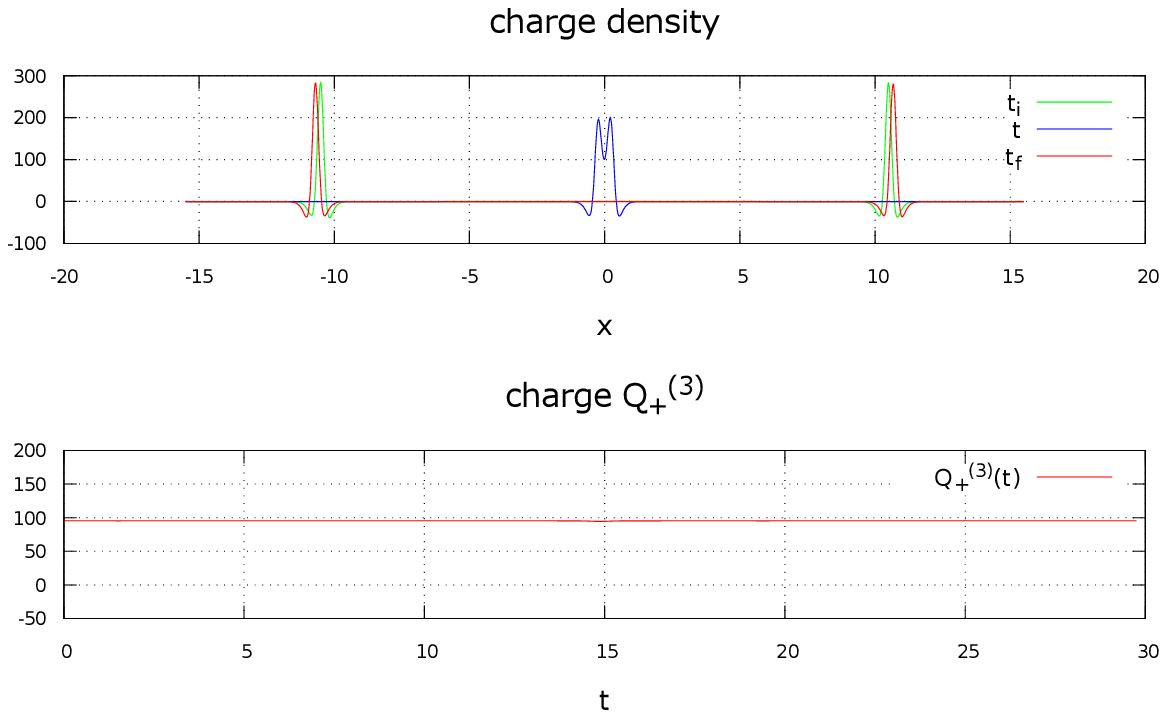}
\parbox{6in}{\caption{(color online)   Top Fig. shows the charge density of $Q_{+}^{(3)}(t)$ vs $x$ for the kink-kink collision with equal and opposite velocities $v_2=-v_1=0.7$ for $q = 2.1$, with initial (green), collision (blue) and final (red) densities of the kink-kink scattering. Bottom Fig. shows the conserved charge $Q_{+}^{(3)}(t)$ of the kink-kink solution.}}
\end{figure}

\begin{figure}
\centering
\label{fig4}
\includegraphics[width=4cm,scale=7, angle=0, height=8cm]{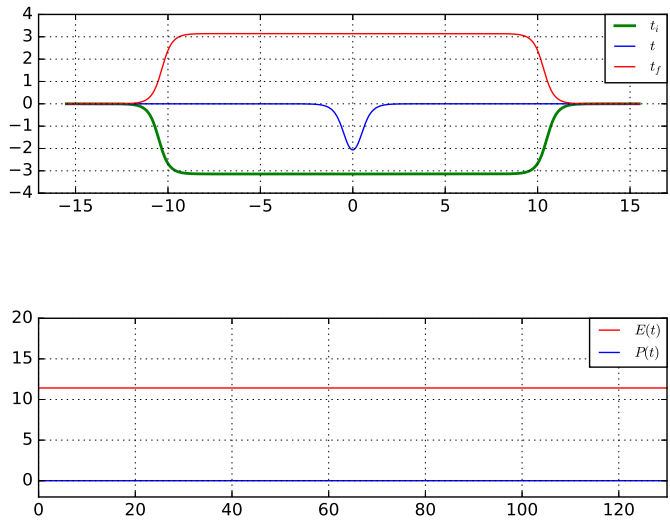}
\parbox{6in}{\caption{(color online) Top Fig. shows kink-antikink collision with velocities $v_2=-v_1= 0.15$ for $q = 2.01$, with initial (green), collision (blue) and final (red) time configurations. Bottom Fig. shows the conserved energy ($E$) and momentum ($P$) charges of the kink-antikink configuration.}}
\end{figure}

\begin{figure}
\centering
\label{fig5}
\includegraphics[width=2cm,scale=6, angle=0,height=6cm]{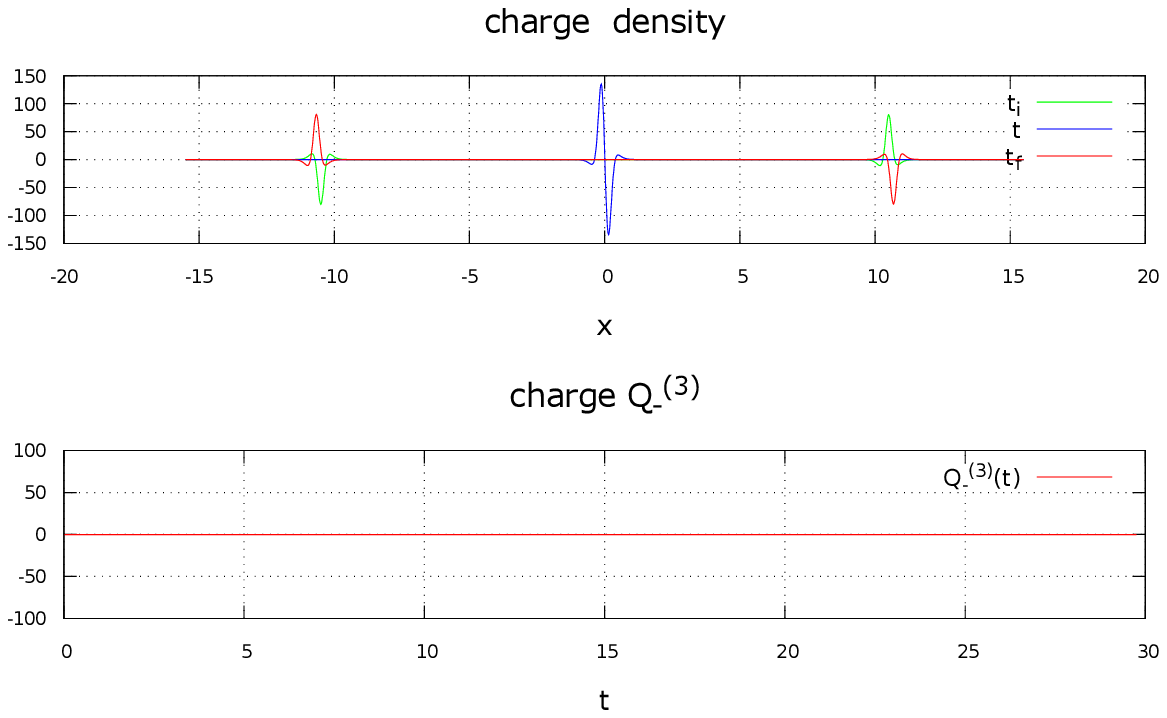} 
\parbox{6in}{\caption{(color online)   Top Fig. shows the charge density of $Q_{-}^{(3)}(t)$ vs $x$  for the kink-antikink collision with equal and opposite velocities $v_2=-v_1=0.7$ for $q = 1.9$, with initial (green), collision (blue) and final (red) densities of the kink-antikink scattering. Bottom Fig. shows the   conserved charge $Q_{-}^{(3)}(t)$ of the kink-antikink solution.}}
\end{figure}

\begin{figure}
\centering
\label{fig6} 
\includegraphics[width=2cm,scale=6, angle=0,height=6cm]{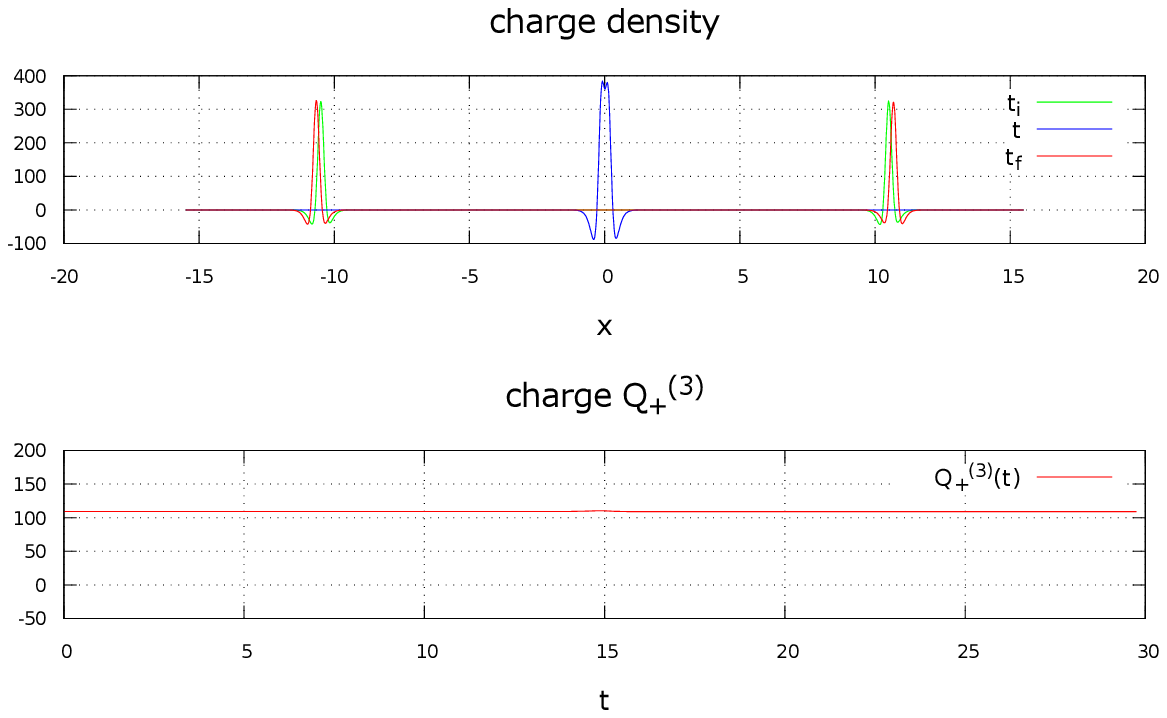}
\parbox{6in}{\caption{(color online)   Top Fig. shows the charge density of $Q_{+}^{(3)}(t)$ vs $x$ for  the kink-antikink collision with equal and opposite velocities $v_2=-v_1=0.7$ for $q = 1.9$, with initial (green), collision (blue) and final (red) densities of the kink-antikink scattering. Bottom Fig. shows the    conserved charge $Q_{+}^{(3)}(t)$ of the kink-antikink solution.}}
\end{figure}

\begin{figure}
\centering
\label{fig7}
\includegraphics[width=2cm,scale=6, angle=0, height=6cm]{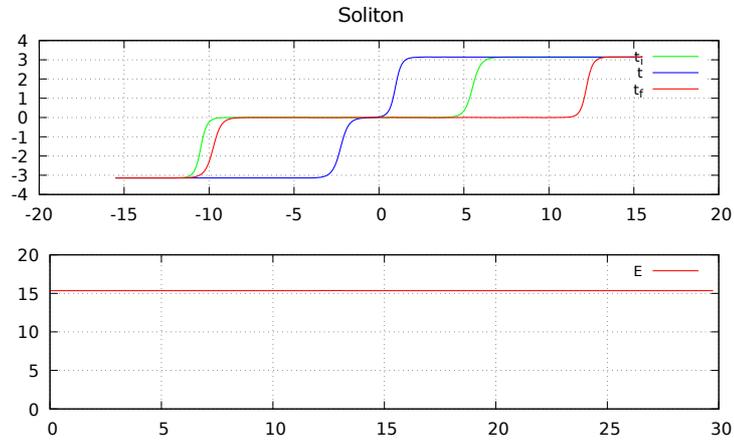}
\parbox{6in}{\caption{(color online) Top Fig. shows kink-kink collision with different velocities $v_2=0.75,\,v_1=-0.5$ for $q = 1.9$, with initial (green), collision (blue) and final (red) time configurations. Bottom Fig. shows the conserved energy of the kink-kink.}}
\end{figure}

\begin{figure}
\centering
\label{fig8}
\includegraphics[width=2cm,scale=6, angle=0,height=6cm]{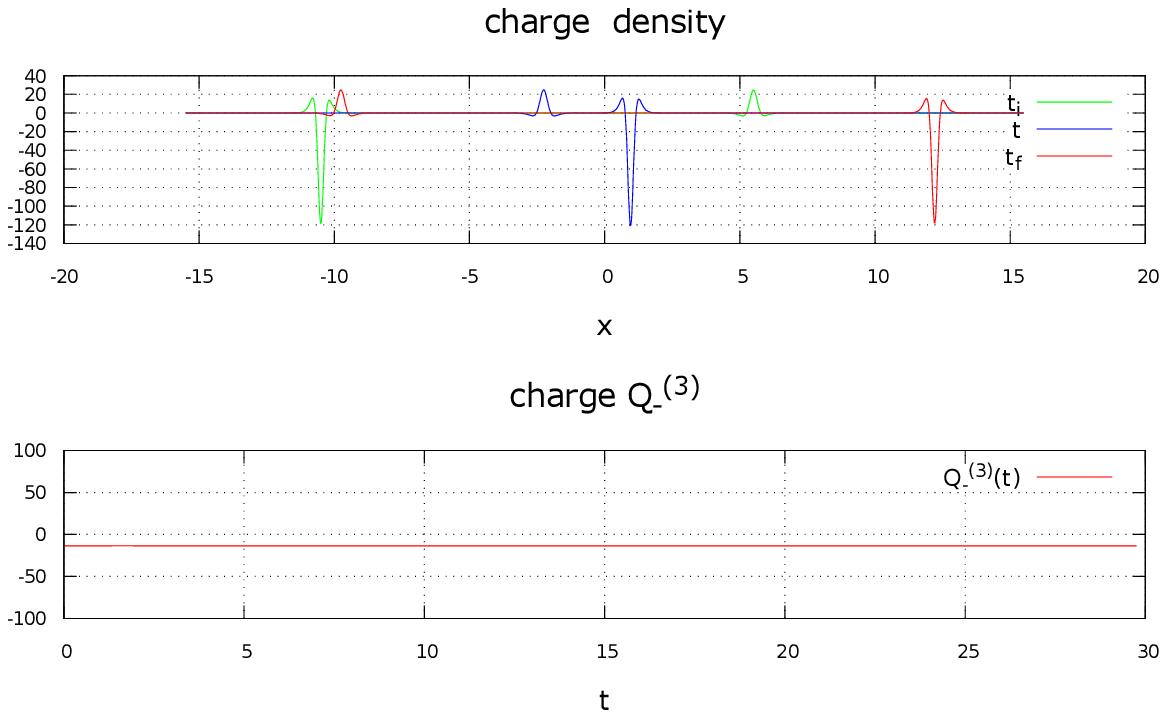} 
\parbox{6in}{\caption{(color online)   Top Fig. shows the charge density of $Q_{-}^{(3)}(t)$ vs $x$ for the kink-kink collision with $v_2=0.75,\,v_1=-0.5$ for $q = 1.9$, with initial (green), collision (blue) and final (red) time  densities of the kink-kink scattering. Bottom Fig. shows the conserved charge $Q_{-}^{(3)}(t)$ of the kink-kink solution.}}
\end{figure}

\begin{figure}
\centering
\label{fig9} 
\includegraphics[width=2cm,scale=6, angle=0,height=6cm]{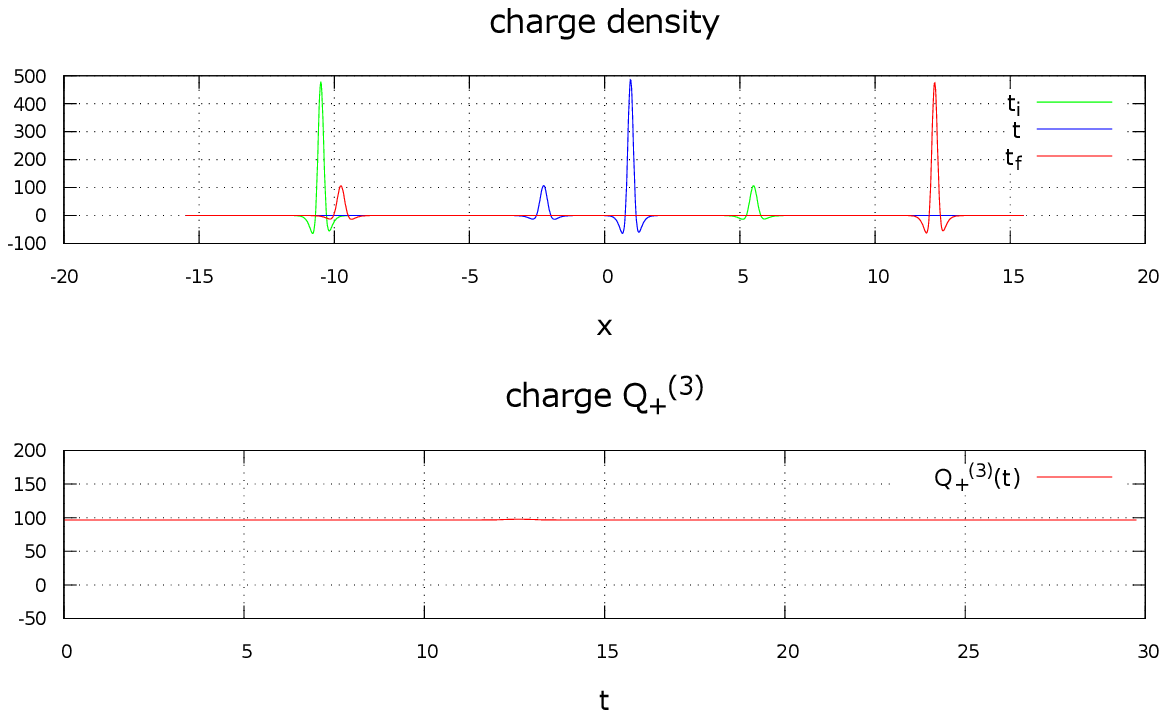}
\parbox{6in}{\caption{(color online)   Top Fig. shows the charge density of $Q_{+}^{(3)}(t)$ vs $x$ for the kink-kink collision with $v_2=0.75,\,v_1=-0.5$ for $q = 1.9$, with initial (green), collision (blue) and final (red) densities of the kink-kink scattering. Bottom Fig. shows the conserved charge $Q_{+}^{(3)}(t)$ of the kink-kink solution.}}
\end{figure}

\begin{figure}
\centering
\label{fig10}
\includegraphics[width=2cm,scale=6, angle=0, height=6cm]{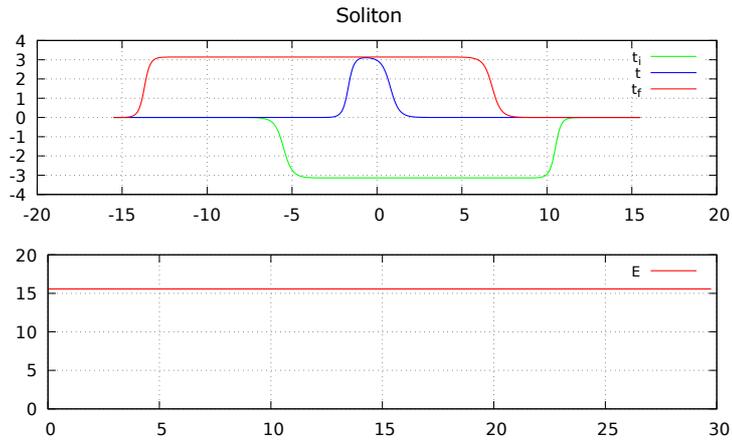}
\parbox{6in}{\caption{(color online) Top Fig. shows kink-antikink collision with different velocities $v_2=0.4,\,v_1=-0.8$ for $q = 2.01$, with initial (green), collision (blue) and final (red) configurations. Bottom Fig. shows the conserved energy of the kink-antikink.}}
\end{figure}

\begin{figure}
\centering
\label{fig11}
\includegraphics[width=2cm,scale=6, angle=0,height=6cm]{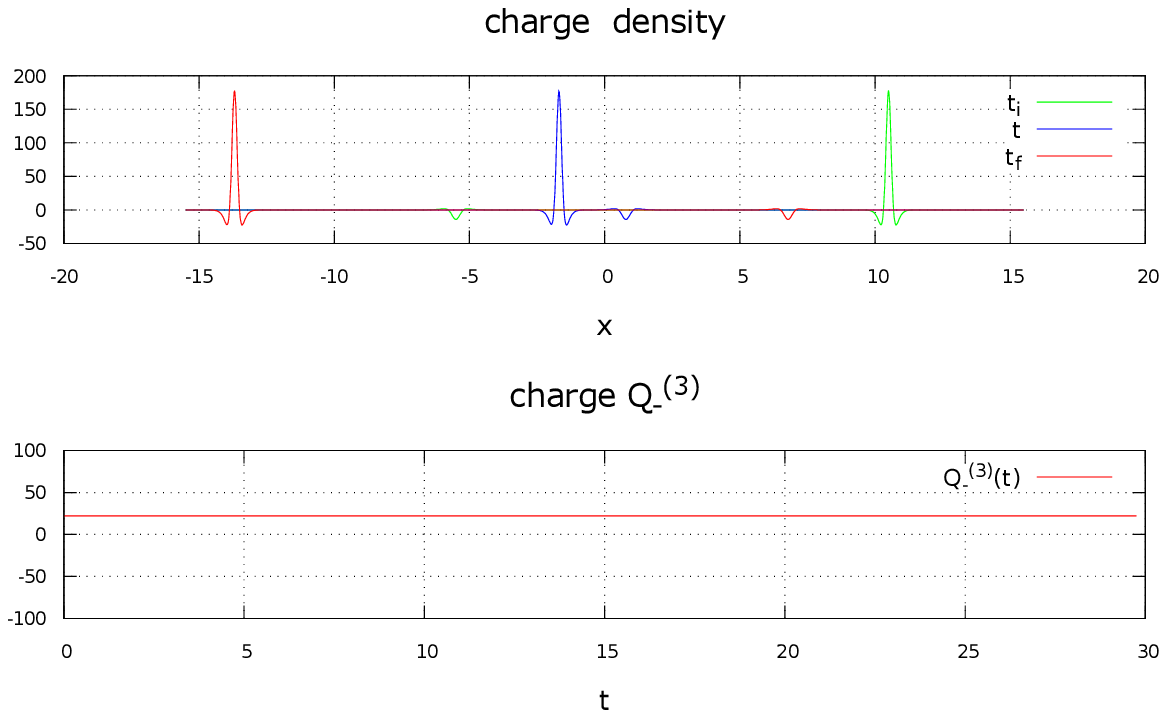} 
\parbox{6in}{\caption{(color online)   Top Fig. shows the charge density of $Q_{-}^{(3)}(t)$ vs $x$ for the kink-antikink collision with $v_2=0.4,\,v_1=-0.8$ for $q = 2.01$, with initial (green), collision (blue) and final (red) densities of the kink-kink scattering. Bottom Fig. shows the  conserved charge $Q_{-}^{(3)}(t)$ of the kink-antikink solution.}}
\end{figure}

\begin{figure}
\centering
\label{fig12} 
\includegraphics[width=2cm,scale=6, angle=0,height=6cm]{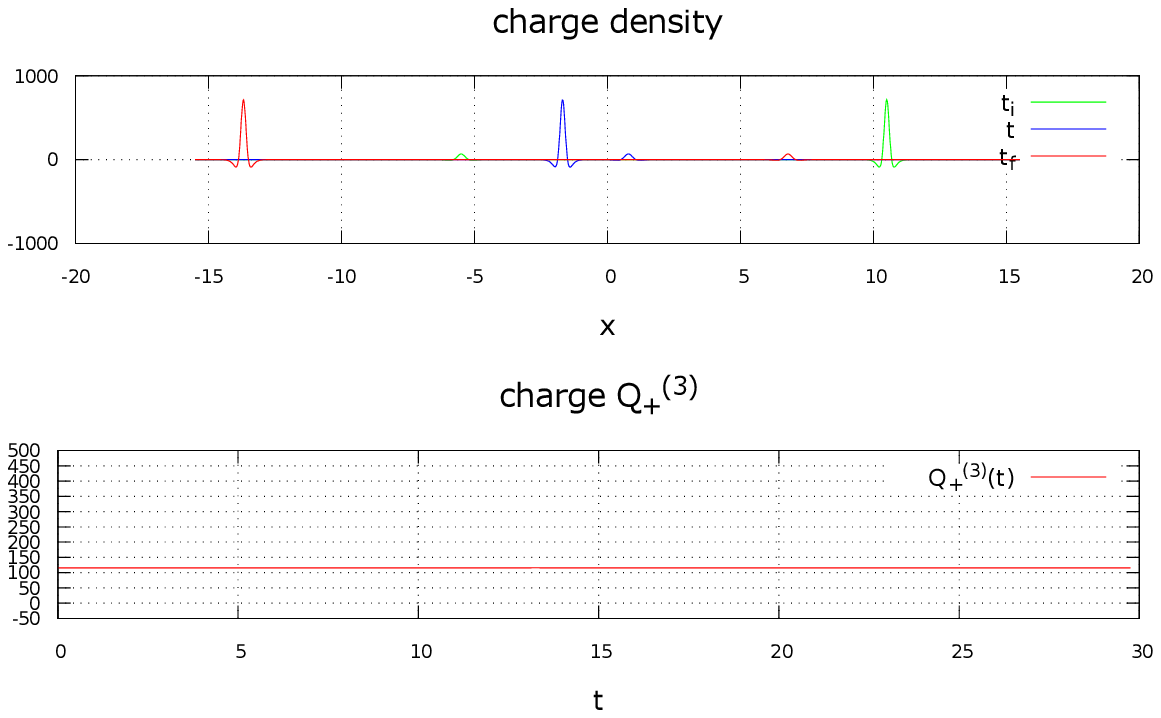}
\parbox{6in}{\caption{(color online)   Top Fig. shows the charge density of $Q_{+}^{(3)}$ vs $x$ for the kink-antikink collision with $v_2=0.4,\,v_1=-0.8$ for $q = 2.01$, with initial (green), collision (blue) and final (red) densities of the kink-kink scattering. Bottom Fig. shows the conserved charge $Q_{+}^{(3)}(t)$ of the kink-antikink solution.}}
\end{figure}

\begin{figure}
\centering
\label{fig13}
\includegraphics[width=3cm,scale=6, angle=0, height=8cm]{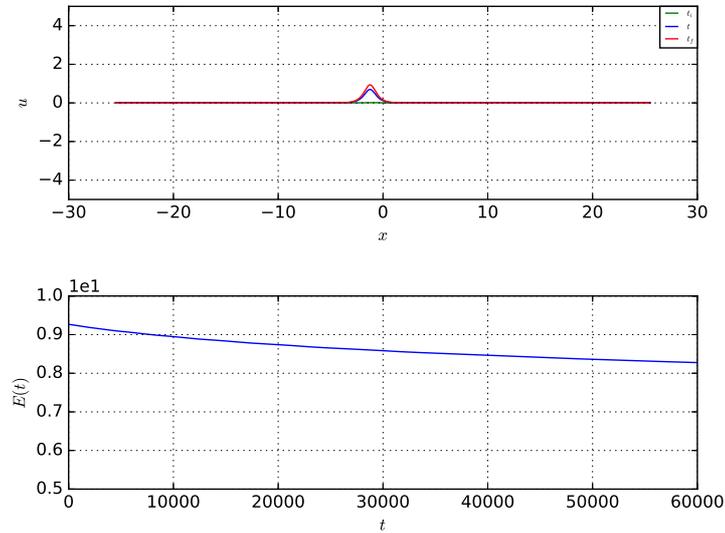}
\parbox{6in}{\caption{(color online) Top Fig. shows the breather oscillation with period $T = 3.87$ for three successive times and $q=1.97$. Bottom Fig. shows the energy vs time. Notice that the energy takes thousands of  units of time to stabilize.}}
\end{figure}

\begin{figure}
\centering
\label{fig14}
\includegraphics[width=3cm,scale=6, angle=0,height=8cm]{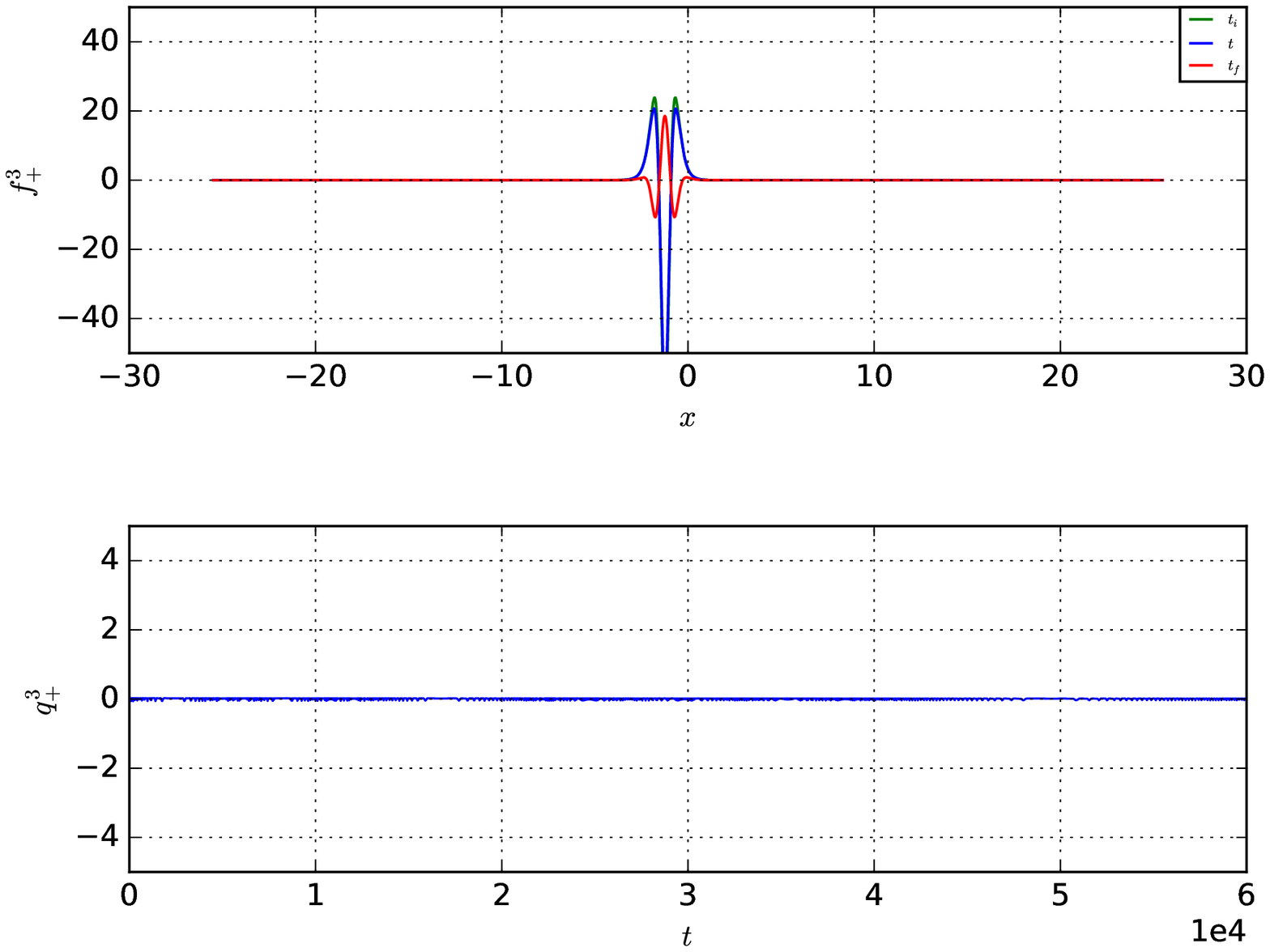} 
\parbox{6in}{\caption{(color online)   Top Fig. shows the charge density vs $x$ corresponding to $Q_{+}^{(3)}(t)$ and the oscillating breather for three successive times with period $T = 3.87$ and $q=1.97$. Bottom Fig. shows the  conserved charge $Q_{+}^{(3)}(t)$ of the breather  in the interval $t=[0\, ,\,  6\times 10^{4}]$.}}
\end{figure}

\begin{figure}
\centering
\label{fig15}
\includegraphics[width=3cm,scale=6, angle=0,height=8cm]{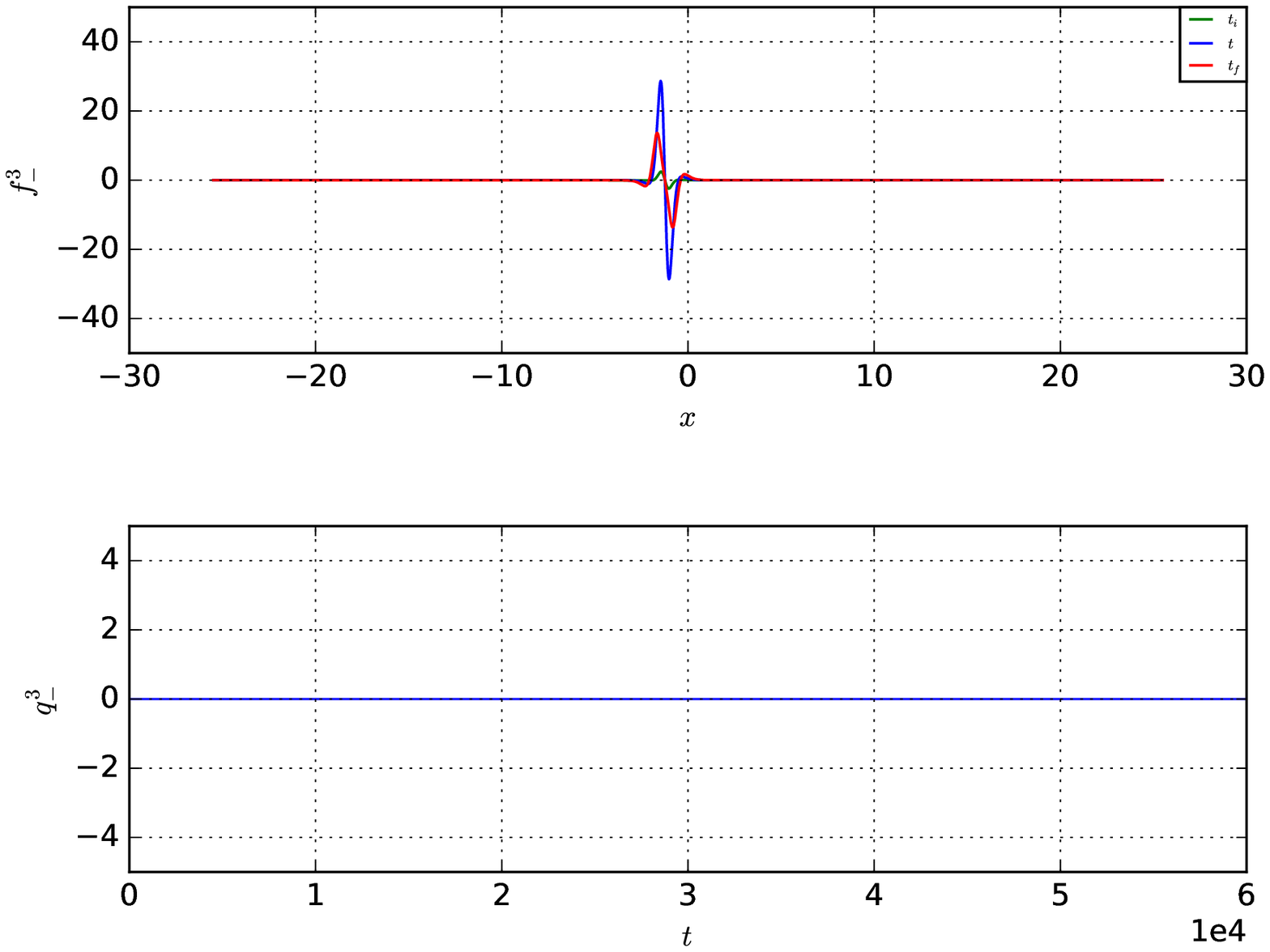} 
\parbox{6in}{\caption{(color online)   Top Fig. shows the charge density vs $x$  corresponding to $Q_{-}^{(3)}(t)$ and the oscillating breather for three successive times with period $T = 3.87$ and $q=1.97$. Bottom Fig. shows the  conserved charge $Q_{-}^{(3)}(t)$ of the breather in the interval $t=[0\, ,\, 6\times 10^{4}]$.}}
\end{figure}

As expected from the symmetry property of the relevant charge density of $Q_{-}^{(3)}$ in (\ref{qq32})-(\ref{qq322}); i.e. it is an odd function for definite parity field configurations, one notices the vanishing of the numerically simulated charge component $Q_{-}^{(3)}$ as presented in the Figs. 2, 5 and 15, corresponding to kink-kink, kink-antikink and breather configurations, respectively. Of course, this implies the vanishing of the related charge $q_{a\,-}^{(3)}$ in (\ref{qq32}). However, for asymmetric kink-kink   (Fig. 7) and kink-antikink (Fig. 10) configurations one gets a non vanishing constant charge component $Q_{-}^{(3)}$ for kink-kink (Fig. 8) and for kink-antikink (Fig. 11) solitons, respectively. These last results for $Q_{-}^{(3)}$ arise from the dynamics of the system and they can not be foreseen from the symmetry considerations discussed above.            
    
Therefore, our numerical simulations show that the charges $q_{a\,\pm}^{(3)}$ are exactly conserved, within  numerical accuracy, as shown in the Figs. 2-3, 5-6, 8-9, 11-12 and 14-15 for the various kink-kink, kink-antikink and breather configurations, respectively, i.e. the anomaly term in (\ref{asymp}) vanishes. In the previous literature \cite{jhep1, cnsns}, by simulating the behaviour of the time-integrated anomalies of type (\ref{asymp}), i.e. $\int dt \int dx \beta^{(3)}$,  these charges have been regarded as merely asymptotically conserved ones. Despite of this fact, here we have established through numerical simulations the exact conservation of these charges associated to the various soliton configurations. It seems to be that the earlier simulations in the literature have been plagued with some numerical artifacts due to the discretization of the higher order derivatives present in the anomaly density and the numerical errors introduced in the space-time $(x,t)$ integration process.

However, more numerical tests (e.g. for multiple $n-$kinks ($n>2$) and wobble-type solutions) and the corresponding analytical results (a formulation of a proper conservation law) are needed in order to establish definitely the exact conservation of the charges $q_{a\,\pm}^{(3)}$. Moreover, it would be a very interesting result if the DSG theories supported wobble-type solutions (configurations of a breather and a kink) as suggested in \cite{jhep1}, with associated higher order conserved charges.

\subsection{Fifth order conserved charges}
\label{sec:fifth}

Following similar reasoning in the simulation of the third order charges above, in order to minimize the fictitious discretization effects into the charge densities, we will make relevant decomposition of the fifth order charges as an addition of certain components. The explicit form of the charges $q^{(5)}_{a\,\pm}$ (\ref{fifthpm}) and their components, written in laboratory coordinates,  become
\br
q^{(5)}_{a\,+} &=& q^{(5;0)}_{a\,+} + \frac{d}{dt}q^{(5;1)}_{a\,+}+ \frac{d^2}{dt^2}q^{(5;2)}_{a\,+}  \label{qq5p}\\
\nonumber
q^{(5;0)}_{a\,+} &\equiv & \frac{1}{8} \int  dx \, \Big\{-12 [(w^{(0,1)})^4+6 (w^{(0,1)})^2 (w^{(1,0)})^2+(w^{(1,0)})^4] +[(w^{(0,1)})^2+(w^{(1,0)})^2]\times \\
&& [(w^{(0,1)})^4+14 (w^{(0,1)})^2 (w^{(1,0)})^2+(w^{(1,0)})^4]+12 [4 (w^{(1,1)})^2+(w^{(0,2)}+w^{(2,0)})^2] -\\ \nonumber
&& 5 \Big[ 8 w^{(0,1)} w^{(1,0)} w^{(1,1)} (w^{(0,2)}+w^{(2,0)})+(w^{(0,1)})^2 [4 (w^{(1,1)})^2+(w^{(0,2)}+w^{(2,0)})^2]+\\
\nonumber
&& (w^{(1,0)})^2 [4 (w^{(1,1)})^2+(w^{(0,2)}+w^{(2,0)})^2] \Big] +
\\ \nonumber
&& 2 V \Big[3 (w^{(0,1)})^4+18 (w^{(0,1)})^2 (w^{(1,0)})^2+3 (w^{(1,0)})^4+4 (w^{(1,1)})^2+(w^{(0,2)}+w^{(2,0)})^2+\\ \nonumber
&&4 w^{(0,1)} (w^{(0,3)}+3 w^{(2,1)})+4 w^{(1,0)} (3 w^{(1,2)}+w^{(3,0)}) \Big]+ \label{qq5p0} \\ 
\nonumber
&& \frac{1}{2} [(w^{(0,3)}+3 w^{(2,1)})^2+(3 w^{(1,2)}+w^{(3,0)})^2]-V^{(1)} (w^{(0,4)}+6 w^{(2,2)}+w^{(4,0)}) \Big\} \nonumber \\
\label{qq5p1}
q^{(5;1)}_{a\,+} & \equiv  &\frac{1}{8} \int  dx \, \Big\{- 4 V [ 2 w^{(1,0)} w^{(1,1)}  + w^{(0,1)} (w^{(0,2)}+w^{(2,0)}) ]+ V^{(1)} (3 w^{(2,1)}+w^{(0,3)}) \Big\}  \\
\label{qq5p2}
q^{(5;2)}_{a\,+} & \equiv  & \frac{1}{2} \int dx\, \Big[ (w^{(0,1)})^2+(w^{(1,0)})^2  \Big]
\er 
and 
\br
q^{(5)}_{a\,-} &=& q^{(5;0)}_{a\,-} + \frac{d}{dt}q^{(5;1)}_{a\,-}+ \frac{d^2}{dt^2}q^{(5;2)}_{a\,-}  \label{qq5m}\\
\nonumber
q^{(5;0)}_{a\,-} &\equiv & \frac{1}{8} \int  dx \, \Big\{48 w^{(0,1)} w^{(1,0)} [(w^{(0,1)})^2+(w^{(1,0)})^2] -2 w^{(0,1)} w^{(1,0)} [3 (w^{(0,1)})^2+(w^{(1,0)})^2] \times \\ \nonumber
&& [(w^{(0,1)})^2+3 (w^{(1,0)})^2] - 48 w^{(1,1)} (w^{(0,2)}+w^{(2,0)}) +\\ \nonumber
&& 10 [2 w^{(1,0)} w^{(1,1)}+w^{(0,1)} (w^{(0,2)}+w^{(2,0)})] [2 w^{(0,1)} w^{(1,1)}+w^{(1,0)} (w^{(0,2)}+w^{(2,0)})]-\\
&& 8 V \Big[3 (w^{(0,1)})^3 w^{(1,0)}+w^{(1,1)} (w^{(0,2)}+w^{(2,0)})+w^{(1,0)} (w^{(0,3)}+3 w^{(2,1)})+ \label{qq5m0} \\ 
\nonumber
&& w^{(0,1)} (3 ((w^{(1,0)})^3+w^{(1,2)})+w^{(3,0)})\Big]  -   (w^{(0,3)}+3 w^{(2,1)}) (3 w^{(1,2)}+w^{(3,0)}) +\\
&& 4 V^{(1)} [w^{(1,3)}+w^{(3,1)}] \Big\} \nonumber \\
\label{qq5m1}
q^{(5;1)}_{a\,-} & \equiv  &\frac{1}{8} \int  dx \, \Big\{  4 V [ 2 w^{(0,1)} w^{(1,1)}   + w^{(1,0)} (w^{(0,2)}+w^{(2,0)}) ]- V^{(1)} (3 w^{(1,2)}+w^{(3,0)}) \Big\} \\
\label{qq5m2}
q^{(5;2)}_{a\,-} & \equiv  & -\frac{1}{2} \int dx\,  w^{(0,1)}  w^{(1,0)}  \\
&=&  -\frac{1}{2}  P\nonumber
\er 
where the notation ($w^{(a , b )} \equiv \frac{\pa^{a + b}}{\pa x^a \pa t^b} w(x,t)$) has been used.  Notice that the charges $q^{(5)}_{a\,\pm}$ have been decomposed as an addition, with successively increasing time-derivatives, of the charge components $q^{(5;a)}_{a\,\pm},\,a=0,1,2$.
 
These components exhibit some remarkable properties. The densities of the components exhibit definite parities provided that the field $w$ and the potential $V$ exhibit the symmetry (\ref{px})-(\ref{evenpot}). In fact, for soliton configurations  with space-reflection symmetry each charge density in the integrands of   $q^{(5;a)}_{a\,+},\,a=0,1,2$ in (\ref{qq5p})-(\ref{qq5p2}) will be even, whereas the integrands of   $q^{(5;a)}_{a\,-},\,a=0,1,2$  in  (\ref{qq5m})-(\ref{qq5m2}) will be odd functions. 

So, we will compute numerically the relevant components $q^{(5;a)}_{\pm}\,( a=0,1,2,3,4)$ and sum them up in order to compute the fifth order charges (\ref{fifthpm}). As in the third order case above, by symmetry arguments only, one expects the vanishing of the charge  $q_{-}^{(5)}$ for all soliton configurations possessing definite parity under space-reflection. This will also be verified through our numerical simulations.

In the next Figs. 16-19 we present the simulations of the charges $q_{\pm}^{(5)}(t)$, their relevant  components and densities, respectively,  for the kink-kink collision. We consider the  charges for definite parity (kink-kink with odd parity) in Figs. 16-17, and the charges for asymmetric kink-kink configuration, in Figs. 18-19.    

\begin{figure}[htbp!]
\begin{minipage}[b]{0.475\linewidth} 
\centering
\includegraphics[scale=0.54]{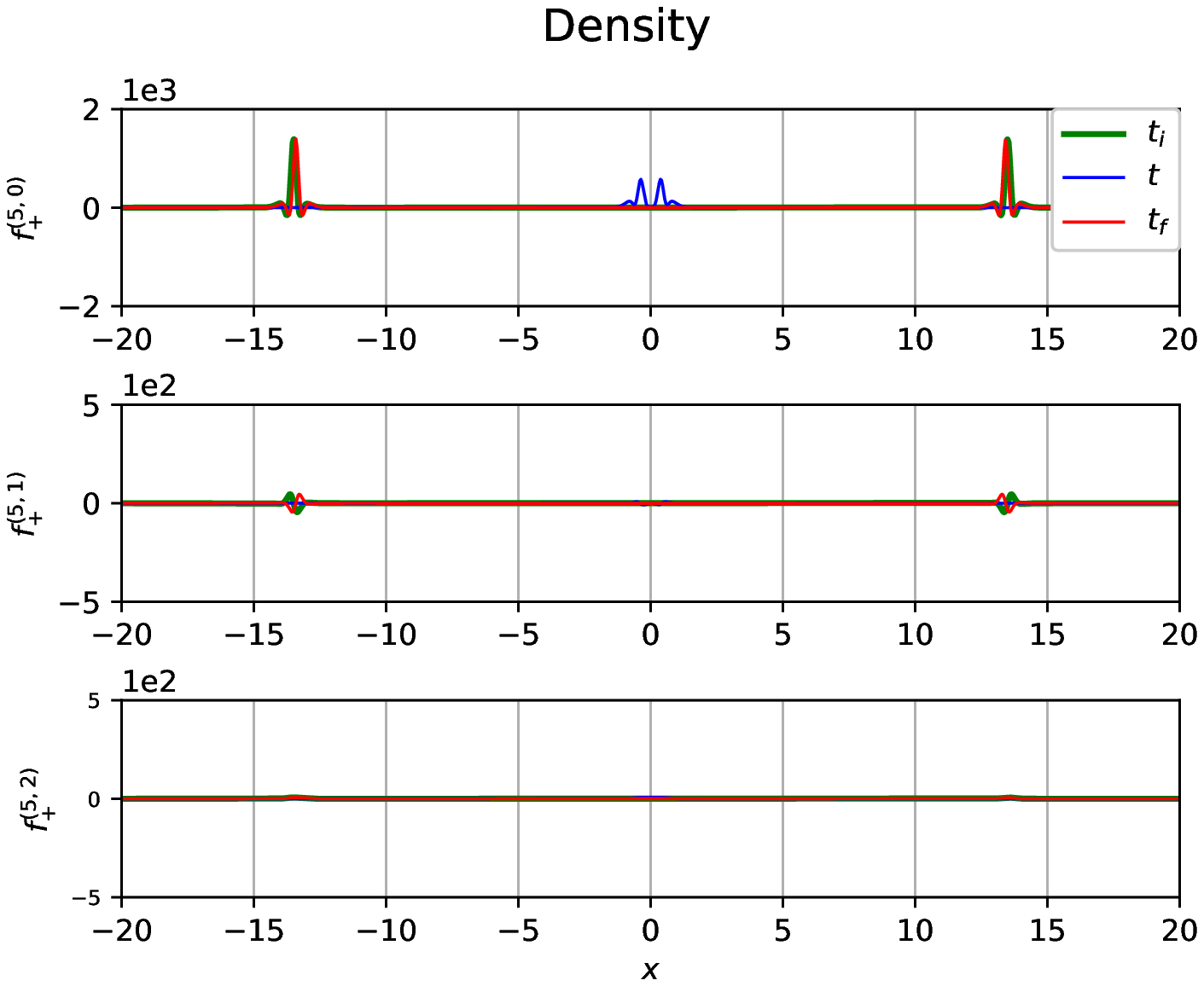}
\end{minipage}
\hspace{0.000001cm} 
\begin{minipage}[b]{0.475\linewidth}
\centering
\includegraphics[scale=0.54]{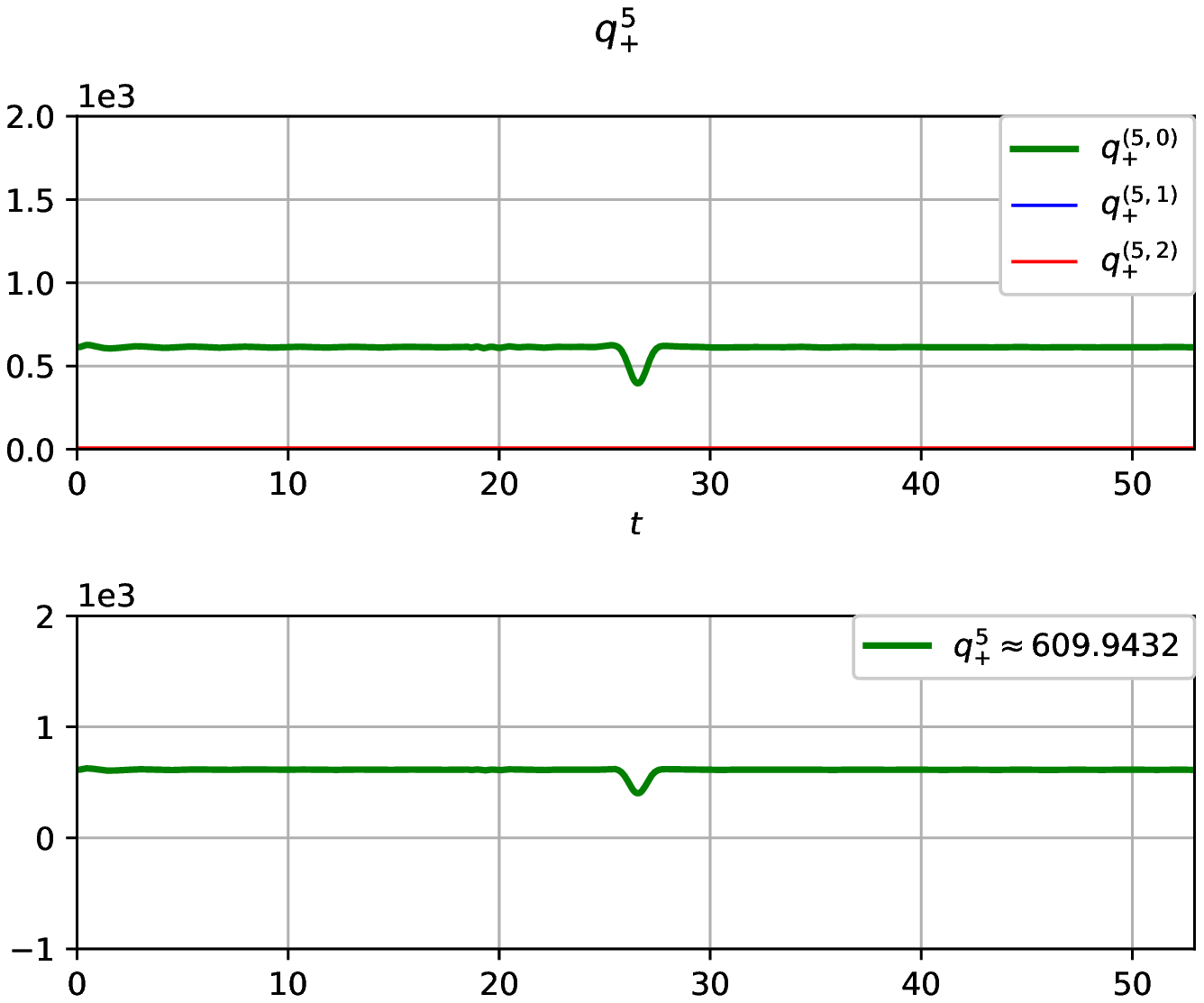}
\end{minipage}
\caption{ Left Figs. show the charge densities  $f_{+}^{(5, a)},\, a=0,1,2$ vs $x$ for the kink-kink collision with $v_2=v_1=-0.5$ for $q = 1.97$, with initial (green), collision (blue) and final (red) densities of the kink-kink scattering. Right Figs. show the conserved charge componensts $q_{+}^{(5,a)}(t),\,a=0,1,2$ and the total charge $q_{+}^(5)(t)$ of the kink-kink solution. }\label{fig16}
\end{figure}

\begin{figure}[htbp!]
\begin{minipage}[b]{0.475\linewidth} 
\centering
\includegraphics[scale=0.54]{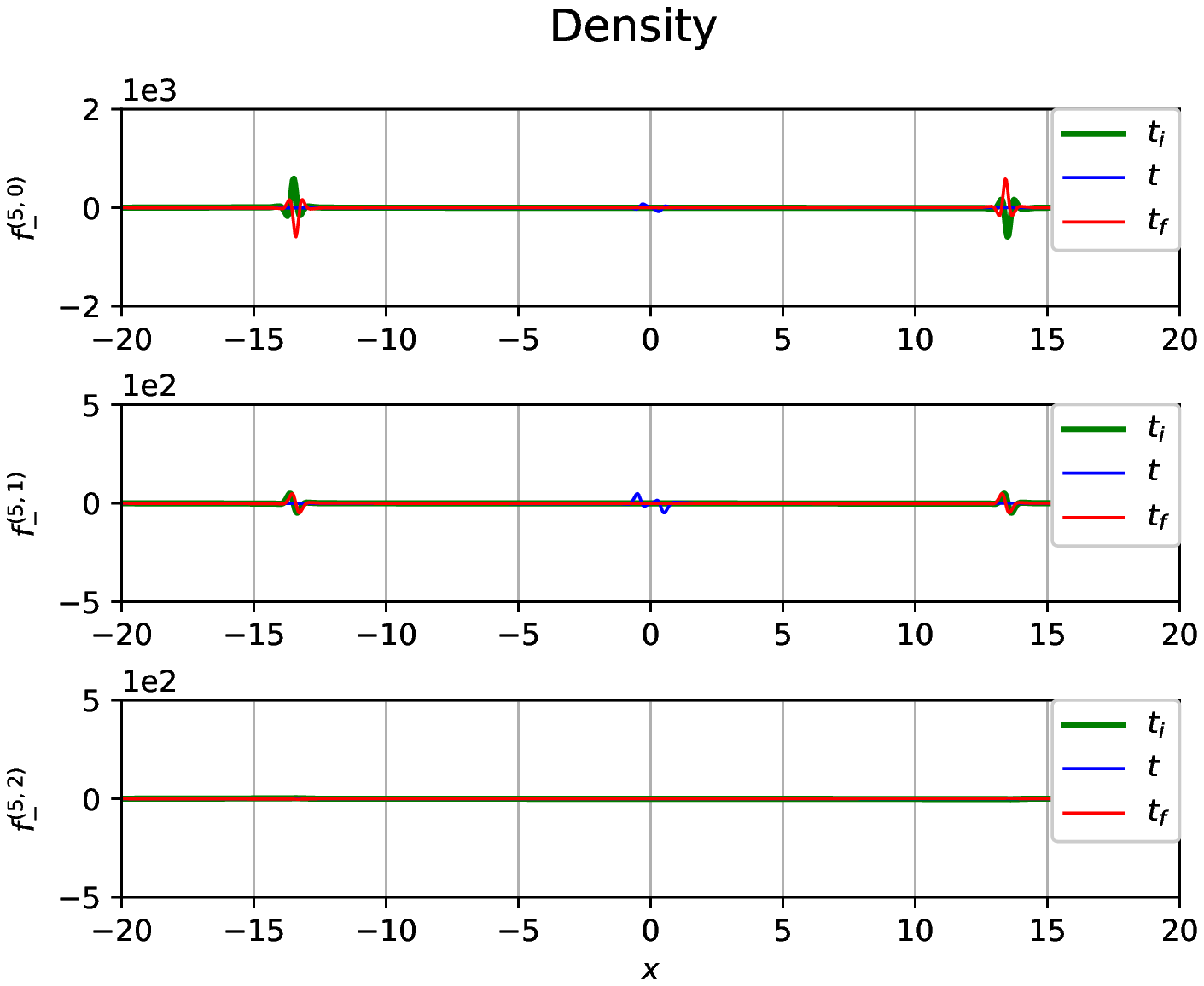}
\end{minipage}
\hspace{0.000001cm} 
\begin{minipage}[b]{0.475\linewidth}
\centering
\includegraphics[scale=0.54]{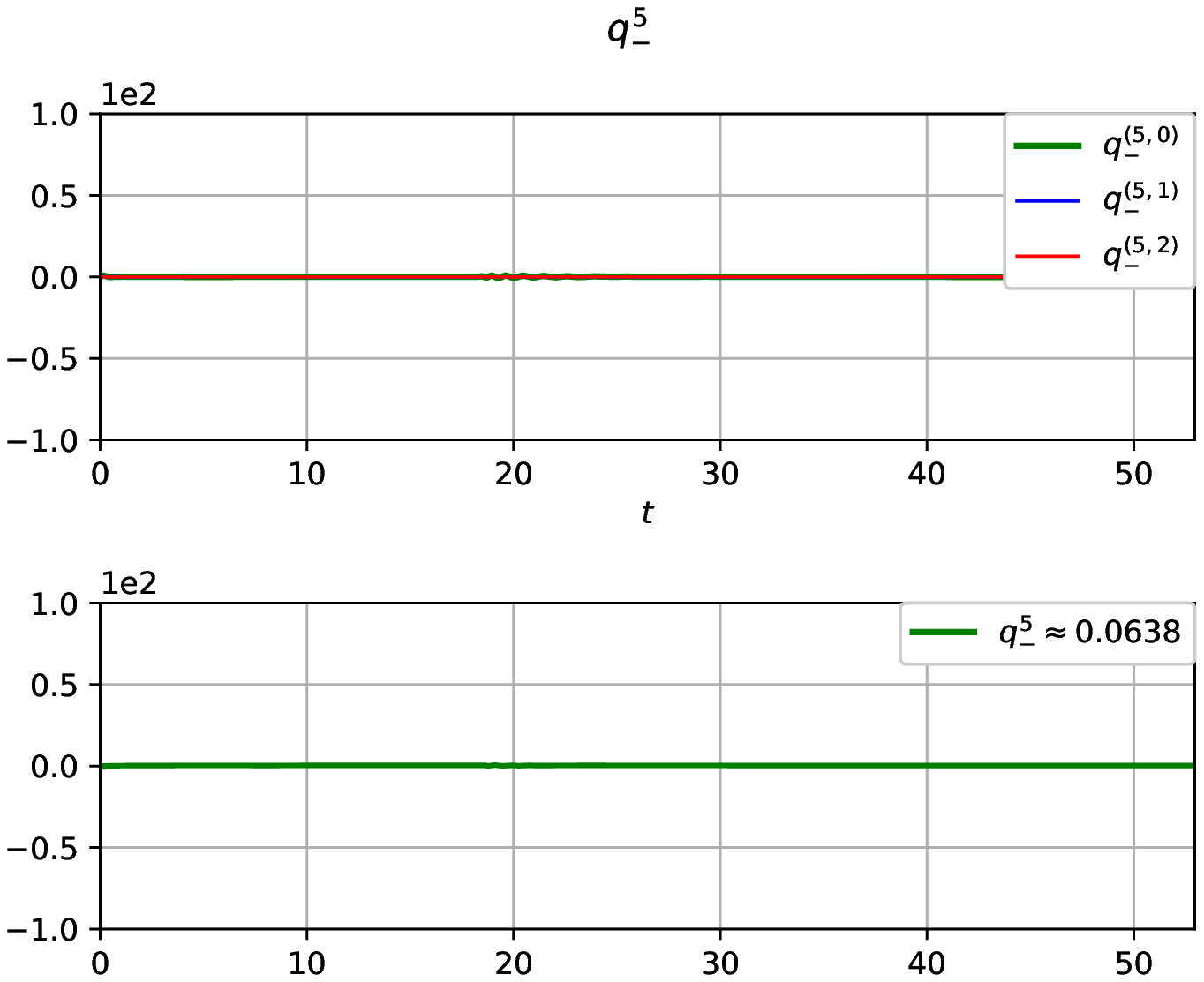}
\end{minipage}
\caption{ Left Figs. show the charge densities  $f_{-}^{(5, a)},\, a=0,1,2$ vs $x$ for the kink-kink collision with $v_2=v_1=-0.5$ for $q = 1.97$, with initial (green), collision (blue) and final (red) densities of the kink-kink scattering. Right Figs. show the conserved charge components $q_{-}^{(5,a)}(t),\,a=0,1,2$ and the total charge $q_{-}^(5)(t)$ of the kink-kink solution. }\label{fig17}
\end{figure}

\begin{figure}[htbp!]
\begin{minipage}[b]{0.475\linewidth} 
\centering
\includegraphics[scale=0.54]{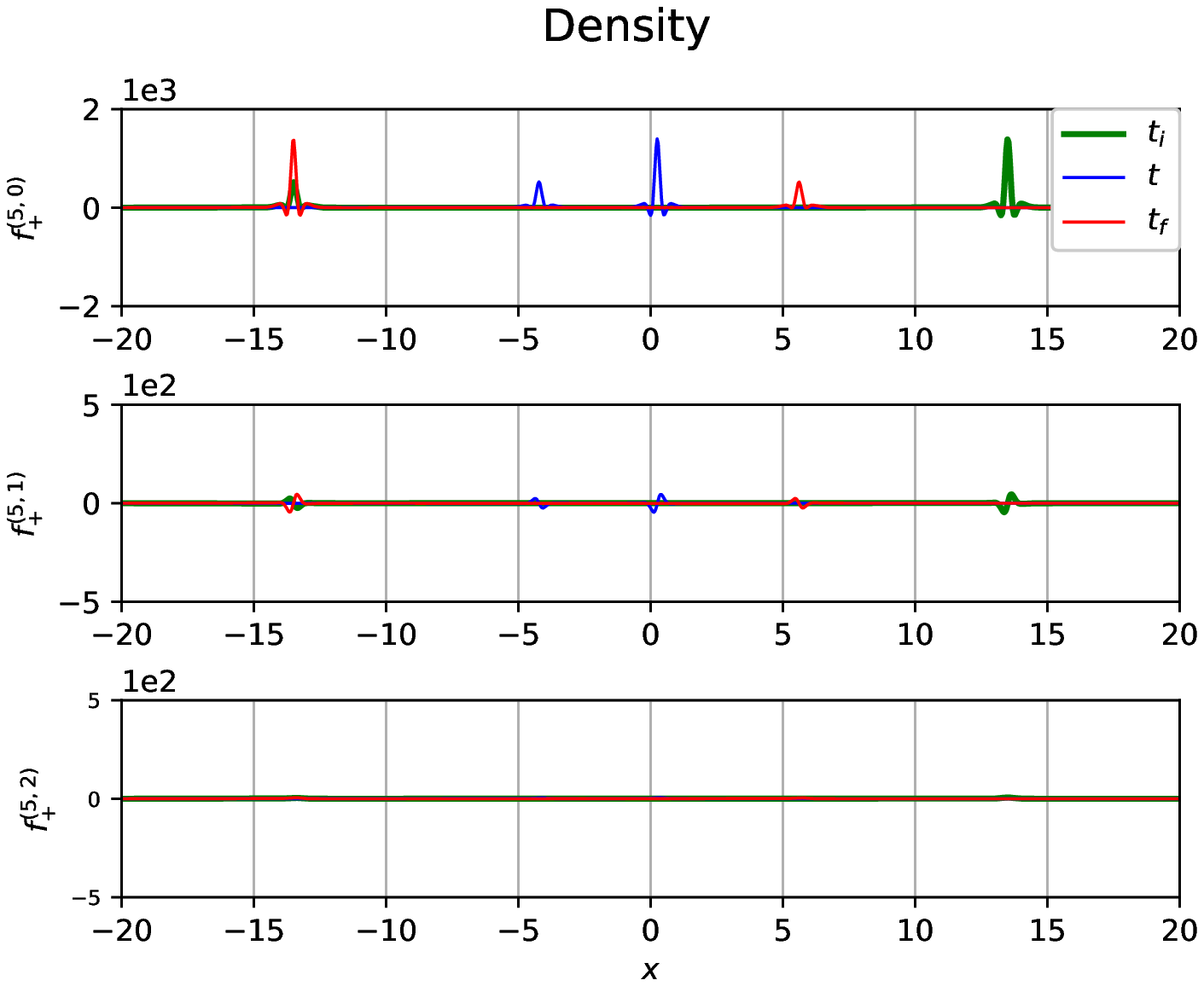}
\end{minipage}
\hspace{0.000001cm} 
\begin{minipage}[b]{0.475\linewidth}
\centering
\includegraphics[scale=0.54]{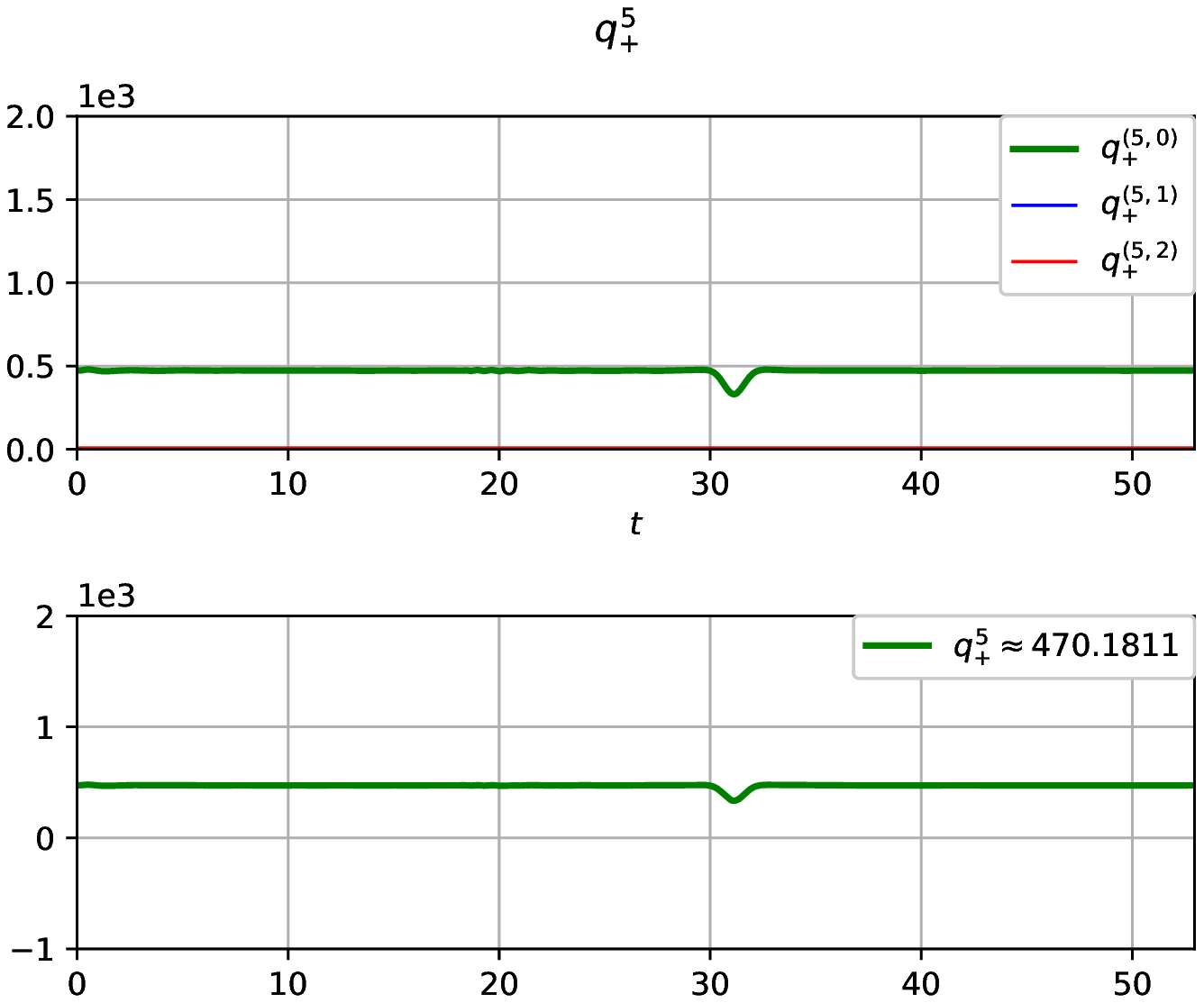}
\end{minipage}
\caption{Left Figs. show the charge densities  $f_{+}^{(5, a)},\, a=0,1,2$ vs $x$ for the kink-kink collision with $v_2=0.35, v_1=-0.5$ for $q = 1.97$, with initial (green), collision (blue) and final (red) densities of the kink-kink scattering. Right Figs. show the conserved charge components $q_{+}^{(5,a)}(t),\,a=0,1,2$ and the total charge $q_{+}^(5)(t)$ of the kink-kink solution.}\label{fig18}
\end{figure}

\begin{figure}[htbp!]
\begin{minipage}[b]{0.475\linewidth} 
\centering
\includegraphics[scale=0.54]{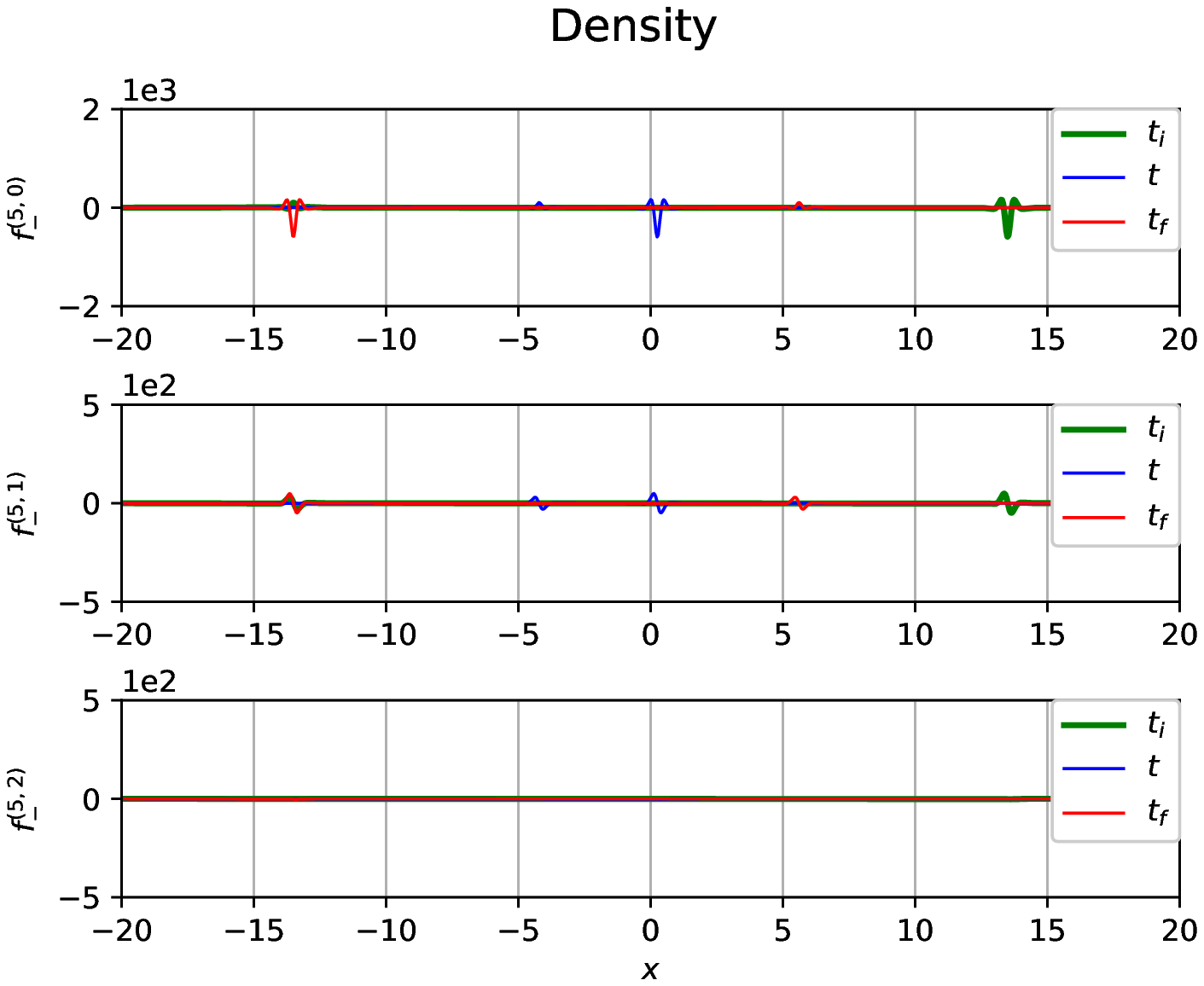}
\end{minipage}
\hspace{0.000001cm} 
\begin{minipage}[b]{0.475\linewidth}
\centering
\includegraphics[scale=0.54]{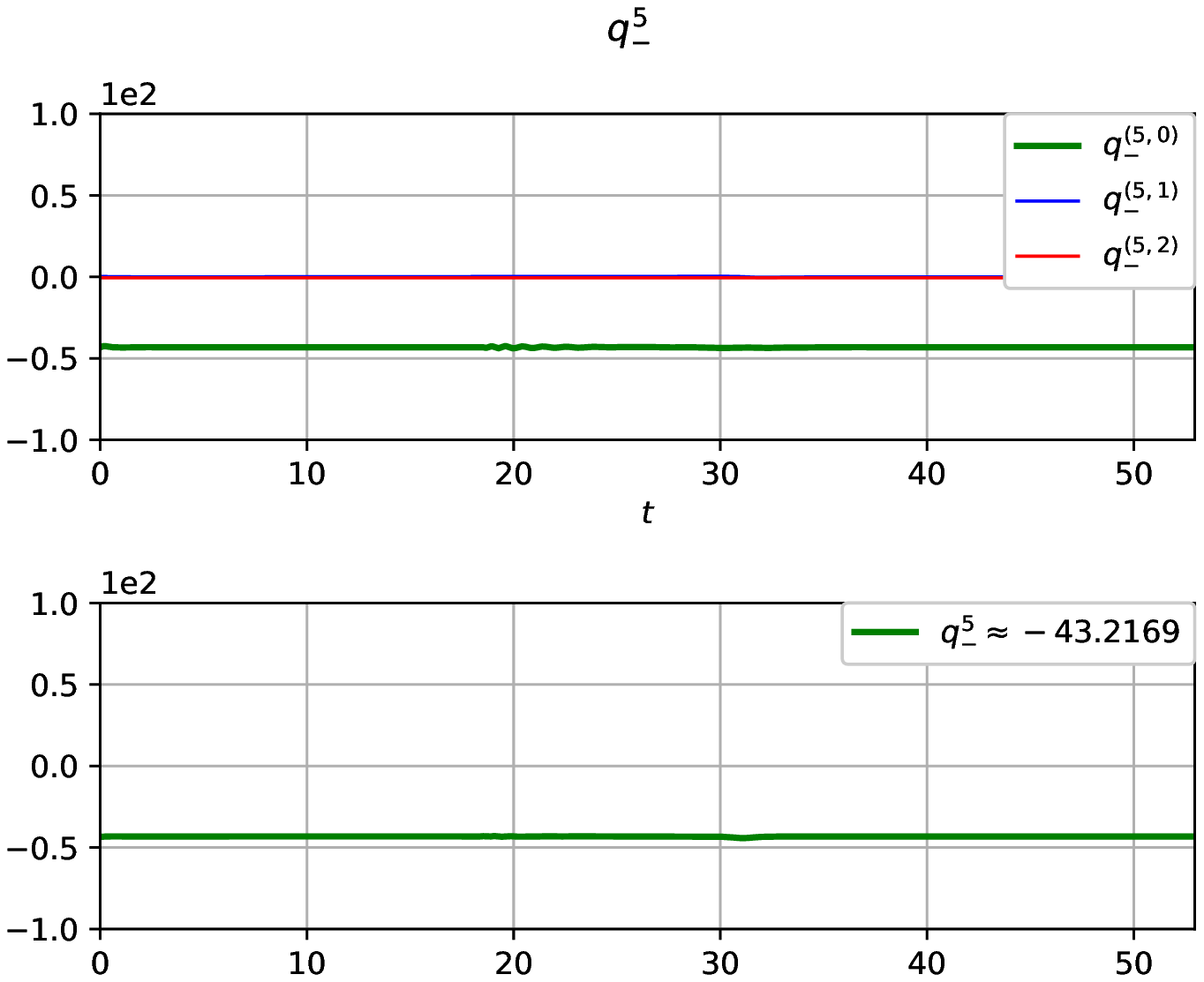}
\end{minipage}
\caption{Left Figs. show the charge densities  $f_{-}^{(5, a)},\, a=0,1,2$ vs $x$ for the kink-kink collision with $v_2=0.35, v_1=-0.5$ for $q = 1.97$, with initial (green), collision (blue) and final (red) densities of the kink-kink scattering. Right Figs. show the conserved charge componensts $q_{-}^{(5,a)}(t),\,a=0,1,2$ and the total charge $q_{-}^(5)(t)$ of the kink-kink solution.}\label{fig19}
\end{figure}

Similarly, in the Figs. 20-23 we present the simulations of the charges $q_{\pm}^{(5)}(t)$, their relevant  components and densities, respectively,  for the kink-antikink collision. We consider the  charges for definite parity (kink-antikink with even parity) in Figs. 20-21, and the charges for asymmetric kink-antikink configuration, in Figs. 22-23.    

\begin{figure}[htbp!]
\begin{minipage}[b]{0.475\linewidth} 
\centering
\includegraphics[scale=0.54]{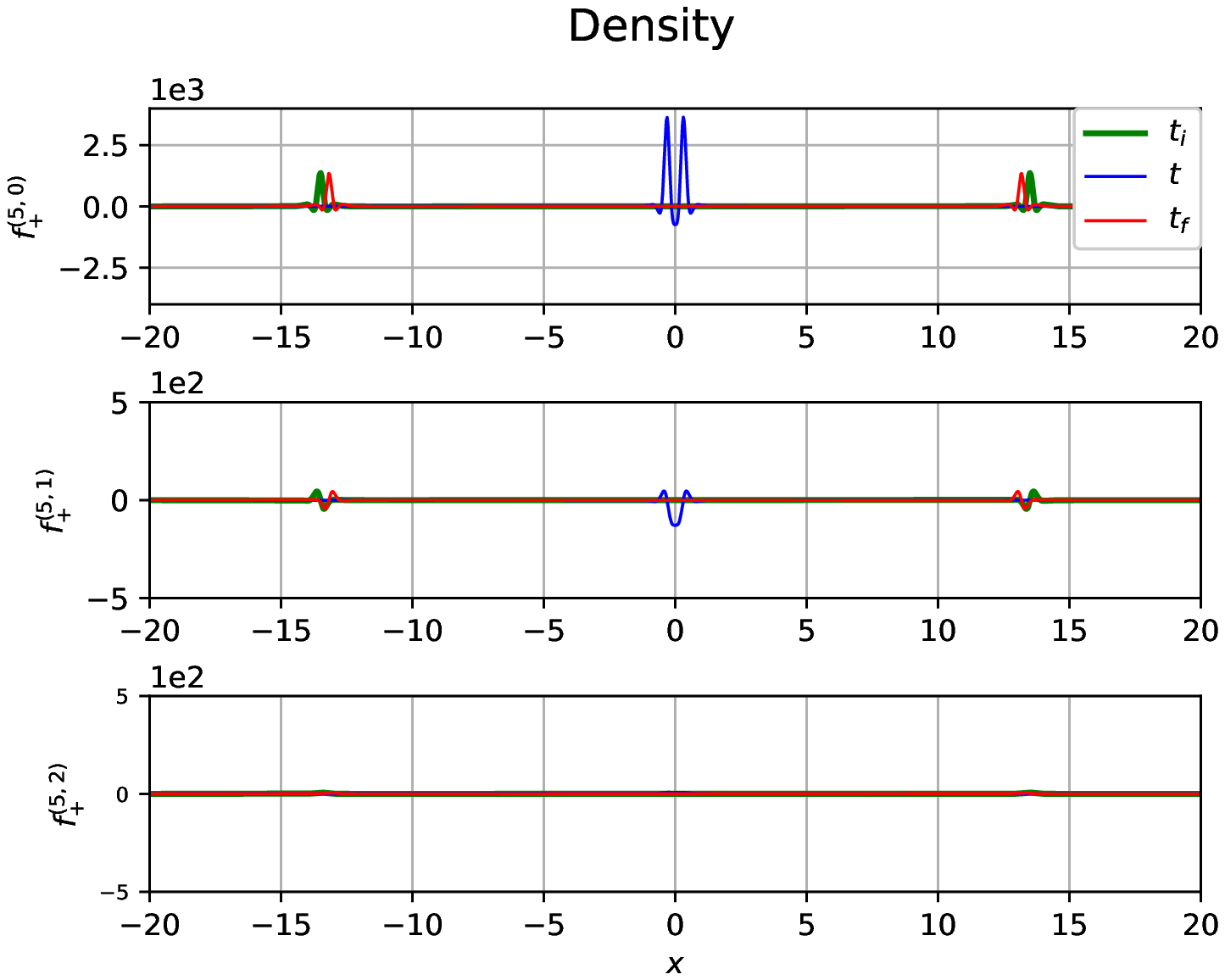}
\end{minipage}
\hspace{0.000001cm} 
\begin{minipage}[b]{0.475\linewidth}
\centering
\includegraphics[scale=0.54]{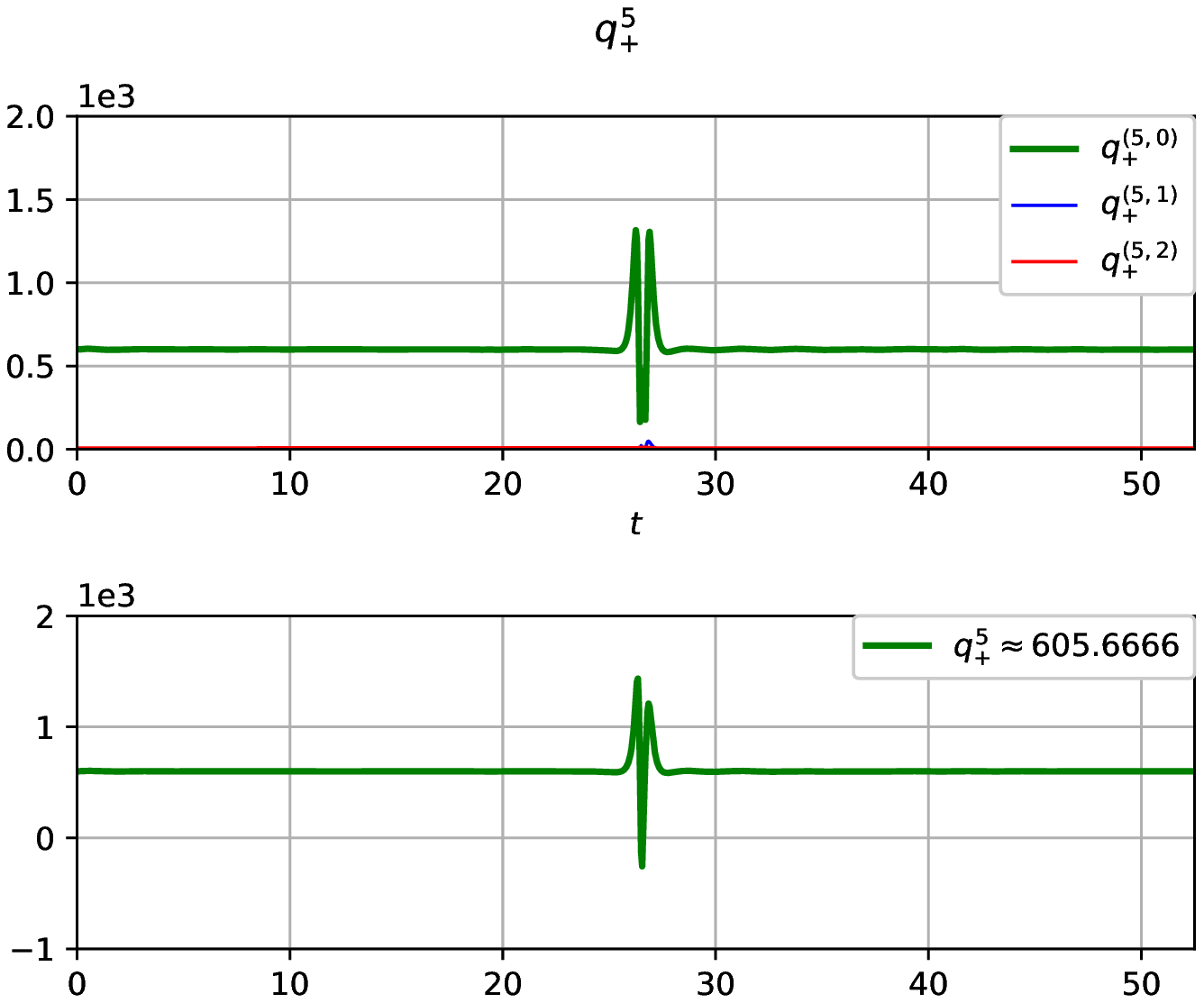}
\end{minipage}
\caption{ Left Figs. show the charge densities  $f_{+}^{(5, a)},\, a=0,1,2$ vs $x$ for the kink-antikink collision with $v_2=v_1=-0.5$ for $q = 2.01$, with initial (green), collision (blue) and final (red) densities of the kink-antikink scattering. Right Figs. show the conserved charge components $q_{-}^{(5,a)}(t),\,a=0,1,2$ and the total charge $q_{-}^(5)(t)$ of the  kink-antikink solution. }\label{fig20}
\end{figure}

\begin{figure}[htbp!]
\begin{minipage}[b]{0.475\linewidth} 
\centering
\includegraphics[scale=0.54]{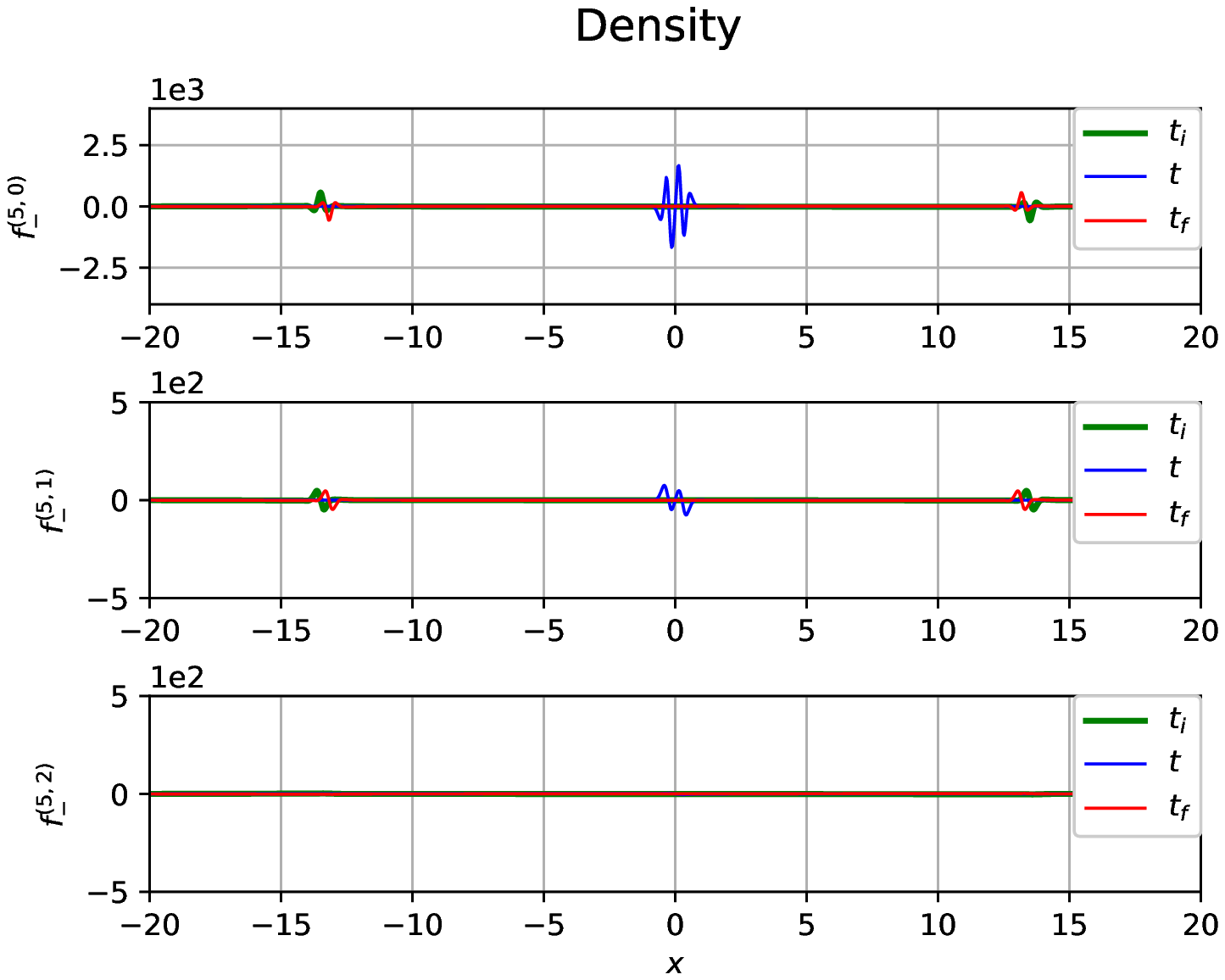}
\end{minipage}
\hspace{0.000001cm} 
\begin{minipage}[b]{0.475\linewidth}
\centering
\includegraphics[scale=0.54]{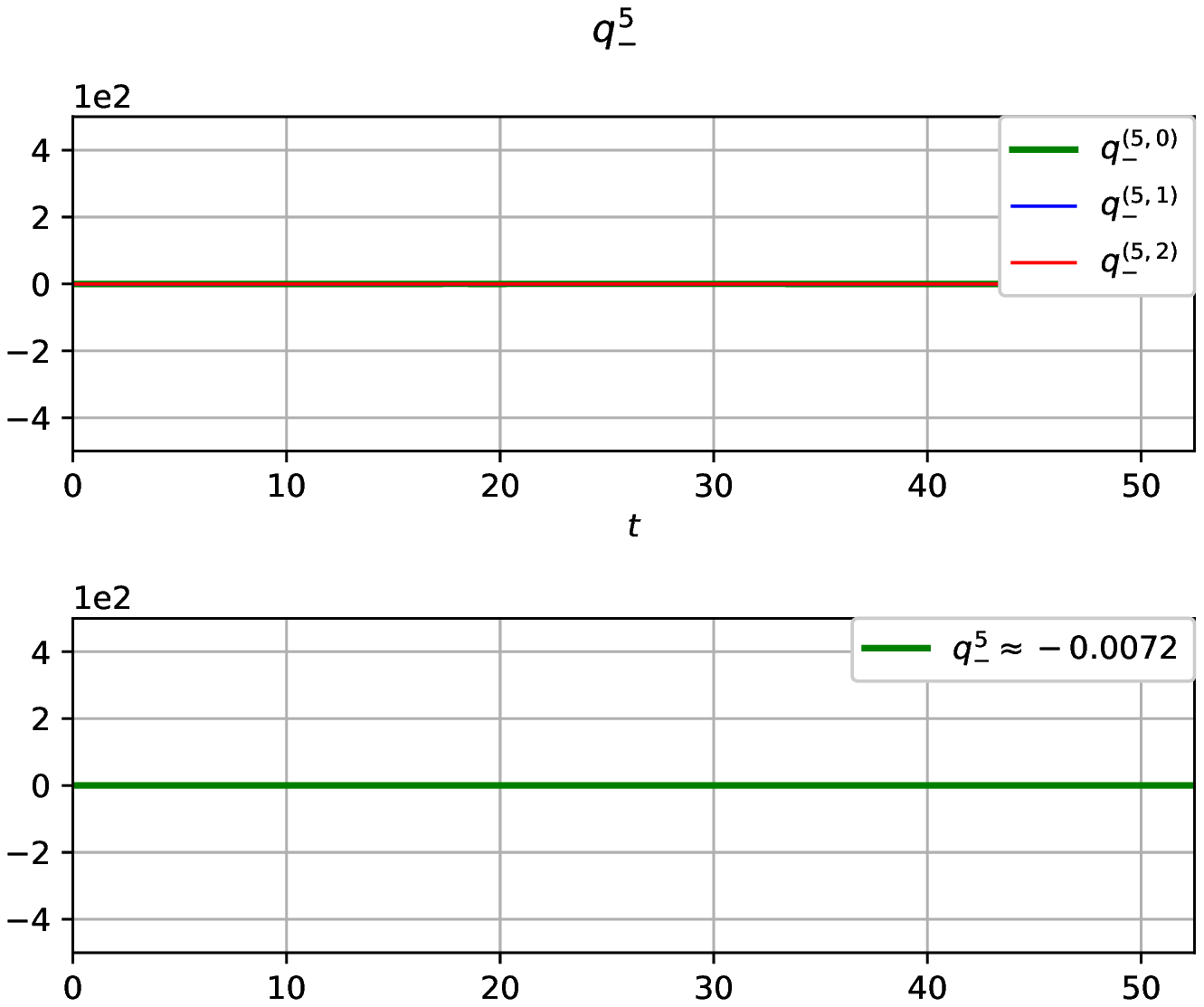}
\end{minipage}
\caption{  Left Figs. show the charge densities  $f_{-}^{(5, a)},\, a=0,1,2$ vs $x$ for the kink-antikink collision with $v_2=v_1=-0.5$ for $q = 2.01$, with initial (green), collision (blue) and final (red) densities of the kink-antikink scattering. Right Figs. show the conserved charge components $q_{-}^{(5,a)}(t),\,a=0,1,2$ and the total charge $q_{-}^(5)(t)$ of the  kink-antikink solution. }\label{fig21}
\end{figure}

\begin{figure}[htbp!]
\begin{minipage}[b]{0.475\linewidth} 
\centering
\includegraphics[scale=0.54]{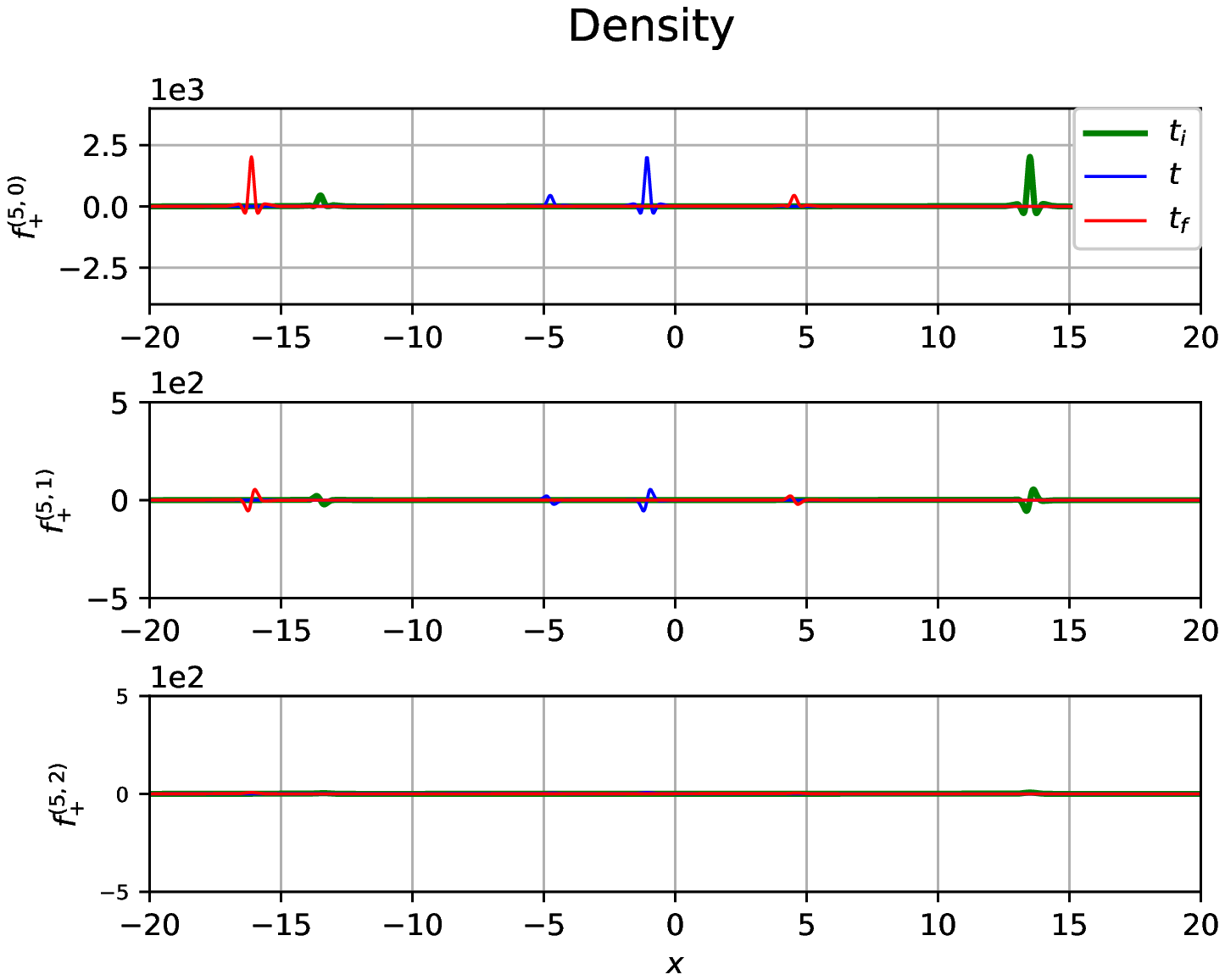}
\end{minipage}
\hspace{0.000001cm} 
\begin{minipage}[b]{0.475\linewidth}
\centering
\includegraphics[scale=0.54]{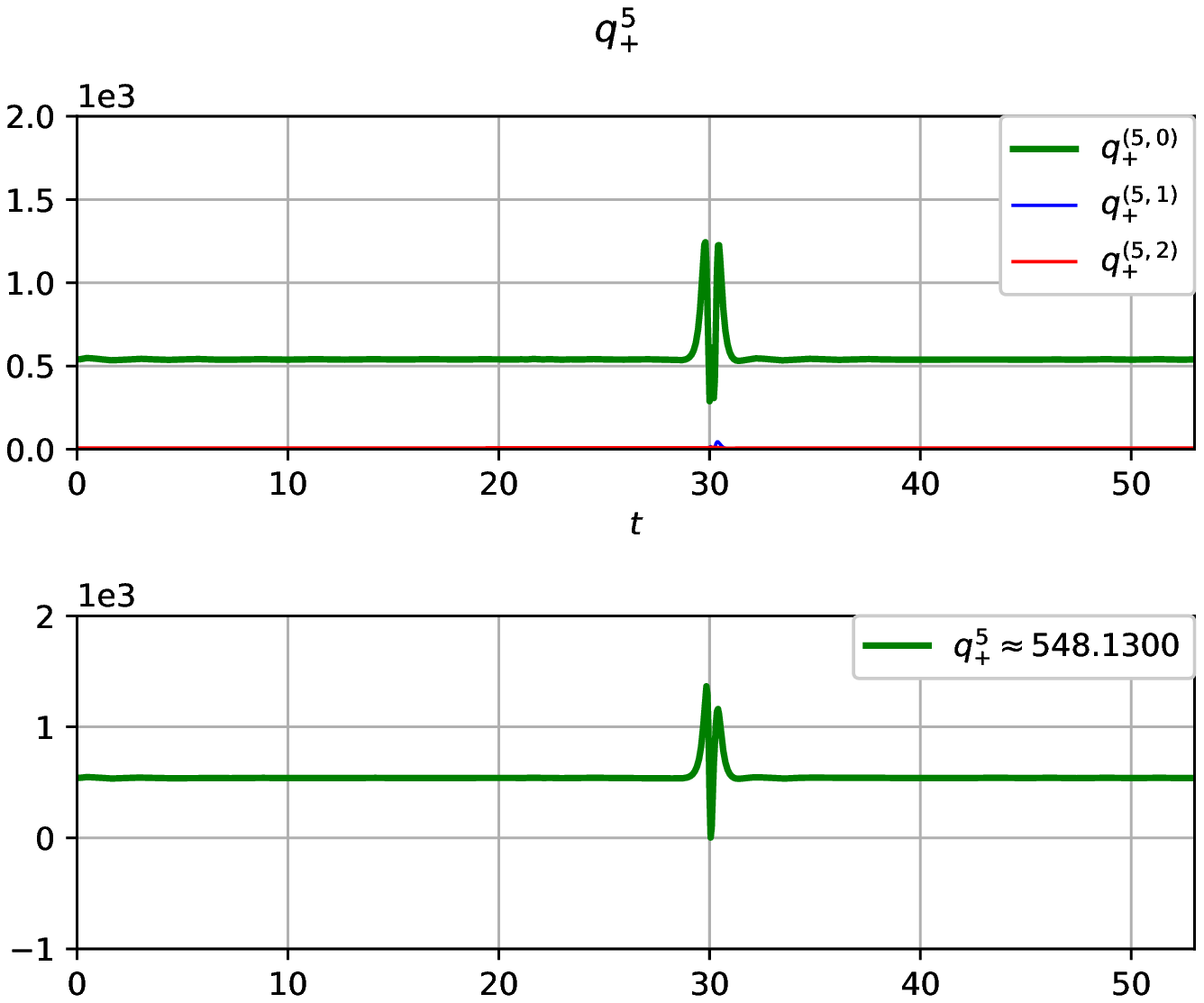}
\end{minipage}
\caption{  Left Figs. show the charge densities  $f_{+}^{(5, a)},\, a=0,1,2$ vs $x$ for the kink-antikink collision with $v_2=0.33, v_1=-0.55$ for $q = 2.015$, with initial (green), collision (blue) and final (red) densities of the kink-antikink scattering. Right Figs. show the conserved charge components $q_{+}^{(5,a)}(t),\,a=0,1,2$ and the total charge $q_{+}^(5)(t)$ of the  kink-antikink solution.}\label{fig22}
\end{figure}

\begin{figure}[htbp!]
\begin{minipage}[b]{0.475\linewidth} 
\centering
\includegraphics[scale=0.54]{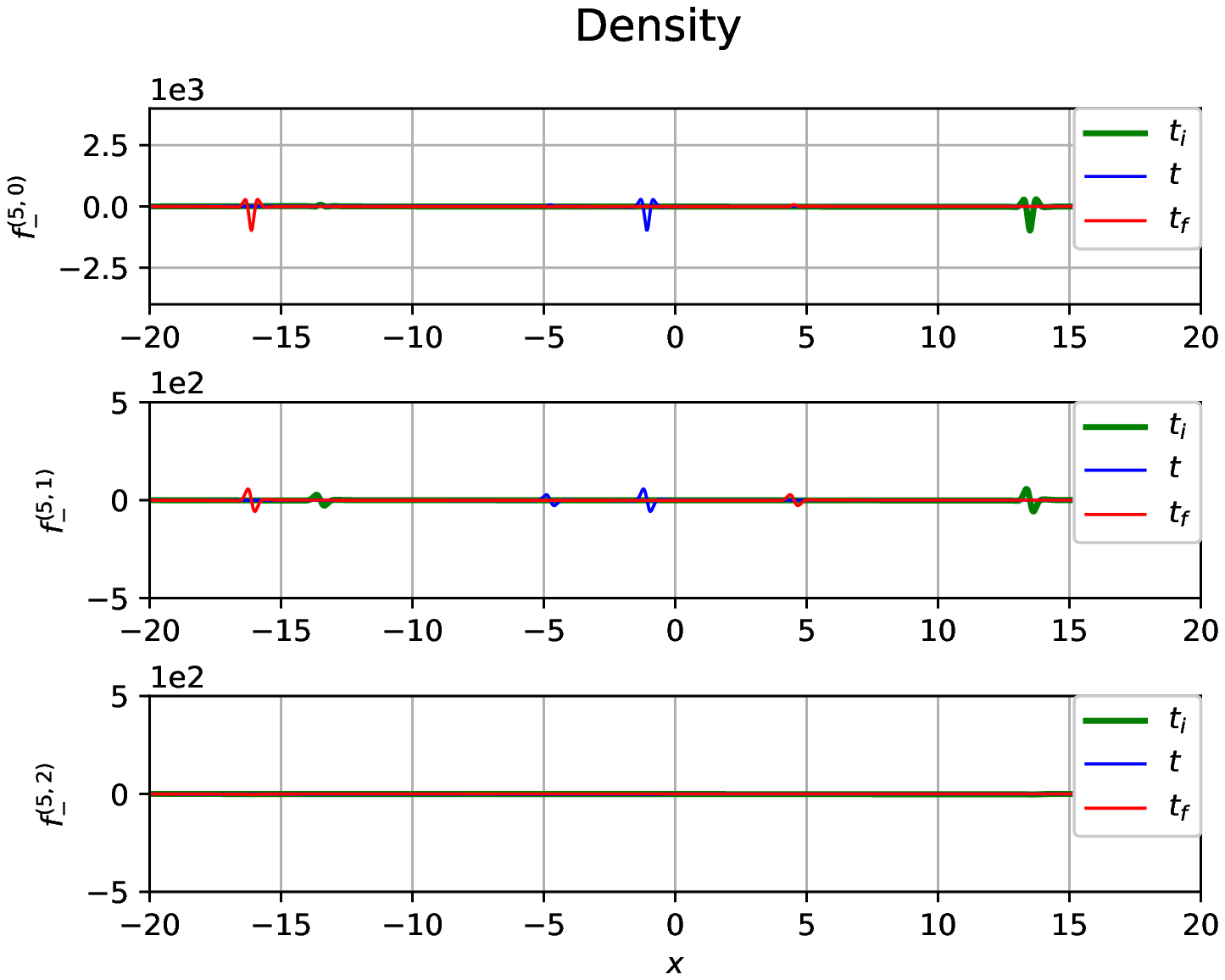}
\end{minipage}
\hspace{0.000001cm} 
\begin{minipage}[b]{0.475\linewidth}
\centering
\includegraphics[scale=0.54]{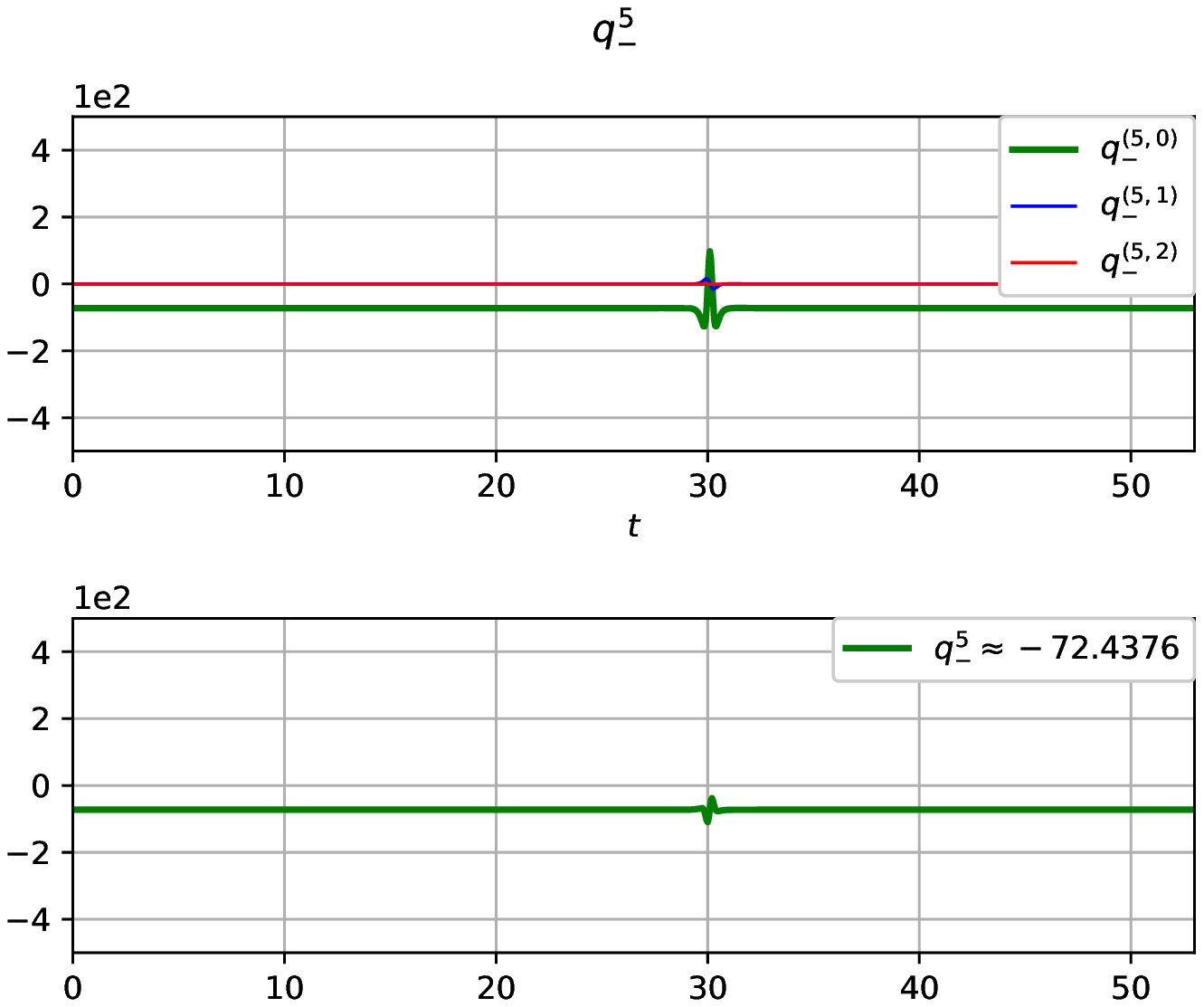}
\end{minipage}
\caption{ Left Figs. show the charge densities  $f_{-}^{(5, a)},\, a=0,1,2$ vs $x$ for the kink-antikink collision with $v_2=0.33, v_1=-0.55$ for $q = 2.015$, with initial (green), collision (blue) and final (red) densities of the kink-antikink scattering. Right Figs. show the conserved charge components $q_{-}^{(5,a)}(t),\,a=0,1,2$ and the total charge $q_{-}^(5)(t)$ of the  kink-antikink solution.}\label{fig23}
\end{figure}

In the Figs. 24-25 we present the simulations of the charge densities  $f_{+}^{(5, a)},\, a=0,1,2$ vs $x$ for the breather with period $T=4.967$ and $q = 2.025$, for three successive times $t_i$(green), $t $(blue) and $t_f$ (red), as well as the conserved charge components $q_{+}^{(5,a)}(t),\,a=0,1,2$ and the total charge $q_{+}^(5)(t)$ of the deformed SG breather. 
 
\begin{figure}[htbp!]
\begin{minipage}[b]{0.475\linewidth} 
\centering
\includegraphics[scale=0.54]{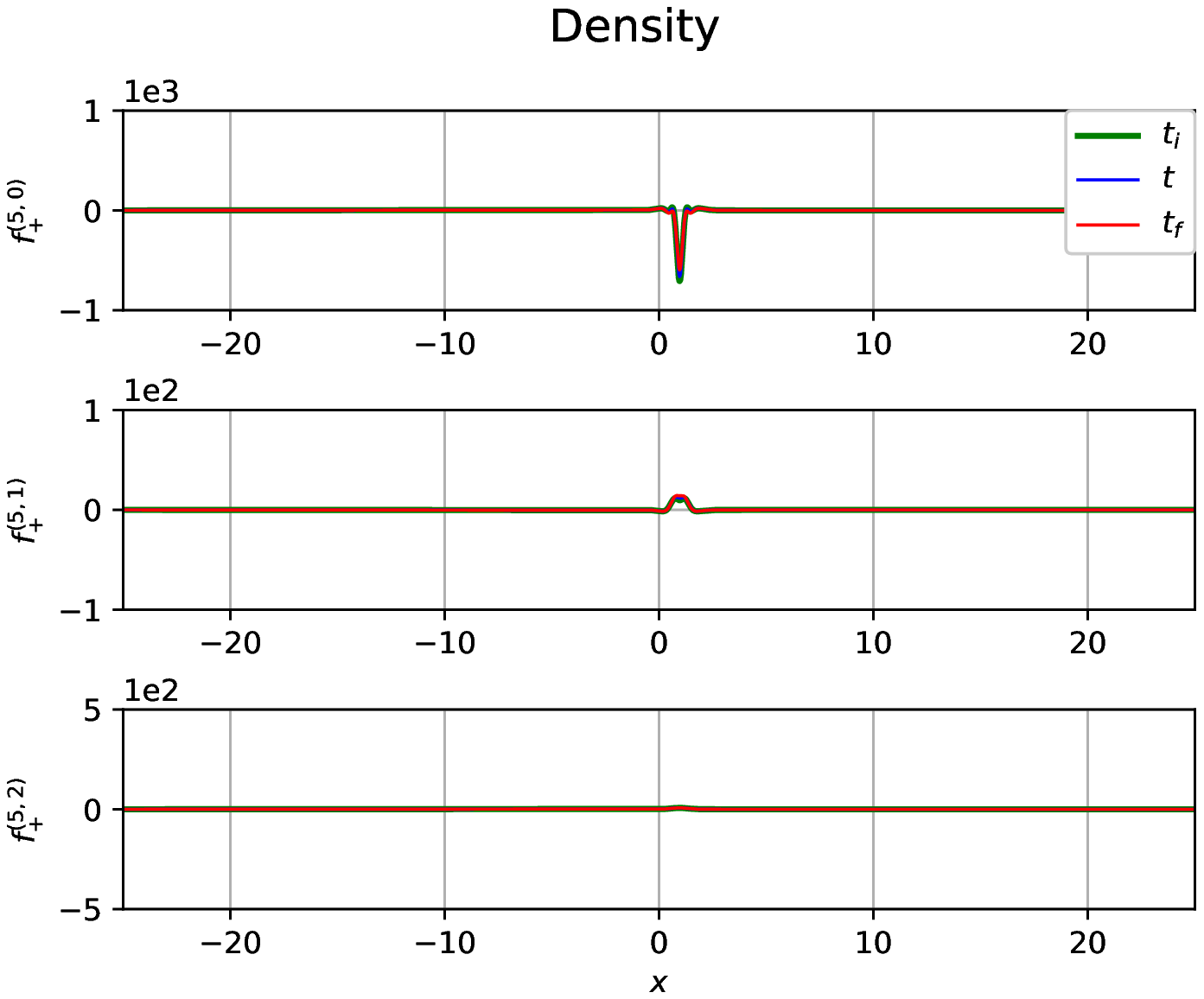}
\end{minipage}
\hspace{0.000001cm} 
\begin{minipage}[b]{0.475\linewidth}
\centering
\includegraphics[scale=0.54]{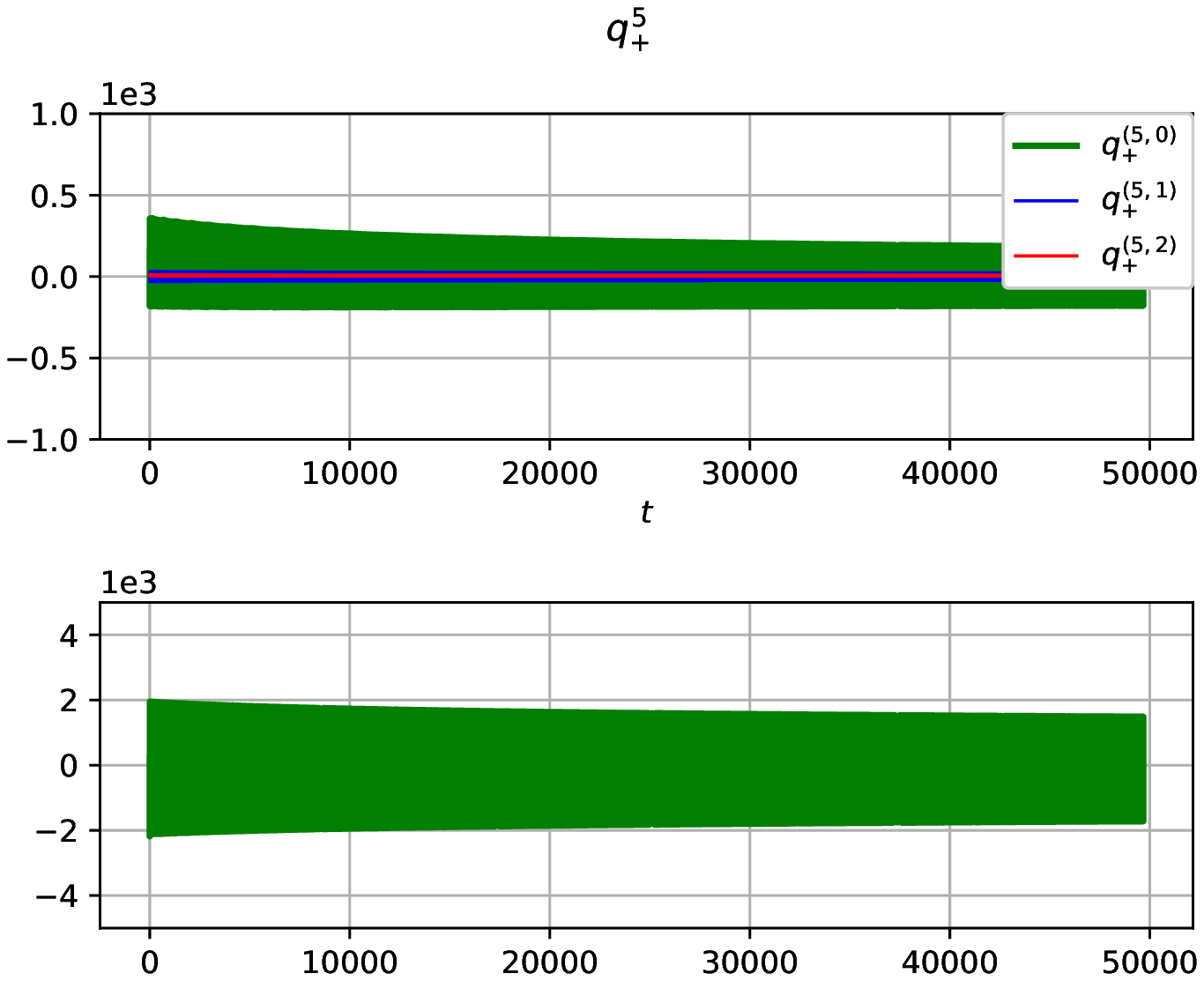}
\end{minipage}
\caption{ Left Figs. show the charge densities  $f_{+}^{(5, a)},\, a=0,1,2$ vs $x$ for the breather with period $T=4.967$ and $q = 2.025$, for three successive times $t_i$(green), $t $(blue) and $t_f$ (red). Right Figs. show the conserved charge components $q_{+}^{(5,a)}(t),\,a=0,1,2$ and the total charge $q_{+}^(5)(t)$ of the breather.}\label{fig24}
\end{figure}

\begin{figure}[htbp!]
\begin{minipage}[b]{0.475\linewidth} 
\centering
\includegraphics[scale=0.54]{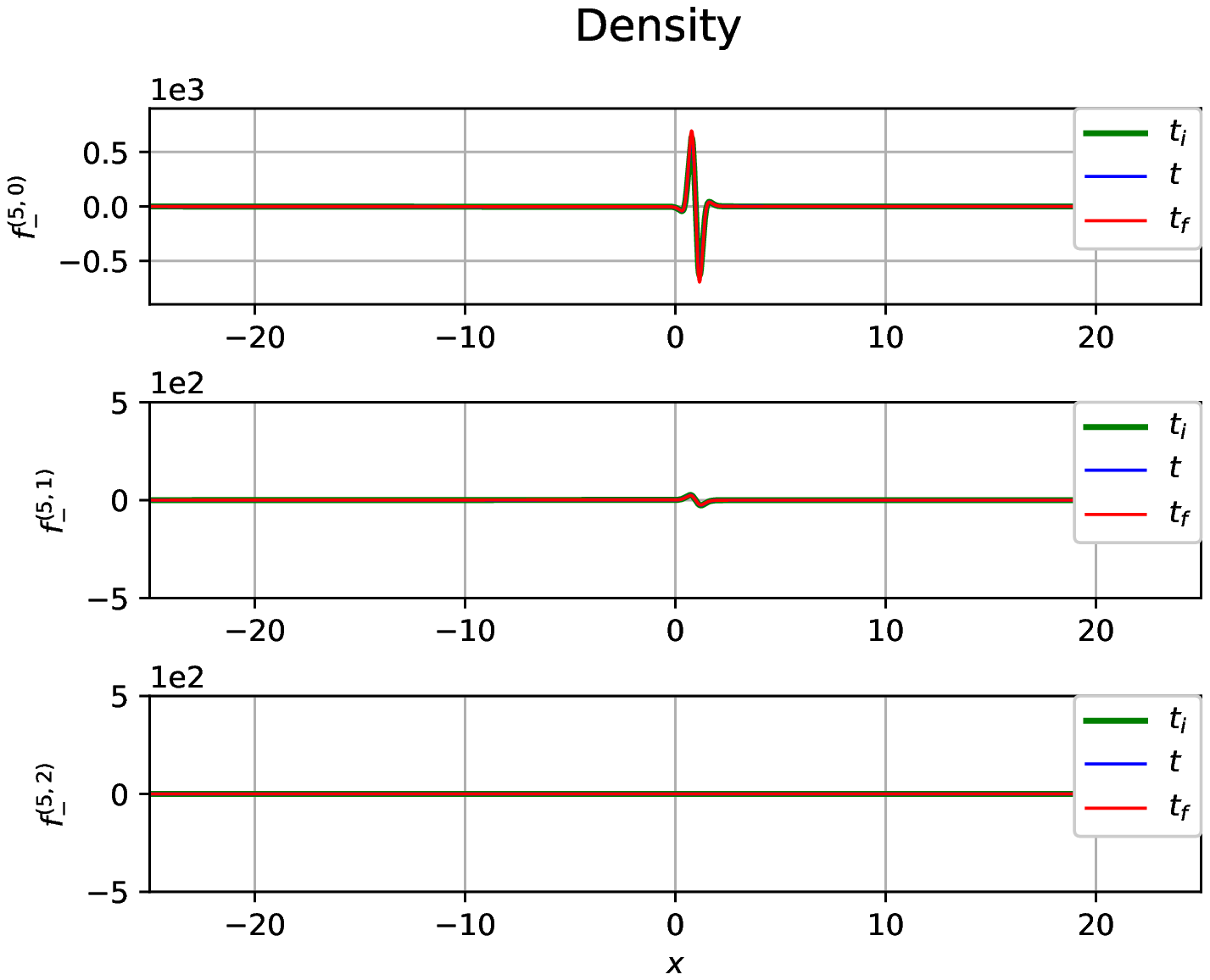}
\end{minipage}
\hspace{0.000001cm} 
\begin{minipage}[b]{0.475\linewidth}
\centering
\includegraphics[scale=0.54]{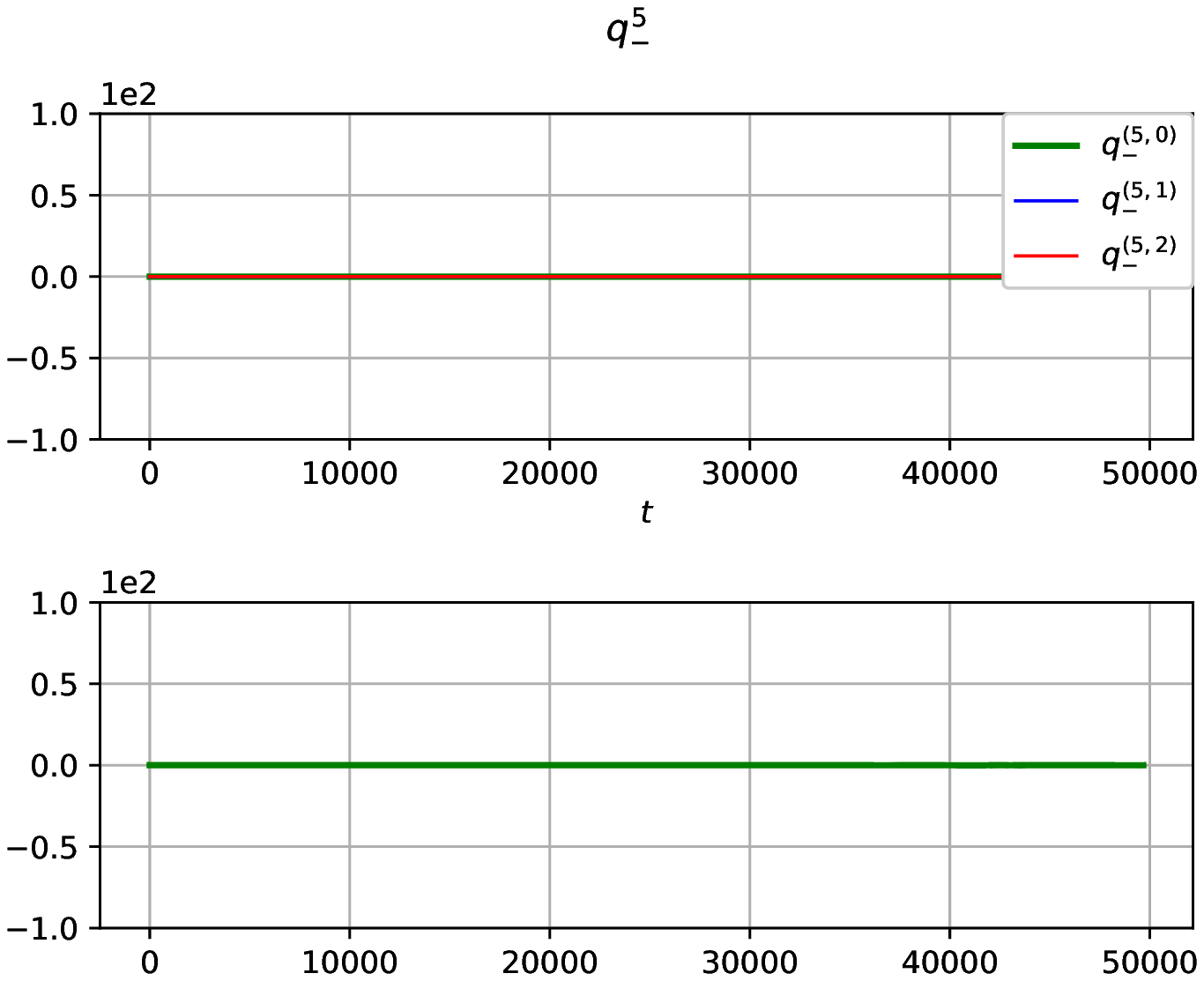}
\end{minipage}
\caption{ Left Figs. show the charge densities  $f_{-}^{(5, a)},\, a=0,1,2$ vs $x$ for the breather with period $T=4.967$ and $q = 2.025$, for three successive times $t_i$(green), $t $(blue) and $t_f$ (red). Right Figs. show the conserved charge components $q_{-}^{(5,a)}(t),\,a=0,1,2$ and the total charge $q_{-}^(5)(t)$ of the breather.}\label{fig25}
\end{figure}

Therefore, our numerical simulations show the exact conservation of the charge $q_{-}^{(5)}$, within  numerical accuracy, for the kink-antikink (kink-kink) with even (odd) parity (see Figs. 17, 21) and for the breather configurations (Fig. 25). Our simulations of the charges $q_{\pm}^{(5)}$ for the kink-kink and  kink-antikink configurations merely show a behaviour similar to an asymptotically conserved charge. In fact, in the regions of soliton collision the charge $q_{+}^{(5)}$ varies considerably, then it  returns to its initial value after collision.  In addition, for the breather, one has that the charge  $q_{+}^{(5)}$ oscillates periodically. This behavior is in contradistinction  to the third order charge $q_{+}^{(3)}$, which vanishes for the breather. These results were obtained in the conventional discretization in which the $x-$derivatives were represented by the symmetric differences of the fields and the potential evaluated uniformly on each lattice point. 

Our results above show that for definite parity configurations the $q_{-}^{(5)}$ charge is exactly conserved, within numerical accuracy. In fact, this charge, as well as the momentum, vanishes for these type of configurations. This behavior is in accordance with the expectations due to  the odd parity symmetries exhibited by their relevant charge densities under space reflection for the various soliton configurations. Whereas, the charge density of $q_{+}^{(5)}$ exhibits even parity under symmetry transformation, it is not possible to infer its value based solely on symmetry arguments. So, the question then arises as to whether our numerical results are `robust' with respect to another discretisation scheme, and if the  results might have been contaminated by some numerical artifacts during the soliton collision process and when the breather oscillation approaches periodically the instantaneous vanishing value. This might be a consequence of discretising the higher order space-time derivatives present in the charge density components $q_{+}^{(5;a)},\,a=0,1,2$ in (\ref{qq5p})-(\ref{qq5p2}). For example, in order to assure that the kinks behave much as they do in the continuum and reduce certain numerical artifacts, such as the Peierls-Nabarro (PN) barrier,  it has been considered a topological discretisation scheme \cite{ward}. We believe that these  issues deserve careful consideration in order to compute the charge $q_{+}^{(5)}$ for the configurations mentioned above.

\subsection{Second and third types of towers and lowest order anomalies}
\label{sec:secth} 

In order to numerically simulate the behavior of the first two anomalies of the second type of towers (\ref{asy1})-(\ref{asy11}) and (\ref{asy2})-(\ref{asy21}), let  us consider
\br
\label{ap1}
\mbox{a}_{+} & = & \int dx \, 2 [\pa_t^2 w + \pa_x^2 w] V\\
\label{am1}
\mbox{a}_{-} & = & \int dx \,  4 [\pa_t \pa_x w ]\, V,
\er 
where the anomalies $\mbox{a}_{\pm}$ were defined in (\ref{apm}).

Similarly, for the  first two anomalies of the third type of towers (\ref{asy3})-(\ref{asy331}) and (\ref{asy4})-(\ref{asy41}), let  us write 
\br
\label{gp1}
\g_{+} & = & \int dx \,  [(\pa_t w)^2 - (\pa_x w)^2] \pa_t V\\
\label{gm1}
\g_{-} & = & -\int dx \, [(\pa_t w)^2 - (\pa_x w)^2] \pa_x V,
\er 
where the anomalies $\g_{\pm}$ were defined in (\ref{gpm}).

Moreover, under the space-reflection transformation (\ref{px})-(\ref{evenpot}), the densities of the above anomalies $\mbox{a}^{(3)}_{\pm}$ and $\g^{(2)}_{\pm}$, respectively, present definite parities, such that some of them vanish upon space integration. Therefore, in such cases one can have exact conserved charges. These results  will be verified for certain solutions as we will see below in the numerical simulations of the anomalies  $\mbox{a}^{(3)}_{\pm}$ and $\g^{(2)}_{\pm}$ for the  kink-kink and  kink-antikink configurations.

\begin{figure}[htbp!]
\begin{minipage}[b]{0.475\linewidth} 
\centering
\includegraphics[scale=0.54]{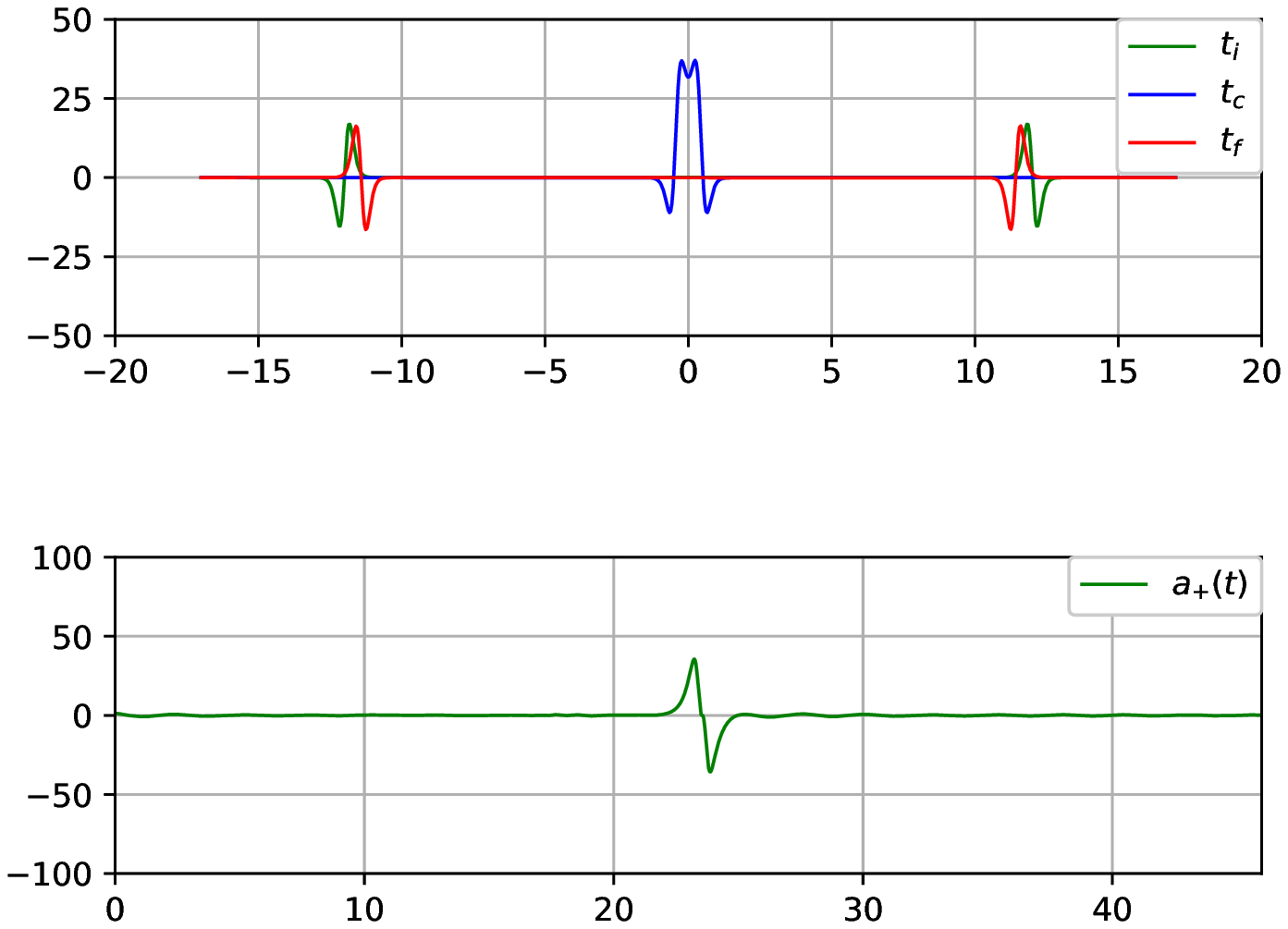}
\end{minipage}
\hspace{0.000001cm} 
\begin{minipage}[b]{0.475\linewidth}
\centering
\includegraphics[scale=0.54]{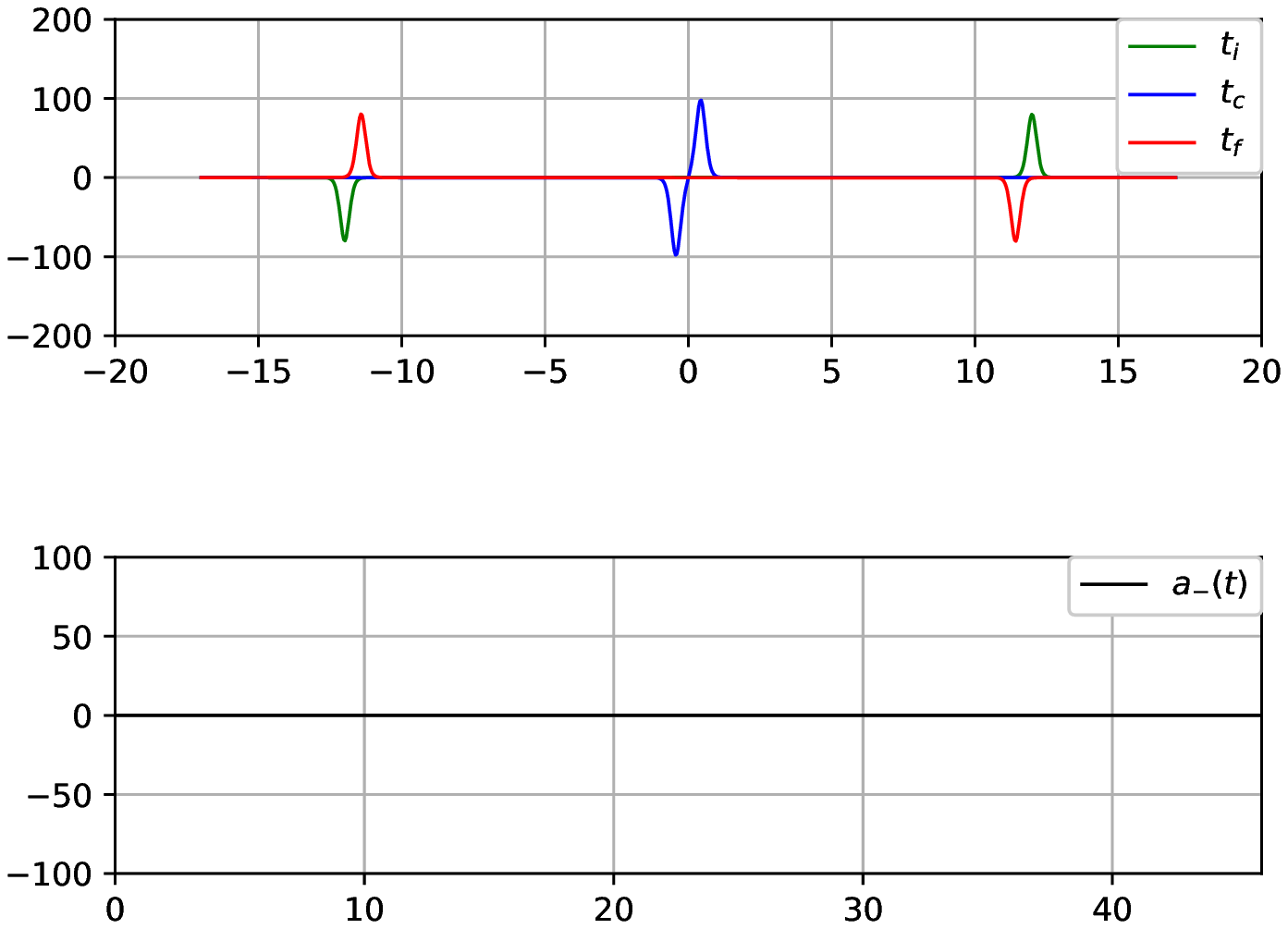}
\end{minipage}
\caption{ Top Figs. show  the anomaly  densities of  (\ref{ap1})-(\ref{am1}), respectively,  plotted in $x-$coordinate for three successive times $t_i$(green), $t $(blue) and $t_f$ (red). Bottom Figs. show the relevant anomalies $\mbox{a}_{\pm}\, \mbox{vs}\, \, t $, for kink-antikink collision shown in Fig. 4.}\label{fig26}
\end{figure}

\begin{figure}[htbp!]
\begin{minipage}[b]{0.475\linewidth} 
\centering
\includegraphics[scale=0.54]{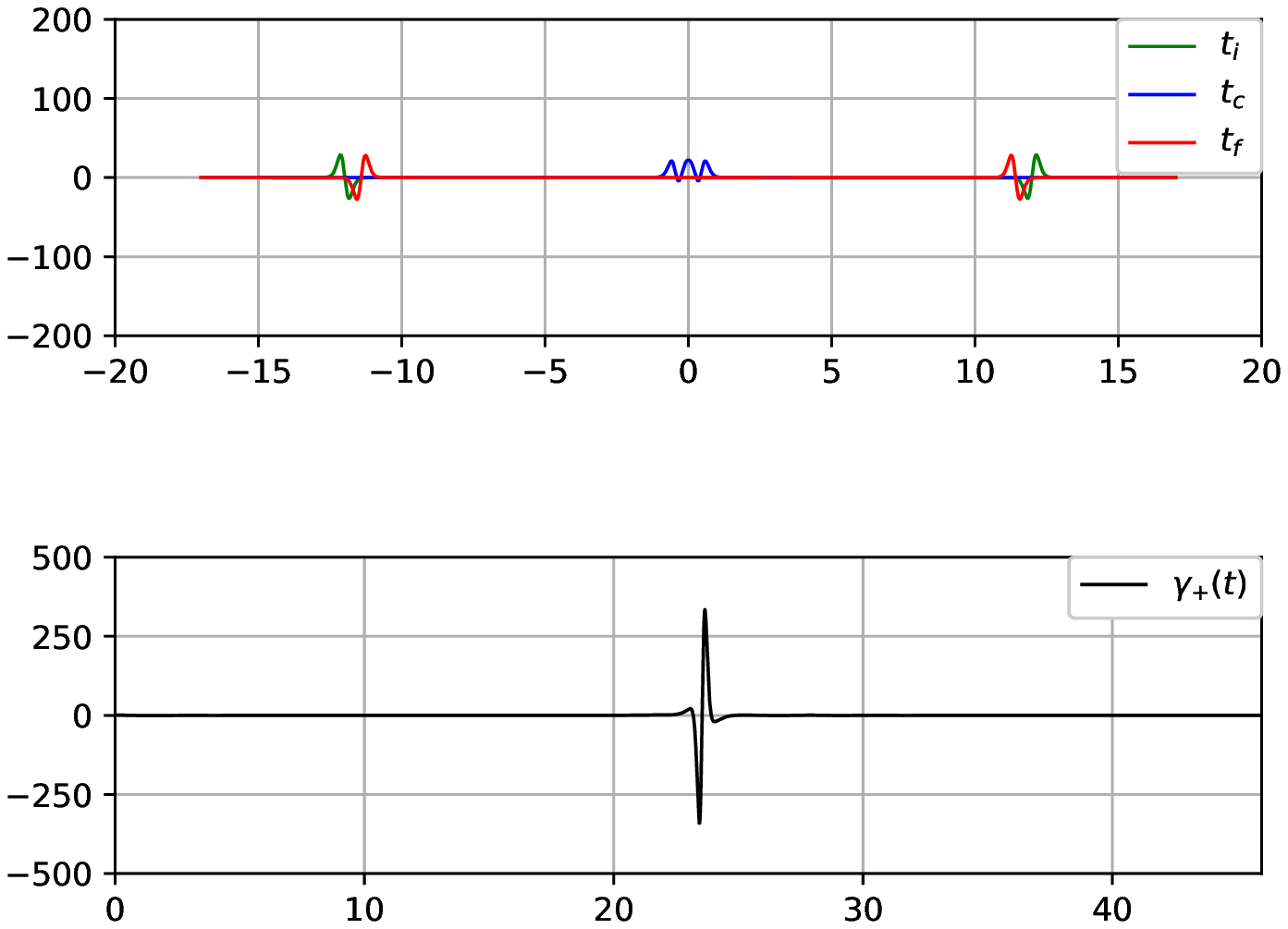}
\end{minipage}
\hspace{0.000001cm} 
\begin{minipage}[b]{0.475\linewidth}
\centering
\includegraphics[scale=0.54]{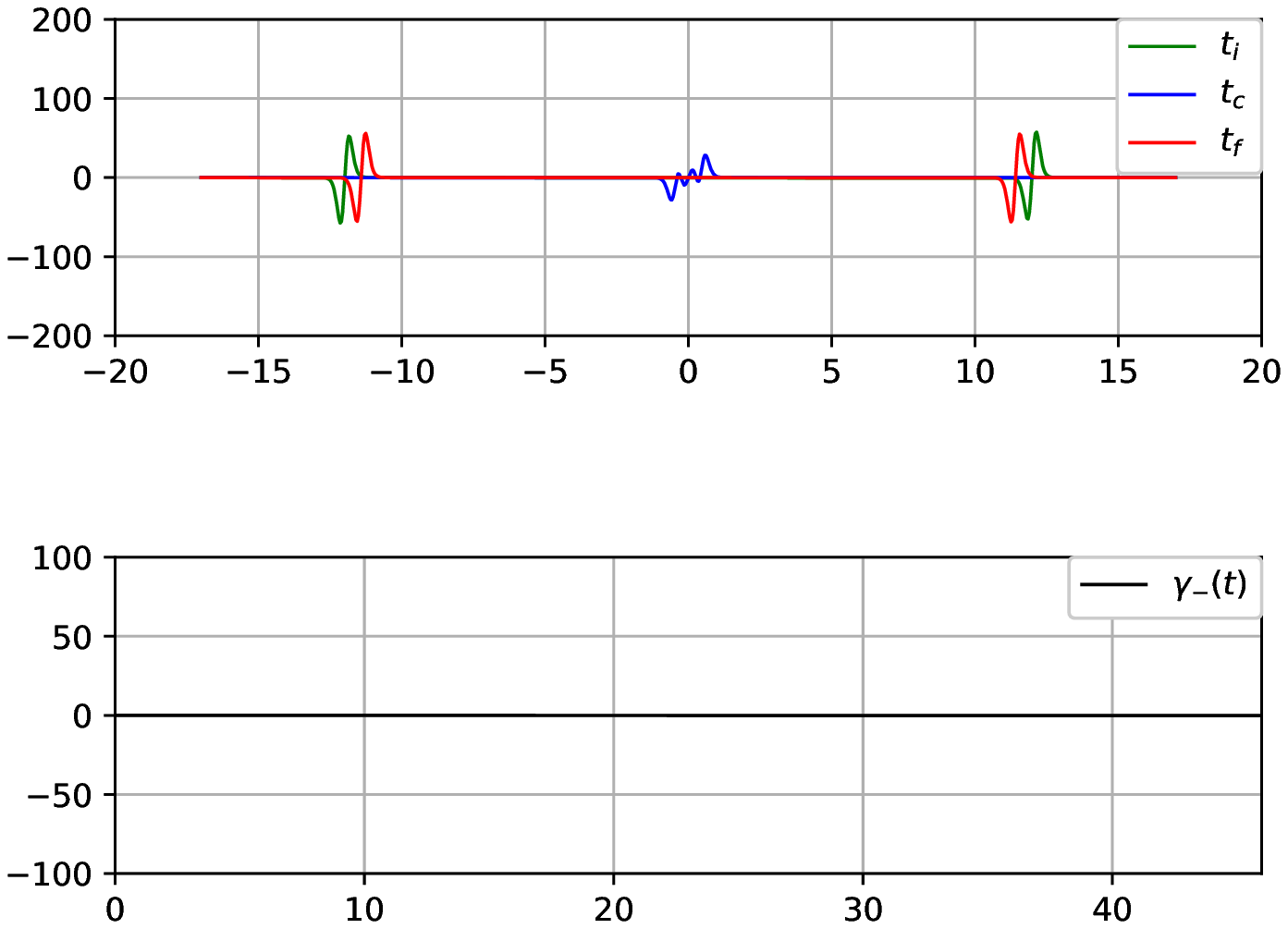}
\end{minipage}
\caption{  Top Figs. show  the anomaly  densities of  (\ref{gp1})-(\ref{gm1}), respectively,  plotted in $x-$coordinate for three successive times $t_i$(green), $t $(blue) and $t_f$ (red). Bottom Figs. show the relevant anomalies $\g_{\pm}\, \mbox{vs}\, \, t $, for kink-antikink collision shown in Fig. 4.}\label{fig27}
\end{figure}

\begin{figure}[htbp!]
\begin{minipage}[b]{0.475\linewidth} 
\centering
\includegraphics[scale=0.54]{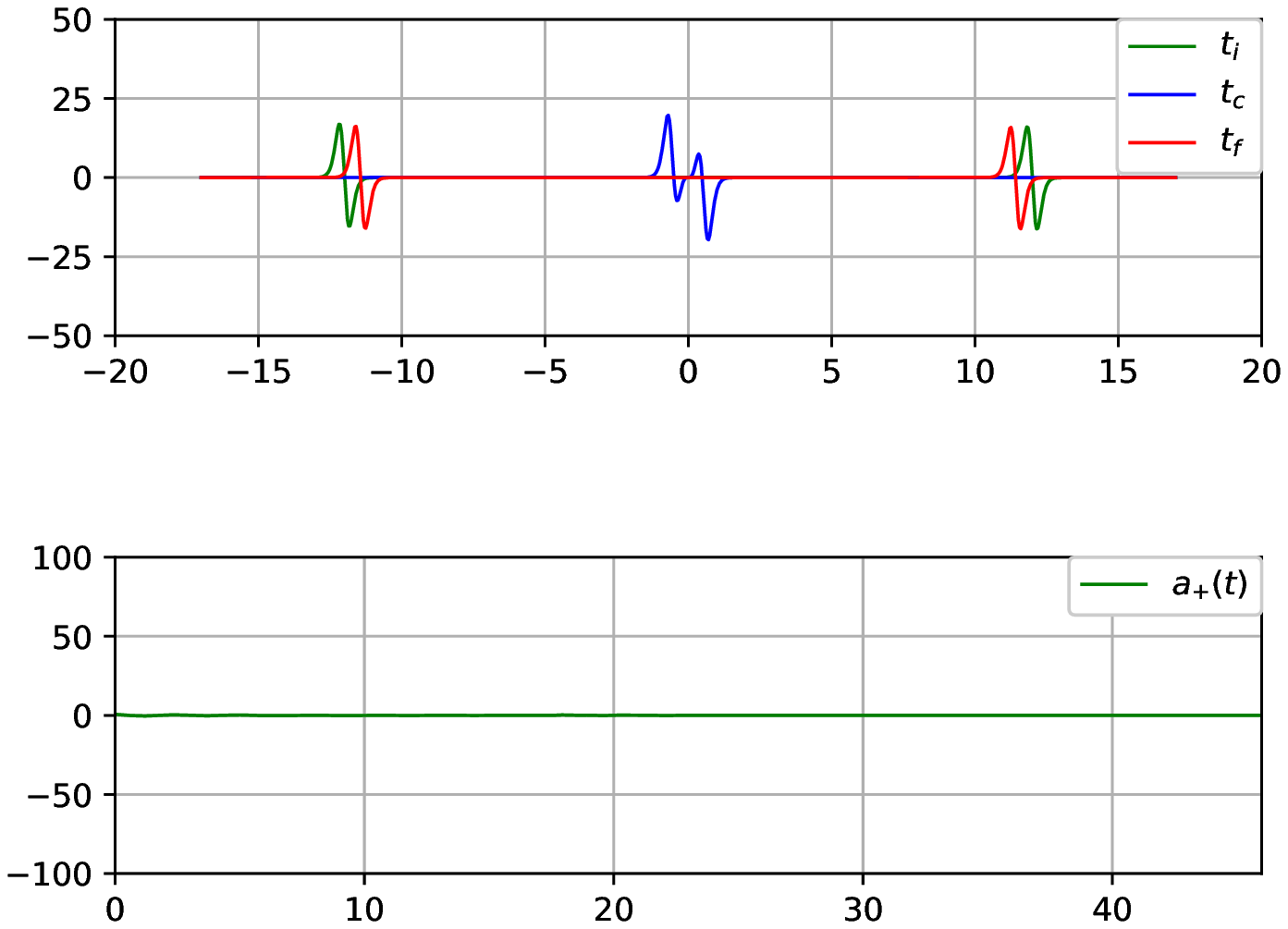}
\end{minipage}
\hspace{0.000001cm} 
\begin{minipage}[b]{0.475\linewidth}
\centering
\includegraphics[scale=0.54]{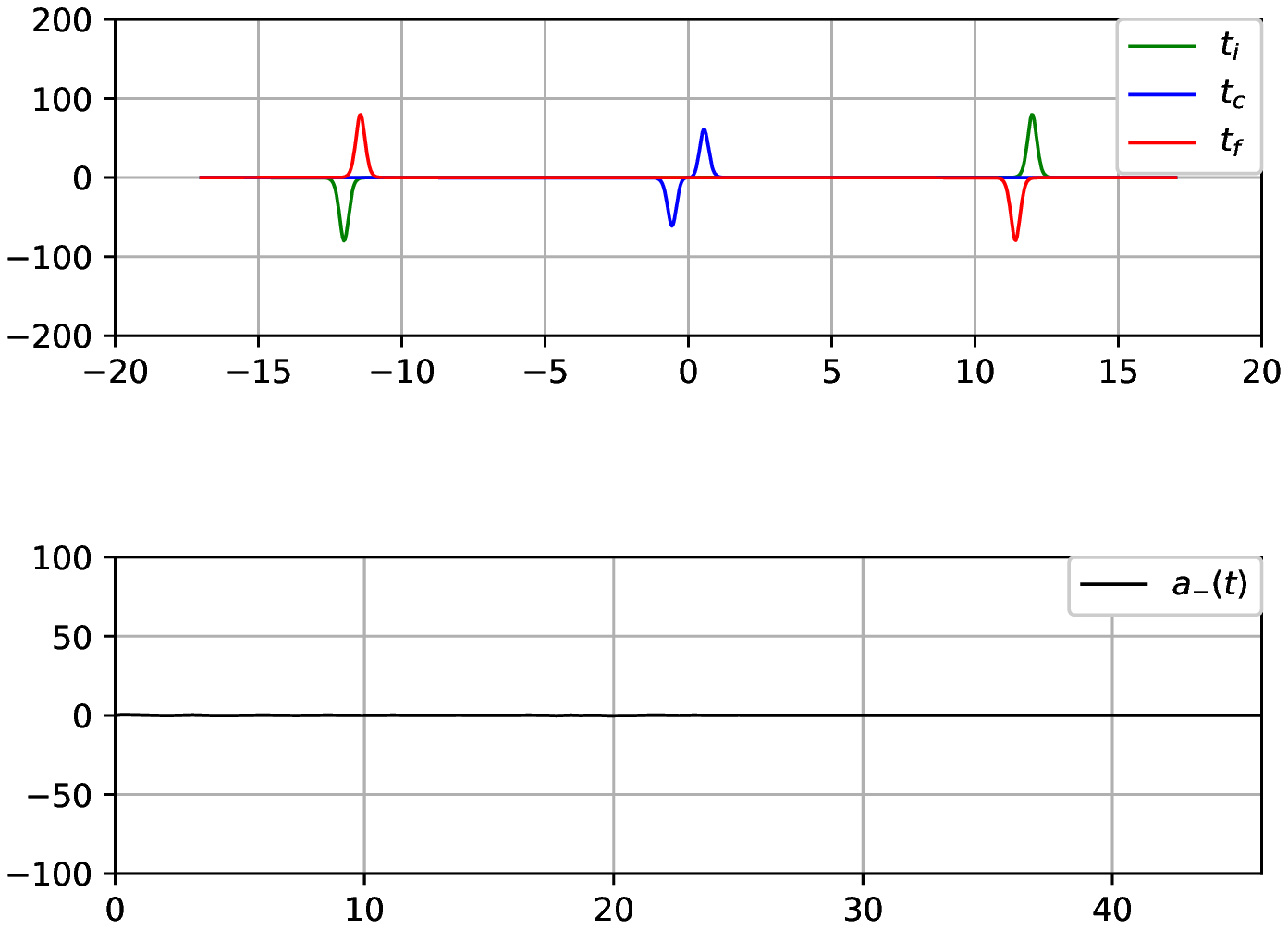}
\end{minipage}
\caption{  Top Figs. show  the anomaly  densities of  (\ref{ap1})-(\ref{am1}), respectively,  plotted in $x-$coordinate for three successive times $t_i$(green), $t $(blue) and $t_f$ (red). Bottom Figs. show the relevant anomalies $\mbox{a}_{\pm}\, \mbox{vs}\, \, t $, for kink-kink collision shown in Fig. 1.}\label{fig28}
\end{figure}

\begin{figure}[htbp!]
\begin{minipage}[b]{0.475\linewidth} 
\centering
\includegraphics[scale=0.54]{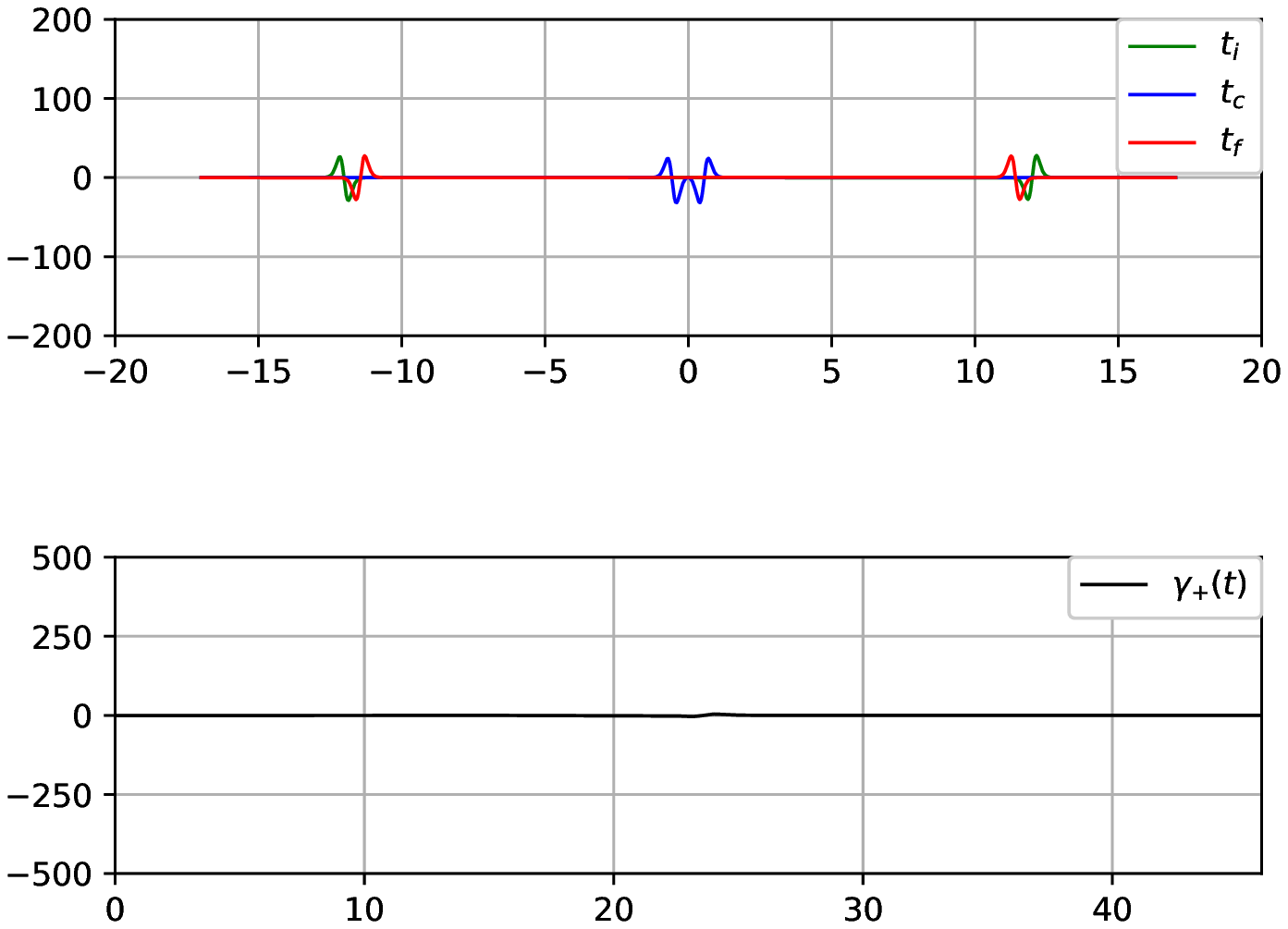}
\end{minipage}
\hspace{0.000001cm} 
\begin{minipage}[b]{0.475\linewidth}
\centering
\includegraphics[scale=0.54]{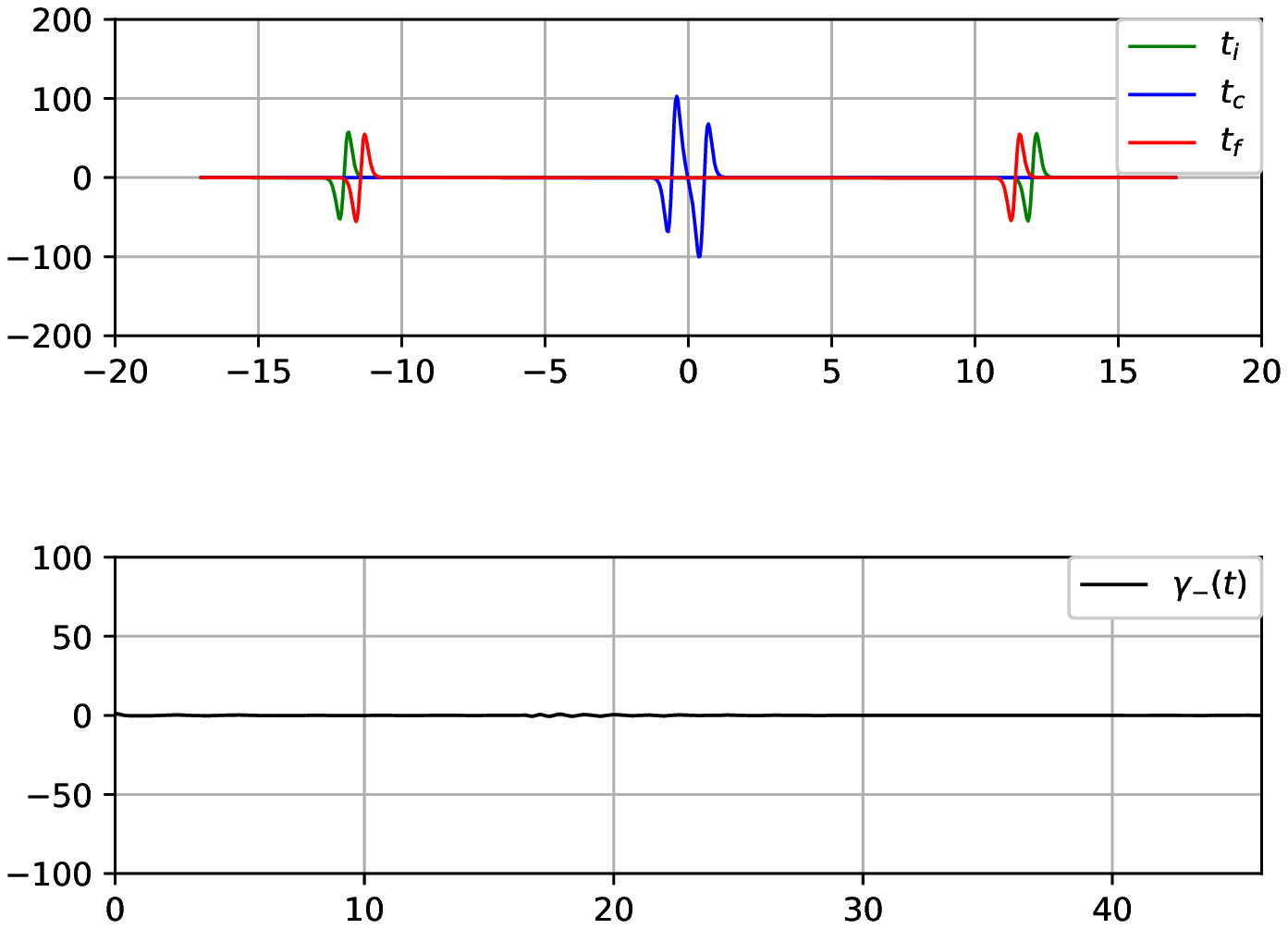}
\end{minipage}
\caption{ Top Figs. show  the anomaly  densities of  (\ref{gp1})-(\ref{gm1}), respectively,  plotted in $x-$coordinate for three successive times $t_i$(green), $t $(blue) and $t_f$ (red). Bottom Figs. show the relevant anomalies $\g_{\pm}\, \mbox{vs}\, \, t $, for kink-kink collision shown in Fig. 1.}\label{fig29}
\end{figure} 

Our numerical simulations in the Figs. 26-29 show the anomalies $\mbox{a}_{\pm}$ and $\g_{\pm}$ and their relevant densities. Notice that the anomalies $\mbox{a}_{-}$ and $\g_{-}$ vanish  for symmetric kink-antikink collision (see Fig. 4), within numerical accuracy,  since their densities possess odd parity under space reflection. Similarly,  for anti-symmetric kink-kink collision (see Fig. 1) the anomalies $\mbox{a}_{+}$ and $\g_{-}$ vanish, since their densities possess odd parity.

These developments strongly suggest that the quasi-integrable models set forward in the literature \cite{jhep1, jhep2,jhep3,jhep4, jhep5, jhep6, arxiv1}, and in particular the model (\ref{sg1}),  would possess more specific integrability structures, such as an infinite set of exactly conserved  charges, and some type of Lax pairs (or linear formulations) for certain deformed potentials. So, in the next section we will tackle the problem of extending the Riccati-type pseudo-potential formalin, which has been used for a variety of well known integrable systems, to the deformed sine-Gordon model (\ref{sg1}). Then, in the next subsections  we formulate the dual Riccati-type representations and then we discuss the conservation laws associated to the equation of motion (\ref{sg1}).

\section{Riccati-type pseudo-potential and conservation laws}
 
\label{sec:riccati}

In \cite{nucci} it has been generated the both Lax equations and Backlund transformations 
for well-known non-linear evolution equations using the concept of pseudo-potentials and the related properties of the Riccati equation. These applications have been done in the context of a variety of  integrable systems (sine-Gordon, KdV, NLS, etc), and allow the Lax pair formulation, the construction of  conservation laws and the Backlund transformations for them \cite{nucci, prl1}. 

So, in the next steps we consider a convenient deformation of the usual pseudo-potential approach to integrable field theories. Let us consider the system of Riccati-type equations 
\br
\label{ricc1}
\pa_\xi u &=& -2 \lambda^{-1} \, u +\, \pa_\xi w + \, \pa_\xi w \,\, u^2,\\
\label{ricc2}
\pa_\eta u &=&-  2 \lambda\,  (V-2)\, u - \frac{1}{2} \,\lambda\, V^{(1)} + \frac{1}{2} \lambda\, V^{(1)}\, u^2 + r - u s, 
\er
where $V(w)$ is the deformed sine-Gordon potential with $V^{(1)} \equiv \frac{d}{d w} V(w)$ and $\lambda$ is the spectral parameter. We consider the following equations for the auxiliary fields $r(\xi,\eta)$ and  $s(\xi,\eta)$
\br
\label{ricc1r}
\pa_\xi r &=& -2 \lambda^{-1} \, r +\, \pa_\xi w  (u \, r + s)+ \lambda X,\\
\label{ricc2s}
\pa_\xi s &=& \pa_\xi w  (u\, s - r)+ \lambda X u,\\
X & \equiv & \pa_{\xi} w \( \frac{V^{(2)}}{2}+2 V -4\),\,\,\,\,\,\,V^{(2)} \equiv \frac{d^2}{d w^2} V(w).  \label{Xanom}
\er  
So, one has a set of two deformed Riccati-type equations for the pseudo-potential $u$ (\ref{ricc1})-(\ref{ricc2}) and a system of equations (\ref{ricc1r})-(\ref{ricc2s}) for the auxiliary fields $r$ and  $s$. 

Notice that, for the integrable SG model potential 
\br
\label{usg}
V = -2 \cos{(2 w)} + 2,
\er
one has that $X=0$, and so the auxiliary system of  eqs. (\ref{ricc1r})-(\ref{ricc2s}) possesses the trivial solution $r=s=0$. Inserting this trivial solution into the system  (\ref{ricc1})-(\ref{ricc2}) and considering  the potential (\ref{usg}), one has a set of two Riccati equations for the usual SG model and they  play an important role in order to study the properties of the integrable SG model, such as the derivation of the infinite number of conserved charges and the Backlund transformations, relating the field $w$ with another solution $\bar{w}$ \cite{prl1}.   
 
Note that only the $\eta-$component $\pa_{\eta} u$ of the Riccati equation associated to the ordinary sine-Gordon equation has been  deformed away from the SG potential (\ref{usg}), and it carries all the information regarding the deformation of the model which are encoded in the potential $V(w)$ and the auxiliary fields $r(\xi,\eta)$ and  $s(\xi,\eta)$. The form of the $\xi-$component $\pa_{\xi}u $ remains the same as the usual Riccati equation associated to the SG model. 
 
We have computed the compatibility condition   $[\pa_{\eta}\pa_\xi  u - \pa_\xi \pa_{\eta} u ]=0$ for the Riccati-type equations (\ref{ricc1})-(\ref{ricc2}), taking into account  the auxiliary system of equations (\ref{ricc1r})-(\ref{ricc2s}) and then rederived  the eq. of motion of the deformed sine-Gordon model (\ref{sg1}).
 
Let us emphasize that for the usual SG model we have the  trivial solution of the system  (\ref{ricc1r})-(\ref{ricc2s}), i.e. $X=0 \rightarrow r=s=0$, and the existence of the Lax pair of de usual SG model reflects in its equivalent Riccati-type representation, provided by the system (\ref{ricc1})-(\ref{ricc2}) with the well known potential (\ref{usg}) \cite{nucci, prl1}. 

Next, let us discuss the relevant conservation laws in the context of the Riccati-type system (\ref{ricc1})-(\ref{ricc2}) and the auxiliary equations (\ref{ricc1r})-(\ref{ricc2s}). So, substituting the expression for $u^2$ from (\ref{ricc1}) into (\ref{ricc2}) and considering (\ref{ricc2s}), one gets the following relationship\footnote{In fact, there are different expressions of this type, we follow below the construction such that the non-homogeneous r.h.s. terms must contain the deformation variables $\{s, r\}$, such that for $r=s=0$ one must recover in the l.h.s., order by order in $\lambda$, the polynomial conservation laws of the standard SG model.}
\br
\label{qcons}
\pa_{\eta} \(u \, \pa_{\xi} w \) + \pa_{\xi} \Big[ \lambda \, (V-2) - \frac{1}{2}\lambda \, u \, V^{(1)} \Big] = -   \pa_{\xi}  s  . 
\er

This equation will be used to uncover an infinite tower of conservation laws associated to the modified SG model (\ref{sg1}). A truly conservation law character of this equation remains to be clarified. In fact, if the function $s(\xi, \eta)$ in the r.h.s. of (\ref{qcons}) possessed  non-local expressions containing terms such as $\int d\xi (...)$, the conservation law property of the equation would be spoiled. In fact, this would give rise to certain `anomalies', as in the context of the anomalous zero-curvature formulation associated to a deformed Lax pair and its quasi-conservation laws \cite{jhep1}. We will show below that the r.h.s. of (\ref{qcons}) can be written in general as $ [-\pa_{\xi}  s ] \equiv \pa_{\xi} S + \pa_{\eta} R$, with $S$ and $R$ being certain local functions of $w$ and its $\xi$ and  $\eta-$derivatives; i.e. there exists a local expression for $\pa_{\xi} s$, such that the eq. (\ref{qcons}) provides a proper local conservation law.    

Next, let us consider the expansions
\br
\label{expan}
 u = \sum_{n=1}^{\infty} u_n \, \lambda^n,\,\,\,\,  s= \sum_{n=0}^{\infty} s_n \, \lambda^{n+2},\,\,\,\,  r= \sum_{n=0}^{\infty} r_n \, \lambda^{n+2}.
\er
The coefficients $u_n$ of the expansion above  can be determined order by order in powers of $\lambda$ from  the Riccati equation (\ref{ricc1}).  In appendix \ref{fsca1} we provide the recursion relation for the $u_n\, 's$ and the expressions for the first $u_n$. Likewise, using the results for the $u_n\, 's$ we get the relevant expressions for the $r_n\, 's$ and $s_n\, 's$ from (\ref{ricc1r})-(\ref{ricc2s}). The recursive relationships for $s_{n}$ and $r_n$ provided in the appendix \ref{fsca11} allow us to find the explicit expressions  of these fields, order by order in powers of the spectral parameter $\lambda$.  

Then, making use of the $u_n$ components  of the expansion of $u$ provided in (\ref{terms}) or (\ref{comp1})-(\ref{comp2}), one can find the conservation laws, order by order in powers of $\lambda$. So, by inserting those expansions into  the eq. (\ref{qcons}) one has that  the coefficient of the $n'$th order term becomes
\br
\label{anocons}
\pa_{\eta} a_{\xi}^{(n)} &+& \pa_{\xi} a_{\eta}^{(n)} = -   \pa_{\xi} s_{n-2},\,\,\,\,\,n=1,2,3,....;\, s_{-1}\equiv 0\\
a_{\xi}^{(n)} &\equiv & u_n \pa_{\xi} w,\,\,\,\,a_{\eta}^{(n)} \equiv   (V-2) \delta_{1, n}  - \frac{1}{2} u_{n-1} V^{(1)} ,\,\,\,\,\,u_0 \equiv 0.
\er
 
So, the first order ${\cal O}(\lambda^{1})$ term provides
\br
\label{n1}
\pa_{\eta} \( \frac{1}{2} (\pa_{\xi} w)^2\) + \pa_{\xi} \( V-2 \) =0.
\er 
Notice that the r.h.s. of (\ref{anocons}) vanishes at this order, i.e. by definition one has $ s_{-1}\equiv 0$. In fact, the conservation law (\ref{n1}) provides the first conserved charge
\br
\label{nn1}
q^{(1)} &=& \int dx\,\Big[ \frac{1}{2} ( \pa_{\xi} w)^2 + (V-2) \Big].
\er
The equations (\ref{n1})-(\ref{nn1}), together with their duals and the relevant charge $ \widetilde{q}^{(1)} = \int dx\, [ \frac{1}{2} ( \pa_{\eta} w)^2 + (V-2) ]$ which will be provided  below,  give rise to the usual energy and momentum charges written in laboratory coordinates $(x,t)$ as
\br
\label{energy1}
E &=& \frac{1}{2} \Big[ q^{(1)} + \widetilde{q}^{(1)}\Big]\\
  &=& \int dx \Big[ \frac{1}{2} (\pa_x w)^2 + \frac{1}{2} (\pa_t w)^2 + (V-2) \Big]
  \label{energy11}
  \er
and 
\br
\label{momentum1}
P &=&  \frac{1}{2} \Big[\widetilde{q}^{(1)}-  q^{(1)} \Big]\\
  &=&\int dx [ \pa_x w\pa_t w ] .\label{momentum11} 
\er
The next order term ${\cal O}(\lambda^{2})$ becomes
\br
\label{n2}
\frac{1}{4}\pa_{\xi} \Big[\pa_{\eta} \( \frac{1}{2} (\pa_{\xi} w)^2\) + \pa_{\xi} \( V-2 \)\Big] = 0. 
\er
Since from (\ref{s0}) one has $\pa_{\xi} s_{0}=0$ one notices that the r.h.s. of (\ref{anocons}) also vanishes at this order. As usual, we can define the charge 
\br
q^{(2)} &=& \int dx\, \frac{1}{4} \(\pa_{\xi}^2 w \pa_{\xi} w +  \pa_{\xi} w V^{(1)}\)\\
&=&\frac{1}{4} \frac{d}{dt} \( E - P\). \er 
So, the eq. (\ref{n2}) does not provide an independent new charge in laboratory coordinates $(x,t)$. So, there is no an independent new charge at this order. Notice that also the usual SG model does not possess an independent charge at this order \cite{sanuki}. 

From this point forward and for the higher order charges the term encoding the deformation away from the usual SG model, i.e. the r.h.s. of (\ref{anocons}), will play an important role in the construction of the conservation laws. So, the third  order ${\cal O}(\lambda^{3})$ term provides 
\br
\label{n3}
\pa_{\eta} \Big[ \frac{1}{8}\pa_{\xi} w \( (\pa_{\xi} w)^3 + \pa_{\xi}^3 w \)\Big] + \pa_{\xi} \( \frac{1}{8}  \pa_{\xi}^2 w V^{(1)} \) =-\pa_{\xi} s_1.
\er 
    
Remarkably, the r.h.s. of (\ref{n3}) can be written as
\br
\label{s1a}
-\pa_{\xi} s_1 &=& \frac{1}{4} X  \pa_{\xi}^2 w  -\frac{1}{4}  \pa_{\xi} w \, \pa_{\xi} X\\
 &=& \pa_{\eta} R_1 + \pa_{\xi} S_1 \label{s1b}
  \\
\label{s1c}
R_1 &\equiv &-\frac{1}{8} [\pa_{\xi}^2 w]^2 + \frac{1}{8} [\pa_{\xi} w]^4 ,\,\,S_1 \equiv - \frac{1}{8} [\pa_{\xi} w]^2 V^{(2)}.\label{rs22}
\er
In order to write (\ref{s1b}) starting from (\ref{s1a}) we have used the explicit expression for $X$ in (\ref{Xanom}) and the deformed sine-Gordon eq. of motion (\ref{sg1}) and its derived expressions such as $\pa_\eta \pa^2_\xi \, w +  V^{(2)}(w)\pa_{\xi}w  = 0$. Therefore, the conservation law (\ref{n3}) turns out to be  
\br
\label{nn3}
\frac{1}{8}\pa^{2}_{\xi}\Big[ \pa_{\eta} \( \frac{1}{2} (\pa_{\xi} w)^2\) + \pa_{\xi} \( V-2 \) \Big]=0.
\er
Notice that this form of the third order conservation law holds strictly for deformed SG models, i.e. for models such that $X \neq 0$. In the usual SG model the eq. (\ref{n3}) with vanishing r.h.s., since in that case $X\equiv 0$, provides the relevant conservation law at this order. 

Next, the charge which follows from the above conservation law (\ref{nn3}) becomes
\br
q^{(3)} = \frac{1}{8} \frac{d^2}{dt^2} \( E - P\).\label{ch32} \er 

Therefore, in this formulation and at this order, in contradistinction to the ordinary SG model, there is not an independent conserved charge for the deformed SG model (\ref{sg1}). 

However, one can show that the third order charge and anomaly of \cite{jhep1, jhep2, cnsns}   presented in (\ref{asymp}) can be rewritten in our notation, respectively, in the form 
\br
q_{a}^{(3)} &=& \int dx [ \frac{1}{8}(\pa_{\xi} w)^4 +\frac{1}{8} \pa_{\xi} w\pa_{\xi}^3 w+  \frac{1}{8}  \pa_{\xi}^2 w V^{(1)} + \frac{1}{4} X \pa_{\xi}w  ] \er
  and 
\br
\label{betaxieta}
\beta^{(3)} \equiv \frac{1}{2} \pa_{\xi}^2w  X = \pa_{\xi} S_1 + \pa_{\eta} R_1 + \frac{1}{4} \pa_{\xi} \( \pa_{\xi}w\, X\),\er
with $R_1, S_1$ given in (\ref{rs22}). 

In view of the form  that $\beta^{(3)}$ takes in (\ref{betaxieta}), the $x-$integrated `anomaly' term on the r.h.s. of (\ref{asymp}) (written for $n=1$) can be promoted to the l.h.s. of that equation by adding some terms to the relevant charge $q_{a}^{(3)}$. So, the quasi-conservation law (\ref{asymp}), in the case $n=1$, can be rewritten as an exact conservation law
\br
\label{embed1}
\frac{d}{dt} \Big\{ q_{a}^{(3)} - \int dx S_1 - \int dx R_1 - \int dx (\pa_{\xi}w X)  \Big\} = 0. 
\er
In fact, a close examination reveals that the eq. (2.15), for $n=1$, of ref. \cite{jhep1}, turns out be the same as our conservation law (\ref{nn3}), except for an overall constant factor. So, the above results show that the third order anomaly term, which appears on the r.h.s. of an inhomogeneous quasi-conservation law, has been removed and incorporated into the components of a new redefined current which satisfies an exact conservation law.     

Moreover, in the recent paper \cite{cnsns} by two of us  it has been shown, through numerical simulation and analytical method, that the `anomaly' $\beta^{(3)}$, once linearly combined with its dual $\widetilde{\beta}^{(3)}$, gives rise to the both anomalies $\beta^{(3)}_{\pm}$, such that the $x-$integrated ``anomaly" $\alpha_{+}^{(3)}\equiv\int dx \beta^{(3)}_{+}$ vanishes when evaluated on some two-soliton configurations (kink-kink, kink-antikink and breather solitons with definite parity under space-reflection symmetry ), and if the deformed potential, evaluated on such a solution, is even under the parity. In fact, it has been shown analytically that the $\beta^{(2n+1)}_{+}$ anomalies vanish for those special two-soliton configurations, rendering the exact conservation of their associated charges $q^{(2n+1)}_{a\,+} \equiv q^{(2n+1)}+ \widetilde{q}^{(2n+1)} $ \cite{cnsns}. For two-soliton  configurations the space-time integration of  the anomalies  $\beta^{(2n+1)}_{-}$ vanish, allowing the existence of the so-called asymptotically conserved charges \cite{jhep1}.   

In order to see more closely the relationship between the charge $q_{a}^{(3)}$, its `anomaly'  $\beta^{(3)}$ and the exactly conserved charge $q^{(3)}$ one can write the next relationship from (\ref{n3}) or (\ref{nn3}) 
\br
\label{qqt0}
\frac{d}{dt} q_{a}^{(3)} - \frac{d}{dt} q^{(3)} &=& \int \, dx  \beta^{(3)}\\
                                                  &=&  \frac{d}{dt} \int\, dx \Big[S_1 + R_1 + \frac{1}{4} \pa_{\xi}w X\Big], \label{qqt}
\er
where the form of $\beta^{(3)}$ provided in (\ref{betaxieta}) has been used. Therefore, upon integration in $t$ the expression (\ref{qqt}) the charge $q_{a}^{(3)}$ can be written as 
\br
\label{asy31}
q_{a}^{(3)} &=&  q^{(3)} + \int \, dx  \Big[S_1 + R_1 + \frac{1}{4} \pa_{\xi}w X\Big],\\
             &=&  \frac{1}{8} \frac{d^2}{dt^2} [ E - P] + \int \, dx  \Big[ \frac{1}{8} (\pa_{\xi} w)^4- \frac{1}{8} (\pa_{\xi}^2 w)^2 +  \frac{1}{2} (\pa_{\xi} w)^2[V -2] \Big], \label{asy32}
\er   
where the expression  of $q^{(3)}$ in (\ref{ch32}) and the expressions of $R_1$ and $S_1$ given in (\ref{rs22}) have been used. So, in view of the relationships (\ref{embed1}) and (\ref{asy31}) one can argue that the asymptotically conserved charge  $q_{a}^{(3)}$ becomes embedded into the third order conservation law (\ref{nn3}) provided  that the `anomaly' term $\beta^{(3)}$ is considered in the form (\ref{betaxieta}).  So, as discussed above, this `anomaly' term turns out to be removed and incorporated into the conservation law (\ref{nn3}) once the expression (\ref{betaxieta}) is taken into account. 

Even though the conservation of the charge $q_{a}^{(3)}$ has been reported before \cite{cnsns}, it is not guaranteed to happen a priori in the present formulation, and in order to check this property we have computed above (see sec. \ref{sec:third}) this charge by numerical simulations of a variety of two-soliton collisions and the breather oscillations for a particular deformation of the SG model. From the analytical point of view, in sections \ref{sec:symmetry} and \ref{sec:third} above, we discussed the exact conservation of the charge $q_{a}^{(3)}$ for some soliton configurations with definite parities under space-reflection transformations. 
   
The next order  ${\cal O}(\lambda^{4})$ conservation law  becomes
\br
  \label{n4i}
\pa_{\eta} \( \frac{5}{16} \pa_{\xi}^2 w (\pa_{\xi} w)^3 + \frac{1}{16}  \pa_{\xi} w \pa_{\xi}^4 w \) + \pa_{\xi} \( \frac{1}{16} V^{(1)} [ (\pa_{\xi} w)^3 + \pa_{\xi}^3 w] \) =- \pa_{\xi} s_{2} .
\er
The r.h.s. of this equation turns out to be 
\br
\pa_{\xi} s_{2} &=&  \frac{1}{8} X \pa_{\xi}^3 w - \frac{1}{8}  \pa_{\xi} w \pa_{\xi}^2 X\\
 &=& \frac{1}{8} \pa_{\xi} [\pa_\xi^2 w X - \pa_\xi w \pa_\xi X]
\er
Therefore one has the conservation law
\br
\label{n4ii}
\pa_{\eta} \( \frac{5}{16} \pa_{\xi}^2 w (\pa_{\xi} w)^3 + \frac{1}{16}  \pa_{\xi} w \pa_{\xi}^4 w \) + \pa_{\xi} \( \frac{1}{16}  [ (\pa_{\xi} w)^3 + \pa_{\xi}^3 w]V^{(1)} - \frac{1}{8}  [\pa_\xi^2 w X - \pa_\xi w \pa_\xi X]   \) &=& 0   
\er
 
Similarly, from (\ref{n4ii}) one can define the charge 
\br
q^{(4)} &=& \int dx \,\Big\{\frac{5}{16} \pa_{\xi}^2 w (\pa_{\xi} w)^3 + \frac{1}{16}  \pa_{\xi} w \pa_{\xi}^4 w  +   \frac{1}{16}  [ (\pa_{\xi} w)^3 + \pa_{\xi}^3 w]V^{(1)} - \frac{1}{8}  [\pa_\xi^2 w X - \pa_\xi w \pa_\xi X]  \Big\}, \\
& \equiv & 0.
\er
So, it has been shown that this charge vanishes identically for suitable boundary conditions. 

The term of order ${\cal O}(\lambda^{5})$ provides the next conservation law 
\br
\label{n5}
\pa_{\eta} \( \frac{11}{32}(\pa_{\xi}^2 w)^2 (\pa_{\xi} w)^2 + \frac{7}{32}  \pa_{\xi}^3 w (\pa_{\xi} w)^3  +  \frac{1}{16}(\pa_{\xi} w)^6 + \frac{1}{32}\pa_{\xi} w \pa_{\xi}^5 w \)  -  \pa_{\xi} \Big[ \frac{1}{2} V^{(1)} u_4 \Big] = -\pa_{\xi} s_3\er
\br
 -\pa_{\xi} s_3&=&-\frac{3}{16} (\pa_{\xi} w)^3 \pa_{\xi} X+ \frac{1}{16}
\( \pa_{\xi}^4 w + \pa_{\xi}[(\pa_{\xi} w)^3] \) \,  X - \frac{1}{16} \pa_{\xi} w \pa_{\xi}^3 X. \label {n5r}\er 

A remarkable fact is that the r.h.s. of the last eq. can be written as 
\br
\label{nn51}
 -\pa_{\xi} s_3 &=& \pa_{\eta} R_3 + \pa_{\xi} S_3 
 \er
where $R_3$ and $S_3$ are defined in (\ref{nn52})-(\ref{nn53}). Then, a lenghty calculation allow us to write the eq. (\ref{n5})  as the fifth order conservation law 
\br
 \frac{1}{32}\pa_{\xi}^4 \Big\{ \frac{1}{2} \pa_{\eta} (\pa_{\xi} w)^2  + \pa_{\xi} (V-2) \Big\} = 0 \label{cons5}.
\er 
From the above conservation law it follows the fifth order  conserved charge
\br
\label{q5}
q^{(5)} & \equiv &\frac{1}{32} \frac{d^4}{dt^4}  \int dx\,\Big[ \frac{1}{2} ( \pa_{\xi} w)^2 + (V-2) \Big] , \\
 &\equiv& \frac{1}{32}  \frac{d^4}{dt^4}  \( E - P \).
\er 
So, the fifth order charge $q^{(5)}$ in (\ref{q5}) is not an independent charge  of the deformed sine-Gordon model (\ref{sg1}) even though it emerges from a truly conservation law  in the Riccati-type formulation, beyond energy and momentum.  

We will define below  a related  embedded charge  $q^{(5)}_a$ and its relevant anomaly term $\beta^{(5)}$. This charge has been computed above (see sec. \ref{sec:fifth}) by numerical simulations of two-soliton collisions for a particular deformation of the SG model.

Notice that the  `anomaly' term $\beta^{(5)}$ introduced in \cite{jhep1} can be written, in our notation,  as a  term of the r.h.s. of (\ref{n5}). So,  one can define into (\ref{n5r}) the relevant anomaly term as 
\br
\label{anomaly50}
 -2^4 \pa_{\xi} s_3&=& 2\beta^{(5)} - \pa_{\xi}[3 (\pa_{\xi} w)^3 X + \pa^3_{\xi} w X -  \pa_{\xi}^2 w \pa_{\xi} X +  \pa_{\xi} w \pa^2_{\xi} X ]  \\
\beta^{(5)} & \equiv &  \frac{1}{16} [ \pa_{\xi}^4 w + 6 (\pa_{\xi} w)^2 \pa_{\xi}^2 w ] \,  X . \label{anomaly500}
\er 

Therefore, the additional terms $\pa_{\xi} [.\,.\,.]$ appearing in the expression of $(- 2^4 \pa_{\xi} s_3)$ and provided in (\ref{anomaly50}) can be incorporated into the l.h.s. of the  conservation law (\ref{n5}). So,  we have rederived the fifth-order quasi-conservation law of \cite{jhep1} by incorporating the terms  $\pa_{\xi}[3 (\pa_{\xi} w)^3 X + \pa^3_{\xi} w X -  \pa_{\xi}^2 w \pa_{\xi} X +  \pa_{\xi} w \pa^2_{\xi} X ] $ into the l.h.s. of (\ref{n5}), while leaving the anomaly term $\beta^{(5)}$ in the r.h.s. of the same equation. Since the anomaly $\beta^{(5)}$ can be written in the form $\pa_{\eta} [.\,.\,.] + \pa_{\xi} [.\,.\,.] $, one can define the asymptotically conserved charge $q^{(5)}_a$ as
\br
\frac{d}{dt}q^{(5)}_a  &=& 2 \int  dx\,  \beta^{(5)}  \\
 &=& \frac{1}{8} \frac{d}{dt} \int  dx \, \Big\{  \frac{1}{2}  (\pa_{\xi} w)^6 + \frac{1}{4}  (\pa^3_{\xi} w)^2- \frac{5}{2}  (\pa^2_{\xi} w)^2 (\pa_{\xi} w)^2 - 6  (\pa_{\xi} w)^4 + 6 (\pa^2_{\xi} w)^2 - \frac{1}{2}  \pa^4_{\xi} w  V^{(1)} + \nonumber \\ \nonumber
&& 2 [ 2  \pa_{\xi} w \pa^3_{\xi} w  + \frac{1}{2} (\pa^2_{\xi} w)^2 +  \frac{3}{2} (\pa_{\xi} w)^4 ] V \Big\} + \\
&&  \frac{1}{8} \frac{d^2}{dt^2} \int  dx\,  \Big\{\frac{1}{2}  \pa^3_{\xi} w V^{(1)}  - 2 \pa_{\xi} w\pa^2_{\xi} w V \Big\} - \frac{1}{2}  \frac{d^3}{dt^3}  \int dx \,  \frac{1}{2} ( \pa_{\xi} w)^2. \label{charge5}
\er
Finally, the fifth-order quasi-conservation law of \cite{jhep1} can also be written as an exact conservation law provided the form (\ref{nn51}) is used in order to write an exact conservation law. The outcome will be the conservation law (\ref{cons5}).    
 
The conservation law  of order  ${\cal O}(\lambda^{6})$ becomes
\br
\label{n6}
 \pa_{\eta} \( u_6 \pa_{\xi} w \) - \pa_{\xi} \( \frac{1}{2} u_5 V^{(1)} \) = - \pa_{\xi} s_{4} && \\
 \pa_{\eta} \( u_6 \pa_{\xi} w \) - \pa_{\xi} \Big\{  \frac{1}{2} u_5 V^{(1)} + \frac{1}{16}\Big[ 3 (\pa_{\xi} w)^3\pa_{\xi} X -\frac{5}{2} (\pa_{\xi} w)^2 \pa_{\xi}^2 w X + \frac{1}{2} \pa_{\xi} w \pa_{\xi}^3 X - \frac{1}{2} \pa_{\xi}^2 w \pa_{\xi}^2 X -  \nonumber \\ \frac{1}{2} \pa_{\xi}^4 w X  \Big] \Big\} 
=0 && \label{n61} 
\er 
where the relevant expression for $\pa_{\xi} s_{4}$ has been incorporated in (\ref{n6}) to get the conservation law  (\ref{n61}).

In this way the r.h.s.'s  $[-\pa_{\xi} s_{n} \, (n=1,2,3,4)]$ of the relevant conservation laws have  been written  as $\pa_{\eta} R_n + \pa_{\xi} S_n$. We will show below that this property holds in general for each term  $[(-\pa_{\xi} s_{n}),\,n\geq 1$], of this tower of conservation laws. Therefore, the conservation laws  (\ref{anocons}) in general can be written as   
 \br
\label{anocons1}
\pa_{\eta} [ a_{\xi}^{(n)} - R_{n-2} ] + \pa_{\xi} [ a_{\eta}^{(n)} - S_{n-2}] = 0,\,\,\,\,\,n=1,2,3,...( S_{k}=R_{k}\equiv 0,\,\,\,k=-1,0).
\er
  
Then, the construction above provides an infinite tower of conservation laws (\ref{anocons1}). 

As a byproduct of our construction, we have shown that the third order asymptotically conserved charge defined in \cite{jhep1} becomes embedded into the relevant conserved charge above (\ref{asy31})-(\ref{asy32}). The concept of quasi-integrability and its asymptotically conserved charges, as introduced in \cite{jhep1} and further extended in \cite{cnsns, jhep4, jhep5} by introducing subsets of exactly conserved charges, depend on the particular field configurations one applies it to, such as the kink-kink, kink-antikink and breather configurations of the deformed model. This is in contradistinction to the usual integrability concept in which the conserved charges are defined for all fields of the model.

Moreover, we have shown through symmetry property arguments and numerical simulations in sections \ref{sec:symmetry} and \ref{sec:third} above, the exact conservation property of the charge $q_{a}^{(3)}$.  Remember that this charge has been shown to be  embedded into the dependent charge $q^{(3)}$ (\ref{ch32}), as presented in the expressions (\ref{embed1}) and (\ref{asy31})-(\ref{asy32}). The charge $q_{a}^{(3)}$ has been regarded simply as an asymptotically conserved one in \cite{jhep1, cnsns}.   

The presence of an infinite number of conservation laws is among the most important features of integrable models, since they impose strong constraints on their dynamics, and allow the existence of soliton-type solutions. As we have discussed above, in the context of deformed SG models, the set of conservation laws can be constructed directly from some structures such as the deformed Riccati-type equations of the system or the abelianization procedure  in the anomalous Lax pair formulation \cite{jhep1}. However, the rigorous proof of the mutual independence and non-triviality of the charges associated to the conservation laws (\ref{anocons1}) is often a non-trivial task. So, in the Riccaty-type pseudo-potential formulation above the charges  do not match these criteria and one has to examine order by order their non-trivialities and mutual independences.

\subsection{Riccati-type pseudo-potential and dual conservation laws}

\label{sec:riccatidual}

We present below a new formulation of the deformed SG model (\ref{sg1}) in the context of the Riccati-type pseudo-potential approach. This will constitute a dual formulation to the model presented above and play and important role, when combined with the previous constructions, in order to study the infinite towers of conserved charges expressed in laboratory coordinates $(x,t)$.  

Since the deformed SG model (\ref{sg1}) is invariant under the transformation $\eta \leftrightarrow \xi$ there will be naturally another Riccati-type formulation dual to the system (\ref{ricc1})-(\ref{ricc2s}) presented above. So, let us consider the next system of equations for the new pseudo-potential 
$\widetilde{u}$
\br
\label{ricc11}
\pa_\eta \widetilde{u} &=& -2 \lambda \, \widetilde{u} +\, \pa_\eta w + \, \pa_\eta w \,\, \widetilde{u}^2,\\
\label{ricc22}
\pa_\xi \widetilde{u} &=&-  \frac{2}{\lambda}\,  (V-2)\, \widetilde{u} - \frac{1}{2 \lambda}  \, V' + \frac{1}{2 \lambda} \, V'\, \widetilde{u}^2 + \widetilde{r} - \widetilde{u} \, \widetilde{s}. 
\er
Notice that we have performed the changes $\lambda \rightarrow  \lambda^{-1}$ and $\xi \leftrightarrow \eta$ in the linear system (\ref{ricc1})-(\ref{ricc2}) and relabelled the pseudo-potential and the auxiliary fields, while maintaining the field $w$ and the deformed sine-Gordon potential $V(w)$ unchanged. 

Next, we consider the following equations for the auxiliary fields $\widetilde{r}(\xi,\eta)$ and  $\widetilde{s}(\xi,\eta)$
\br
\label{ricc1rd}
\pa_\eta \widetilde{r} &=& -2 \lambda \, \widetilde{r} +\, \pa_\eta w  (\widetilde{u} \,\widetilde{r} + \widetilde{s})+ \lambda^{-1} \widetilde{X},\\
\label{ricc2sd}
\pa_\eta \widetilde{s} &=& \pa_\eta w  (\widetilde{u} \, \widetilde{s} - \widetilde{r})+ \lambda^{-1} \widetilde{X} \widetilde{u},\\
\widetilde{X} & \equiv & \pa_{\eta} w \( \frac{V^{(2)}}{2}+2 V -4\),\,\,\,\,\,\,V^{(2)} \equiv \frac{d^2}{d w^2} V(w).  \label{Xanomd}
\er  
So, one has a set of two deformed Riccati-type equations for the pseudo-potential $\widetilde{u}$ (\ref{ricc11})-(\ref{ricc22}) and a system of equations (\ref{ricc1rd})-(\ref{ricc2sd}) for the auxiliary fields $\widetilde{r}$ and  $\widetilde{s}$. Likewise, for the particular  potential (\ref{usg}) one has that $\widetilde{X}$ vanishes identically, and  the linear system  (\ref{ricc11})-(\ref{ricc22})  will describe the ordinary sine-Gordon integrable model, provided that $\widetilde{s}=\widetilde{r}=0$ in (\ref{ricc22}).  

Similarly, as in the previous subsection, substituting the expression for $\widetilde{u}^2$ from (\ref{ricc1rd}) into (\ref{ricc2sd}) one can get  the following relationship
\br
\label{qcons2}
\pa_{\xi} \(\widetilde{u} \, \pa_{\eta} w \) + \pa_{\eta} \( \lambda^{-1} \, (V-2) - \frac{1}{2}\lambda^{-1} \, \widetilde{u} \, V^{(1)} \) = - \pa_{\eta} \widetilde{s}   . 
\er

This equation can be used to uncover an infinite number of new conservation laws associated to the modified SG model (\ref{sg1}). So, let us consider the expansions
\br
\label{expan2}
\widetilde{u} = \sum_{n=1}^{\infty} \widetilde{u}_n \, \lambda^{-n},\, \widetilde{s}= \sum_{n=0}^{\infty} \widetilde{s}_n \, \lambda^{-(n+2)},\,\,\,\,  \widetilde{r}= \sum_{n=0}^{\infty} \widetilde{r}_n \, \lambda^{-(n+2)}.
\er
The components $\widetilde{u}_n$ can be determined recursively by substituting the above expression into (\ref{ricc11}). Whereas, the components $\widetilde{s}_n$ and $\widetilde{r}_n$ can be obtained from the system of eqs. (\ref{ricc1rd})-(\ref{ricc2sd}). In appendices C and D we provide the expressions for the first $\widetilde{u}_n,\,\widetilde{s}_n, \widetilde{r}_n$. Then, making use of these components, one can find the conservation laws, order by order in powers of $\lambda^{-1}$, by inserting those expansions into  the eq. (\ref{qcons2}).  So the $(-n)'$th order conservation law becomes
\br
\label{anocons2}
\pa_{\xi} \widetilde{a}_{\eta}^{(n)} &+& \pa_{\eta} \widetilde{a}_{\xi}^{(n)} = -   \pa_{\eta} \widetilde{s}_{n-2},\,\,\,\,\,n=1,2,3,....;\, \widetilde{s}_{-1}\equiv 0\\
\widetilde{a}_{\eta}^{(n)} &\equiv & \widetilde{u}_n \pa_{\eta} w,\,\,\,\,\widetilde{a}_{\xi}^{(n)} \equiv   (V-2) \delta_{1, n}  - \frac{1}{2} \widetilde{u}_{n-1} V^{(1)} ,\,\,\,\,\,\widetilde{u}_0 \equiv 0. 
\er 
So, the first order ${\cal O}(\lambda^{-1})$ term provides
\br
\label{n12}
\pa_{\xi} \( \frac{1}{2} (\pa_{\eta} w)^2\) + \pa_{\eta} \( V -2\) =0.
\er 
The last equation furnishes the conserved charge
\br
\widetilde{q}^{(1)} = \int dx\, [ \frac{1}{2} (\pa_{\eta} w)^2 + (V-2)].
\er
This charge combined to its dual  in the last subsection has been used to write the energy and momentum charges as in eqs. (\ref{energy1})-(\ref{momentum11}).

The next order term ${\cal O}(\lambda^{-2})$ becomes
\br
\label{n2d}
\pa_{\xi} \( \frac{1}{4} \pa_{\eta}^2 w \pa_{\eta} w\) + \pa_{\eta} \( \frac{1}{4} \pa_{\eta} w  V^{(1)}  \) = 0. 
\er

Notice that the r.h.s. vanishes since $\pa_{\eta} \widetilde{s}_{0}=0$. As usual, we  define the charge 
\br
\widetilde{q}^{(2)} &=& \int dx\, \frac{1}{4} \(\pa_{\eta}^2 w \pa_{\eta} w +  \pa_{\eta} w V^{(1)}\)\\
&=&\frac{1}{4} \frac{d}{dt} \( E + P\). \er 
So, the eq. (\ref{n2d}) does not provide an independent new charge. So, as in the dual case of last subsection, there is no an independent new charge at this order. 

As in the construction of preceding subsection, from this point forward and for the higher order charges the terms encoding the deformation away from the usual SG model, i.e. $(-\pa_{\eta} \widetilde{s}_{n-2})$ in the r.h.s. of  (\ref{anocons2}), will play an important role. 

The third  order ${\cal O}(\lambda^{-3})$ term provides
\br
\label{n32}
\pa_{\xi} \Big[ \frac{1}{8}\pa_{\eta} w \( (\pa_{\eta} w)^3 + \pa_{\eta}^3 w \)\Big] + \pa_{\eta} \( \frac{1}{8}  \pa_{\eta}^2 w V^{(1)} \) =  -\pa_{\eta} \widetilde{s}_1,
\er 
The r.h.s. of (\ref{n32}) can be written as
\br
\label{s1ad}
-\pa_{\eta} \widetilde{s}_1 &=& \frac{1}{4} \widetilde{X}  \pa_{\eta}^2 w  -\frac{1}{4}  \pa_{\eta} w \, \pa_{\eta} \widetilde{X}\\
 &=& \pa_{\xi} \widetilde{R}_1 + \pa_{\eta} \widetilde{S}_1 \label{s1bd}
  \\
\label{s1cd}
\widetilde{R}_1 &\equiv &-\frac{1}{8} [\pa_{\eta}^2 w]^2 + \frac{1}{8} [\pa_{\eta} w]^4 ,\,\,\widetilde{S}_1 \equiv - \frac{1}{8} [\pa_{\eta} w]^2 V^{(2)}.\label{rs22d}
\er
In order to write (\ref{s1bd}) starting from (\ref{s1ad}) we have used the explicit expression for $\widetilde{X}$ in (\ref{Xanomd}) and the deformed sine-Gordon eq. of motion (\ref{sg1}). Therefore, the conservation law (\ref{n32}) turns out to be  
\br
\label{nn3d}
\pa_{\xi} \Big[  \frac{1}{8}  \pa_{\eta} w \pa_{\eta}^3 w + \frac{1}{8} [\pa_{\eta}^2 w]^2   \Big] + \pa_{\eta} \Big[ \frac{1}{8}  \pa_{\eta}^2 w V^{(1)} +  \frac{1}{8} [\pa_{\eta} w]^2 V^{(2)}  \Big] = 0.
\er

Notice that this form of the third order conservation law holds strictly for deformed SG models, i.e. for models such that $\widetilde{X} \neq 0$. In the usual SG model the eq. (\ref{n32}) furnishes a conservation law at this order, provided that the r.h.s. is set to zero, since in that case $\widetilde{X}\equiv 0$. So, the charge which follows from the above conservation law (\ref{nn3d}) becomes
\br
\label{ch31d}
\widetilde{q}^{(3)} &=& \int dx\, \Big[  \frac{1}{8}  \pa_{\eta} w \pa_{\eta}^3 w + \frac{1}{8} [\pa_{\eta}^2 w]^2   + \frac{1}{8}  \pa_{\eta}^2 w V^{(1)} +  \frac{1}{8} [\pa_{\eta} w]^2 V^{(2)}  \Big] \\
&=&\frac{1}{8} \frac{d^2}{dt^2} \( E + P\).\label{ch32d} \er 

Therefore, also in this dual formulation and at this order, in contradistinction to the ordinary SG model, there is not an independent conserved charge for the deformed SG model (\ref{sg1}). 

Moreover, as in (\ref{asympd}) one can show that the third order charge and anomaly in \cite{jhep1, jhep2, cnsns}, in our notation, can be rewritten, respectively, in the form 
\br
\widetilde{q}_{a}^{(3)} &=& \int dx [ \frac{1}{8}(\pa_{\eta} w)^4 +\frac{1}{8} \pa_{\eta} w\pa_{\eta}^3 w+  \frac{1}{8}  \pa_{\eta}^2 w V^{(1)} + \frac{1}{4} \widetilde{X} \pa_{\eta}w  ] \er
  and 
\br
\label{betaxietad}
\widetilde{\beta}^{(3)} \equiv \frac{1}{2} \pa_{\eta}^2 w  \widetilde{X} = \pa_{\eta} \widetilde{S}_1 + \pa_{\xi} \widetilde{R}_1 + \frac{1}{4} \pa_{\eta} \( \pa_{\eta} w\, \widetilde{X}\),\er
with $\widetilde{R}_1, \widetilde{S}_1$ given in (\ref{rs22d}). In view of the form  that $\widetilde{\beta}^{(3)}$ takes in (\ref{betaxietad}) this `anomaly' term on the r.h.s. of (\ref{asympd}) can be promoted to the l.h.s. of that equation such that the quasi-conservation law (see eq. (2.31) of \cite{jhep1} for $n=1$) can be rewritten as a proper conservation law
\br
\label{embed1d}
\frac{d}{dt} \Big\{ \widetilde{q}_{a}^{(3)} - \int dx \widetilde{S}_1 - \int dx \widetilde{R}_1 - \int dx (\pa_{\eta} w \widetilde{X})  \Big\} = 0. 
\er
In fact, a close examination reveals that the eq. (2.31), for $n=1$, of ref. \cite{jhep1}, turns out be the same as our conservation law (\ref{nn3d}), except for an overall constant factor.     

It is possible to write  a relationship between the charge $\widetilde{q}_{a}^{(3)}$, its `anomaly'  $\widetilde{\beta}^{(3)}$ and the exactly conserved charge $\widetilde{q}^{(3)}$, so  from (\ref{n32}) or (\ref{nn3d}) one has 
\br
\label{asy31d}
\widetilde{q}_{a}^{(3)} &=&  \widetilde{q}^{(3)} + \int \, dx  \Big[\widetilde{S}_1 + \widetilde{R}_1 + \frac{1}{4} \pa_{\eta}w \widetilde{X}\Big],\\
             &=&  \frac{1}{8} \frac{d^2}{dt^2} [ E + P] + \int \, dx  \Big[ \frac{1}{8} (\pa_{\eta} w)^4- \frac{1}{8} (\pa_{\eta}^2 w)^2 +  \frac{1}{2} (\pa_{\eta} w)^2[V -2] \Big]. \label{asy32d}
\er  
The charge $\widetilde{q}_{a}^{(3)}$, conveniently combined with its dual $q_{a}^{(3)}$ in (\ref{asy31})-(\ref{asy32}), has been computed above (see sec. \ref{sec:third}) by numerical simulations of two-soliton collisions for a deformed SG model. 

The next order term ${\cal O}(\lambda^{-4})$ becomes
\br
  \label{n4id}
\pa_{\xi} \( \frac{5}{16} \pa_{\eta}^2 w (\pa_{\eta} w)^3 + \frac{1}{16}  \pa_{\eta} w \pa_{\eta}^4 w \) + \pa_{\eta} \( \frac{1}{16} V^{(1)} [ (\pa_{\eta} w)^3 + \pa_{\eta}^3 w] \) = \pa_{\eta} \widetilde{s}_{2} .
\er
The r.h.s. of this equation can be written as 
\br
\pa_{\eta} \widetilde{s}_{2} &=&  \frac{1}{8} \widetilde{X} \pa_{\eta}^3 w - \frac{1}{8}  \pa_{\eta} w \pa_{\eta}^2 \widetilde{X}\\
 &=& \frac{1}{8} \pa_{\eta} [\pa_\eta^2 w \widetilde{X} - \pa_\eta w \pa_\eta \widetilde{X}].
\er
Therefore one has the conservation law
\br
\label{n4iid}
\pa_{\xi} \( \frac{5}{16} \pa_{\eta}^2 w (\pa_{\eta} w)^3 + \frac{1}{16}  \pa_{\eta} w \pa_{\eta}^4 w \) + \pa_{\eta} \( \frac{1}{16}  [ (\pa_{\eta} w)^3 + \pa_{\eta}^3 w]V^{(1)} - \frac{1}{8}  [\pa_\eta^2 w \widetilde{X} - \pa_\eta w \pa_\eta \widetilde{X}]   \) = 0.   
\er 
From (\ref{n4iid}) one can define the charge 
\br
\widetilde{q}^{(4)} &=& \int dx \,\Big\{\frac{5}{16} \pa_{\eta}^2 w (\pa_{\eta} w)^3 + \frac{1}{16}  \pa_{\eta} w \pa_{\eta}^4 w  +   \frac{1}{16}  [ (\pa_{\eta} w)^3 + \pa_{\eta}^3 w]V^{(1)} - \frac{1}{8}  [\pa_\eta^2 w \widetilde{X} - \pa_\eta w \pa_\eta \widetilde{X}]  \Big\}, \\
& \equiv & 0.
\er
So, it has been shown that the  charge at this order also vanishes identically for suitable boundary conditions.

The term of order  ${\cal O}(\lambda^{-5})$ provides the next quasi-conservation law 
\br
\label{n5d}
\pa_{\xi} \( \frac{11}{32}(\pa_{\eta}^2 w)^2 (\pa_{\eta} w)^2 + \frac{7}{32}  \pa_{\eta}^3 w (\pa_{\eta} w)^3  +  \frac{1}{16}(\pa_{\eta} w)^6 + \frac{1}{32}\pa_{\eta} w \pa_{\eta}^5 w \)  -  \pa_{\eta} \Big[ \frac{1}{2} V^{(1)} \widetilde{u}_4 \Big] = -\pa_{\eta} \widetilde{s}_3\er
\br
 -\pa_{\eta} \widetilde{s}_3&=&-\frac{3}{16} (\pa_{\eta} w)^3 \pa_{\eta} \widetilde{X}+\frac{1}{16}
\( \pa_{\eta}^4 w + \pa_{\eta}[(\pa_{\eta} w)^3] \) \,  \widetilde{X}- \frac{1}{16} \pa_{\eta} w \pa^3_{\eta} \widetilde{X}.\er 

The r.h.s. of the last eq. can be written as 
\br
\label{nn51d}
 -\pa_{\eta} \widetilde{s}_3 &=& \pa_{\xi} \widetilde{R}_3 + \pa_{\eta} \widetilde{S}_3 
 \er
where $\widetilde{R}_3$ and $\widetilde{S}_3$ are provided  in (\ref{nn52d})-(\ref{nn53d}).  Then, a lenghty calculation allow us to write the eq. (\ref{n5d})  as the fifth order conservation law 
\br
\frac{1}{32}\pa_{\eta}^4 \Big\{ \frac{1}{2} \pa_{\xi} (\pa_{\eta} w)^2  + \pa_{\eta} (V-2) \Big\} = 0 \label{cons5d}.
\er 
From the above conservation law it follows the fifth order  conserved charge
\br
\label{q5d}
\widetilde{q}^{(5)} & \equiv &\frac{1}{32} \frac{d^4}{dt^4}  \int dx\,\Big[ \frac{1}{2} ( \pa_{\eta} w)^2 + (V-2) \Big] , \\
 &\equiv& \frac{1}{32}  \frac{d^4}{dt^4}  \( E + P \).
\er 
So, the fifth order dual charge $\widetilde{q}^{(5)}$ in (\ref{q5d}) is not an independent charge  of the deformed sine-Gordon model (\ref{sg1}) even though it emerges from a truly conservation law  in the Riccati-type dual formulation, beyond energy and momentum.  

We will define below  a related  embedded charge  $\widetilde{q}^{(5)}_a$ and its relevant anomaly term $\widetilde{\beta}^{(5)}$. This charge has been computed above (see sec.  \ref{sec:fifth})  by numerical simulations of two-soliton collisions for a particular deformation of the SG model.

Notice that the  `anomaly' term $\widetilde{\beta}^{(5)}$ introduced in \cite{jhep1} can be written, in our notation,  as a  term of the r.h.s. of (\ref{n5d}). So,  one can define in (\ref{n5d}) the relevant anomaly term as 
\br
\label{anomaly50d}
 -2^4 \pa_{\eta} \widetilde{s}_3&=& 2\widetilde {\beta}^{(5)} - \pa_{\eta}[3 (\pa_{\eta} w)^3 \widetilde{X} + \pa^3_{\eta} w \widetilde{X} -  \pa_{\eta}^2 w \pa_{\eta} \widetilde{X} +  \pa_{\eta} w \pa^2_{\eta} \widetilde{X} ]  \\
\widetilde{\beta}^{(5)} & \equiv &  \frac{1}{16} [ \pa_{\eta}^4 w + 6 (\pa_{\eta} w)^2 \pa_{\eta}^2 w ] \,  \widetilde{X} . \label{anomaly500d}
\er 

Therefore, the additional terms inside $\pa_{\eta} [.\,.\,.]$ appearing in the expression of $(- 2^4 \pa_{\eta} s_3)$ and provided in (\ref{anomaly50d}) can be incorporated into the l.h.s. of the  conservation law (\ref{n5d}). So,  we have rederived the fifth-order quasi-conservation law of \cite{jhep1} by incorporating the terms  $\pa_{\eta}[3 (\pa_{\eta} w)^3 \widetilde{X} + \pa^3_{\eta} w \widetilde{X} -  \pa_{\eta}^2 w \pa_{\eta} \widetilde{X} +  \pa_{\eta} w \pa^2_{\eta} \widetilde{X} ] $ into the l.h.s. of (\ref{n5d}), while leaving the anomaly term $\widetilde{\beta}^{(5)}$ in the r.h.s. of the same equation. Since the anomaly $\widetilde{\beta}^{(5)}$ can be written in the form $\pa_{\eta} [.\,.\,.] + \pa_{\xi} [.\,.\,.] $, one can define the asymptotically conserved charge $\widetilde{q}^{(5)}_a$ as
\br
\frac{d}{dt}\widetilde{q}^{(5)}_a  &=& 2 \int  dx\,  \widetilde{\beta}^{(5)}  \\
 &=& \frac{1}{8} \frac{d}{dt} \int  dx \, \Big\{  \frac{1}{2}  (\pa_{\eta} w)^6 + \frac{1}{4}  (\pa^3_{\eta} w)^2- \frac{5}{2}  (\pa^2_{\eta} w)^2 (\pa_{\eta} w)^2 - 6  (\pa_{\eta} w)^4 + 6 (\pa^2_{\eta} w)^2 - \frac{1}{2}  \pa^4_{\eta} w  V^{(1)} + \nonumber \\ \nonumber
&& 2 [ 2  \pa_{\eta} w \pa^3_{\eta} w  + \frac{1}{2} (\pa^2_{\eta} w)^2 +  \frac{3}{2} (\pa_{\eta} w)^4 ] V \Big\} + \\
&&  \frac{1}{8} \frac{d^2}{dt^2} \int  dx\,  \Big\{\frac{1}{2}  \pa^3_{\eta} w V^{(1)}  - 2 \pa_{\eta} w\pa^2_{\eta} w V \Big\} - \frac{1}{2}  \frac{d^3}{dt^3}  \int dx \,  \frac{1}{2} ( \pa_{\eta} w)^2. \label{charge5d}
\er

We have computed in sec. 3.3  this charge $\widetilde{q}^{(5)}_a$, combined with its dual (\ref{charge5}), by numerical simulations of two-soliton collisions for a  particular deformation of the SG model.

Finally, the fifth-order dual quasi-conservation law of \cite{jhep1} can also be written as an exact conservation law provided the form (\ref{nn51d}) is used in order to write an exact conservation law. The outcome will be the conservation law (\ref{cons5d}).  

For completeness we provide the next conservation law  of order  ${\cal O}(\lambda^{-6})$ 
\br
\label{n6d}
 \pa_{\xi} \( u_6 \pa_{\eta} w \) - \pa_{\eta} \( \frac{1}{2} \widetilde{u}_5 V^{(1)} \) = - \pa_{\eta} \widetilde{s}_{4} && \\
 \pa_{\xi} \( \widetilde{u}_6 \pa_{\eta} w \) - \pa_{\eta} \Big\{  \frac{1}{2} \widetilde{u}_5 V^{(1)} + \frac{1}{16}\Big[ 3 (\pa_{\eta} w)^3\pa_{\eta} \widetilde{X} -\frac{5}{2} (\pa_{\eta} w)^2 \pa_{\eta}^2 w \widetilde{X} + \frac{1}{2} \pa_{\eta} w \pa_{\eta}^3 \widetilde{X} - \frac{1}{2} \pa_{\eta}^2 w \pa_{\eta}^2 \widetilde{X} -  \nonumber \\ \frac{1}{2} \pa_{\eta}^4 w \widetilde{X}  \Big] \Big\} 
=0 && \label{n61d} 
\er 
where the relevant expression for $\pa_{\eta} \widetilde{s}_{4}$ has been incorporated in (\ref{n6d}) in order to get (\ref{n61d}).
 
In this way,  the r.h.s.'s  $[-\pa_{\eta} \widetilde{s}_{n} \, (n=1,2,3,4)]$ of the relevant dual conservation laws have  been written  as $\pa_{\xi} \widetilde{R}_n + \pa_{\eta} \widetilde{S}_n$. Therefore, the conservation laws  (\ref{anocons2}) can be written as   
 \br
\label{anocons1d}
\pa_{\xi} [ \widetilde{a}_{\eta}^{(n)} - \widetilde{R}_{n-2} ] + \pa_{\eta} [ \widetilde{a}_{\xi}^{(n)} - \widetilde{S}_{n-2}] = 0,\,\,\,\,\,n=1,2,3,...( \widetilde{S}_{k}= \widetilde{R}_{k}\equiv 0,\,\,\,k=-1,0).
\er
  
So, beyond the energy-momentum charges, the above towers of higher order asymptotically conserved charges share the same form as the usual sine-Gordon charges, even though  the dynamics governing their behaviour is related to the deformed  sine-Gordon with potential $V(w)$ supporting solitary waves.

\section{New pseudo-potentials and non-local conservation laws}
 \label{newdiff}

In this section we  provide new towers of conservation laws  by considering other pseudo-potential representations of the deformed SG equation. The procedures will carefully take into account the structures of the deformation encoded in the variable $X$ in (\ref{Xanom}) and the auxiliary  fields $r$ and $s$ of (\ref{ricc1r})-(\ref{ricc2s}), as well as encoded in the dual expressions $\widetilde{X}$ in (\ref{Xanomd}) and the fields $\widetilde{r}$ and $\widetilde{s}$ in (\ref{ricc1rd})-(\ref{ricc2sd}). 
 
Let us define $\psi \equiv r-u s$ and  write the Riccati-type eq. (\ref{ricc2}) as  
\br 
\label{ricc2psi}
\pa_\eta u &=& -  2 \lambda\,  (V-2)\, u - \frac{1}{2} \,\lambda\, V^{(1)} + \frac{1}{2} \lambda\, V^{(1)}\, u^2 + \psi. 
\er

Therefore, imposing the compatibility condition to the pair of eqs. (\ref{ricc1}) and (\ref{ricc2psi}), and taking into account  the auxiliary system of equations (\ref{ricc1r})-(\ref{ricc2s}), one gets a linear  first order equation for $\psi$
\br
\label{psi1}
\pa_{\xi} \psi  +2 \lambda^{-1} \psi  -2 u \pa_{\xi} w\, \psi &=&  (2 \lambda -2 u - \lambda\, \pa_{\xi} u) Y,\,\,\,\,\,\,\\
\label{YY}
Y & \equiv & \frac{1}{2}V^{(2)}(w) + 2 V(w) -4.
\er  

This is a non-homogeneous ordinary differential equation for $\psi$ in the variable $\xi$, which can be  integrated  by quadratures. Its general solution becomes 
\br
\nonumber
\psi (\xi, \eta) &=& C\, e^{-\frac{2}{\lambda} \int^{\xi}\, [1- \lambda u(\xi', \eta) \frac{\pa w(\xi', \eta)}{\pa \xi'} ] d \xi'} - e^{-\frac{2}{\lambda} \int^{\xi}\, [1- \lambda u(\xi', \eta) \frac{\pa w(\xi', \eta)}{\pa \xi'} ] d \xi'} \int^{\xi} e^{\frac{2}{\lambda} \int^{\xi''}\, [1- \lambda u(\xi', \eta) \frac{\pa w(\xi', \eta)}{\pa \xi'}] d \xi'} \times \\
&& Y(\xi'', \eta)  \Big\{ 2 u(\xi'',\eta) + \lambda \Big[ \frac{\pa  u(\xi'', \eta)}{\pa \xi''}  - 2 \frac{\pa w(\xi'', \eta)}{\pa \xi''}  \Big]    \Big\}  d\xi'' .\label{psi11}
\er
Imposing the condition $\psi=0 $ for $Y=0$ to this solution, as it must hold for the usual SG model, one must set $C=0$. In fact, this condition removes the contribution of the  homogeneous sector of the differential equation (\ref{psi1}) to the general solution in (\ref{psi11}) . So, one has 
\br
\nonumber
\psi (\xi, \eta) &=&  - e^{-\frac{2}{\lambda} \int^{\xi}\, [1- \lambda u(\xi', \eta) \frac{\pa w(\xi', \eta)}{\pa \xi'} ] d \xi'} \int^{\xi} e^{\frac{2}{\lambda} \int^{\xi''}\, [1- \lambda u(\xi', \eta) \frac{\pa w(\xi', \eta)}{\pa \xi'}] d \xi'} \times\\
&& Y(\xi'', \eta)  \Big\{ 2 u(\xi'',\eta) + \lambda \Big[ \frac{\pa  u(\xi'', \eta)}{\pa \xi''}  - 2 \frac{\pa w(\xi'', \eta)}{\pa \xi''}  \Big]    \Big\}  d\xi''.\label{psi111}
\er
The expression for $\psi$ in  (\ref{psi111}) is highly non-local and, once inserted into (\ref{ricc2psi}), the new system of  eqs. (\ref{ricc1}) and (\ref{ricc2psi}) will provide a new non-local Riccati-type representation for the DSG model (\ref{sg1}).

However, we are interested in obtaining a new set of local conservation laws associated to the system of Riccati-type equations (\ref{ricc1}) and (\ref{ricc2psi}); so,  let us define a new pseudo-potential as
\br
\label{PPsi2}
\Psi & \equiv & -\pa_{\xi} s\\
\label{PPsi21}
&=&  -\lambda X u + \pa_{\xi} w \, \psi,
\er
where the eq. (\ref{ricc2s}) has been used in order to trade $\pa_{\xi} s$ for $\psi$, i.e. $\pa_{\xi} s = \lambda X u - \pa_{\xi} w \, \psi$. So, the r.h.s. of the conservation law (\ref{qcons}) can be written as 
\br
\label{psi2}
\pa_{\eta} \(u \, \pa_{\xi} w \) + \pa_{\xi} \( \lambda \, (V-2) - \frac{1}{2}\lambda \, u \, V^{(1)} \)   &=& \Psi.
\er
The quantity $\Psi$ satisfies a linear  ordinary differential equation in the independent variable $\xi$
\br
\label{xiPsi}
\pa_{\xi}\Psi = \Big[- 2  \lambda^{-1} + 2 u\, \pa_{\xi}w + \frac{\pa^2_{\xi}w}{\pa_{\xi}w}\Big]\Psi - \lambda u \, \pa_{\xi}w \, \pa_{\xi} Y.
\er
When the components $u_n$ of the quantity $u$ (\ref{expan}) are taken into account from the appendix \ref{fsca1}, the equation (\ref{xiPsi}) can be solved for $\Psi$ by expanding it as a power series in the spectral parameter $\lambda$
\br
\label{seriesPsi}
\Psi &=& \sum_{n=1}^{\infty} \lambda^{2+n} \Psi_{n}.  
\er
The first set of components $\Psi_{n}$ are provided in the appendix \ref{aPsi1}.  
 
The conservation law (\ref{psi2}), taking into account the series expansion for $\Psi$ and its components provided in (\ref{Psin}), reproduces the  set of conservation laws presented in section  \ref{sec:riccati}.

An important observation is that from (\ref{xiPsi}) one can write the Riccati-type pseudo-potential $u$ in terms of the quantity $\Psi$ and the field $w$ as
\br
\label{uu1}
u = \frac{\lambda \Psi \pa_{\xi}^2 w - \pa_{\xi} w ( 2 \Psi + \lambda \pa_{\xi} \Psi)}{\lambda^2 (\pa_{\xi} w)^3  Y^{(1)} - 2\lambda (\pa_{\xi} w)^2\Psi}.
\er
Notice that by introducing (\ref{uu1}) into the l.h.s. of equation (\ref{psi2}) one can get for  $\Psi$  an expression of the form  
\br
\label{psiRS}
\Psi \equiv \pa_{\eta} R + \pa_{\xi} S,\er
where $R$ and  $S$  can be defined as  some functionals of the field $w$ and its derivatives, once the expressions for $\Psi$ and its components $\Psi_n$  from (\ref{Psin}) are substituted into the l.h.s. of (\ref{psi2}). The identity  (\ref{psiRS}) shows that in general one must have  $\Psi_{n-2}  \equiv \pa_{\eta} R_{n-2} + \pa_{\xi} S_{n-2} [=(-\pa_{\xi} s_{n}),\,n\geq 3]$. This property has been verified, order by order in powers of $\lambda$, in the construction of the first set of conservation laws and their associated charges $q^{(n)},\ n=1,2,...,6$, in sec. \ref{sec:riccati}.

Likewise, in order to write a dual set of local conservation laws associated to the system of Riccati-type equations (\ref{ricc11}) and (\ref{ricc22}), one can define the quantity 
\br
\label{PPsi2d}
\widetilde{\Psi} & \equiv & -\pa_{\eta} \widetilde{s}\\
\label{PPsi21d}
&=&  -\lambda^{-1} \widetilde{X}\, \widetilde{u} + \pa_{\eta} w \, \widetilde{\psi},
\er
where the eq. (\ref{ricc2sd}) has been used in order to trade $\pa_{\eta} \widetilde{s}$ for $\widetilde{\psi}$, i.e. $\pa_{\eta} \widetilde{s} = \lambda^{-1} \widetilde{X}\, \widetilde{u} - \pa_{\eta} w \, \widetilde{\psi}$. So, the r.h.s. of the conservation law (\ref{qcons2}) can be written as 
\br
\label{psi2d}
\pa_{\xi} \( \widetilde{u} \, \pa_{\eta} w \) + \pa_{\eta} \( \lambda^{-1} \, (V-2) - \frac{1}{2}\lambda^{-1} \, \widetilde{u} \, V^{(1)} \)   &=& \widetilde{\Psi},
\er
where the pseudo-potential  $\widetilde{\Psi}$ satisfies a linear  ordinary differential equation in the independent variable $\eta$
\br
\label{xiPsid}
\pa_{\eta}\widetilde{\Psi} = \Big[- 2  \lambda + 2 \widetilde{u}\, \pa_{\eta} w + \frac{\pa^2_{\eta} w}{\pa_{\eta} w}\Big]\widetilde{\Psi} - \lambda^{-1} \widetilde{u} \, \pa_{\eta}w \, \pa_{\eta} Y.
\er
The equation (\ref{xiPsid}) can be solved for $\Psi$ by expanding it as a power series in the spectral parameter $\lambda$
\br
\label{seriesPsid}
\widetilde{\Psi} &=& \sum_{n=1}^{\infty} \lambda^{-(2+n)} \widetilde{\Psi}_{n}.  
\er
The first set of components $\widetilde{\Psi}_{n}$ are provided in the appendix \ref{aPsi11}.  
 
One can verify that the conservation law (\ref{psi2d}), taking into account the series expansion of $\widetilde{\Psi}$ and its components provided in the appendix \ref{aPsi11}, reproduces the  set of dual conservation laws presented in subsection  \ref{sec:riccatidual}.

\subsection{Pseudo-potentials and a linear system associated to DSG}
\label{sec:linear}

In this section we tackle the problem of writing a linear system of equations associated to the DSG model.  We will proceed by performing some transformations to the Riccati eq. (\ref{ricc1}) and to the  conservation law (\ref{qcons}), as well as to the eq. (\ref{xiPsi}) written for the the auxiliary field $s$. So, let us consider the transformation
\br
\label{trans1}
u = - \frac{1}{\pa_{\xi} w} \pa_{\xi}\( \log{\phi}\).
\er

Inserting (\ref{trans1}) into (\ref{ricc1}) one gets the  equation
\br
\label{lax1}
\pa^2_{\xi} \phi + (\pa_{\xi} w)^2 \phi +2 \lambda^{-1} \pa_{\xi} \phi - \frac{\pa }{\pa_{\xi}} 
[ \log{\(\pa_{\xi} w\)}] \, \pa_{\xi}\phi \equiv L_1 \phi =0.
\er
Similarly, inserting (\ref{trans1}) into (\ref{qcons}) and performing a $\xi-$integration once, one gets
\br
\label{lax2}
\pa_{\eta} \phi -\lambda [ V(w)- 2] \phi - \frac{1}{2} \lambda \frac{V'(w)}{\pa_{\xi} w} \, \pa_{\xi} \phi - s \, \phi \equiv L_2 \phi =0.
\er

In addition, combining (\ref{trans1}) and (\ref{xiPsi}), and taking into account  the expression $\Psi = - \pa_{\xi} s$ defined in (\ref{PPsi2}), one gets the next equation for the quantity $s$
\br
\label{xis}
\pa_{\xi}^2 s = \Big[- 2  \lambda^{-1} - 2 \frac{\pa_{\xi} \phi}{\phi} + \frac{\pa^2_{\xi}w}{\pa_{\xi}w}\Big] \pa_\xi s  - \lambda  \frac{\pa_{\xi} \phi}{\phi} \, \pa_{\xi} Y.
\er
From (\ref{lax1})-(\ref{lax2}) one can show that the compatibility condition is satisfied, $\pa_{\eta} (\pa^2_{\xi} \phi  )- \pa_{\xi}^2 (\pa_{\eta} \phi) =0$, provided that $s$ satisfies the second order differential equation (\ref{xis}) and $w$ the deformed sine-Gordon equation of motion (\ref{sg1}). In addition, from (\ref{lax2}) and (\ref{xis}) one can write the relevant expressions for $  \pa_{\eta} \log{\phi}$ and   $  \pa_{\xi} \log{\phi}$, such that their compatibility condition reproduces the conservation law (\ref{psi2}). 

Notice that  the pseudo-potential approach has been used in \cite{nucci} in order to obtain the Lax pair of the usual SG model. In fact, in the limit $V \rightarrow V_{SG}$ one has that $Y=0$ (SG limit) and $\Psi=0$ implying   $s=0$ and the set of  operators $\{ L_1\,,\,L_2\}$ in (\ref{lax1})-(\ref{lax2}) turn out to be the Lax representation of the usual SG model, provided that the potential $V$ takes the form (\ref{usg}).  
  
On the other hand, the so-called non-homogeneous nonlinear Lax pair associated to an integrable system has been discussed in \cite{lou} starting from a known Lax pair and the Darboux transformation of the model. Along the same lines, since the system of equations (\ref{lax1})-(\ref{lax2}) satisfy the compatibility condition, as discussed above, the system (\ref{lax1})-(\ref{lax2}) can be regarded as a non-local and non-linear Lax pair  representation of the deformed SG model (\ref{sg1}).

Therefore, following the results above and the same general lines suggested in \cite{lou}, it is worth to pursue a linear formulation of the DSG model. Next, we will undertake this goal by seeking a linear and a first order in $\xi-$derivative ($\eta-$derivative) differential equation for the pseudo-potential  $\phi$ as
\br
\label{lin01}
\pa_{\xi} \phi &=&  {\cal A}_{\xi}  \phi\\
\label{lin02} 
\pa_{\eta} \phi &=&  {\cal A}_{\eta} \phi,
\er
such that the compatibility condition for the system above defines the equation of motion
\br
\label{consist1}
\pa_{\eta}  {\cal A}_{\xi} - \pa_{\xi}  {\cal A}_{\eta} =0.
\er  
Therefore, substituting the above Ansatz (\ref{lin01}) into (\ref{lax1}) one gets a Riccati equation for the quantity ${\cal A}_{\xi} $
\br
\label{riccatiA}
\pa_{\xi} {\cal A}_{\xi} = -2 a_1 + \(\frac{1}{2} \pa_{\xi}\log{a_1}-\frac{2}{\lambda}\) {\cal A}_{\xi}-  {\cal A}_{\xi} ^2,\,\,\,\,\,\,\,\,\,\,\,\, a_1 \equiv \frac{1}{2} (\pa_{\xi} w)^2.
\er
Notice that the linear system (\ref{lin01})-(\ref{lin02}), as well as the equation of motion  (\ref{consist1}) are defined up to a gauge transformation
\br
\label{gauge1}
\phi  &\rightarrow & e^{\Lambda} \phi \\
\label{gauge2}
 {\cal A}_{\xi} &\rightarrow&  {\cal A}_{\xi} + \pa_{\xi} \Lambda,\,\,\,\, {\cal A}_{\eta} \rightarrow  {\cal A}_{\eta} + \pa_{\eta} \Lambda,
\er
for an arbitrary function $\Lambda$.

Taking into account  (\ref{lin01})-(\ref{lin02}) from (\ref{lax2}) one can get an expression for the quantity $s$ 
\br
\label{ss0}
s =  {\cal A}_{\eta}  - \lambda (V-2) - \frac{1}{2} \lambda \frac{V^{(1)}}{\pa_{\xi} w} {\cal A}_{\xi}. 
\er
Since $s$ in (\ref{ss0}) depends on the connection $({\cal A}_{\xi}\, ,\,{\cal A}_{\eta})$, defined in the linear system (\ref{lin01})-(\ref{lin02}), and the quantities $V,\, V^{(1)}$ and $\pa_{\xi} w$, one can argue that $s$ depends only on the field $w$ of the model and its derivatives. The above results and the careful inspection of the terms $\pa_{\xi}  \phi $ and $\pa_{\eta}  \phi$ appearing in (\ref{lax1})-(\ref{xis}), as well as the system (\ref{lin01})-(\ref{lin02}), suggest that the model might possess a linear formulation. In the following, taking into account  the gauge freedom (\ref{gauge1})-(\ref{gauge2}) and a certain amount of guesswork, we will determine the simplest expressions for the quantities ${\cal A}_{\eta}$ and $ {\cal A}_{\xi}$ of the linear system (\ref{lin01})-(\ref{lin02}). In this way, we propose the following system of equations as a linear formulation of the deformed SG model\footnote{We will provide below a gauge transformation between the systems (\ref{2lax1})-(\ref{Aeta}) and (\ref{lin01})-(\ref{lin02}).}  
\br
\label{2lax1}
 {\cal L}_1 \Phi &=&0,\\
\label{l1} {\cal L}_1 & \equiv &\pa_{\xi} - A_{\xi} \\
A_{\xi} &\equiv &   \frac{\lambda}{2} (\pa_{\xi} w)^2 - 2 \frac{ (\pa_{\xi} w)^3}{\pa_{\xi}^2 w} \label{Axi}\\ 
 \label{2lax2}
{\cal L}_2 \Phi & =& 0,\\
\label{l2} {\cal L}_2  &\equiv& \pa_{\eta} - A_{\eta}\\
A_{\eta}& \equiv & -2 \lambda  - \lambda V  
+ \zeta\,     \label{Aeta}
\er
where the auxiliary non-local field $\zeta$ is defined as
\br
\label{zeta}
\zeta &=& \int^{\xi} d\xi' \Big[ 6 V^{(1)}  \frac{(\pa_{\xi'} w)^2}{\pa_{\xi'}^2 w}
- 2 V^{(2)}   \frac{(\pa_{\xi'} w)^4 }{(\pa_{\xi'}^2 w)^2}\Big]. 
\er
In fact, taking into account the expression for the auxiliary field $\zeta$, the compatibility condition of the linear problem defined by the system of eqs. (\ref{2lax1}) and (\ref{2lax2}) provides the equation 
\br
\label{zeroeq}
   \Delta(\xi,\eta)\, \lambda - 6 \frac{\pa_{\xi} w}{\pa_{\xi}^2 w}  \Delta(\xi,\eta) + 2 
\frac{(\pa_{\xi} w)^2}{(\pa_{\xi}^2 w)^2} \pa_{\xi}   \Delta(\xi,\eta)   =0,\\
 \Delta(\xi,\eta) \equiv \pa_{\xi}\pa_{\eta} w + V^{(1)}(w).
\er
The first term in the above equation  is linear in the spectral parameter $\lambda$, and then the quantity  $\Delta(\xi,\eta)$ must vanish, furnishing in this way the deformed SG equation of motion (\ref{sg1}). The remaining terms in (\ref{zeroeq}) must also vanish once $\Delta(\xi,\eta)=0$ is imposed. So, the operators ${\cal L}_1  $ and ${\cal L}_2$ constitute a pair of linear operators related to the deformed SG model (\ref{sg1}).

The linear systems (\ref{lin01})-(\ref{lin02}) with connection $\( {\cal A}_{\xi}\, ,\, {\cal A}_{\eta}\)$ and (\ref{2lax1})-(\ref{Aeta}) with connection $\(A_{\xi}\,,\, A_{\eta}\)$, respectively, are related by the gauge transformation of the type (\ref{gauge1})-(\ref{gauge2}). So, one has
\br
\label{expL}
\phi &=& e^{-\Lambda} \, \Phi\\
\label{gauge11}
A_{\xi} &=& {\cal A}_{\xi} + \pa_{\xi} \Lambda\\
A_{\eta} &=& {\cal A}_{\eta} + \pa_{\eta} \Lambda,
\label{gauge22}
\er
where the quantity $\Omega \equiv \pa_{\xi}\Lambda$ satisfies the Riccati equation
\br
\pa_{\xi} \Omega &=& \frac{2}{\lambda} a_0 + a_0^2 + 8 a_1+ 2\lambda\, a_0 a_1 + \lambda^2 a_1^2- 4 \lambda\, \frac{a_1^2}{a_0} + \pa_{\xi}a_0-\(\frac{2}{\lambda} + 2 a_0 + 2 \lambda\,  a_1  + 4\frac{a_1}{a_0} \) \Omega + \Omega^2,\\
a_0&\equiv& -2\frac{(\pa_{\xi}w)^3}{\pa_{\xi}^2 w},\,\,\,\,\,\,b_1 \equiv -2 - V,
\er
which is obtained from (\ref{riccatiA}) and (\ref{gauge11}).

Next, as a first application of the linear problem above, let us construct the energy and momentum charges. Let us write the linear system (\ref{2lax1})-(\ref{Aeta}) as
\br
\pa_{\xi} \Phi &=& A_{\xi} \Phi; \label{lin1}\\
\pa_{\eta} \Phi &=& A_{\eta} \Phi \label{lin2}\\
A_{\xi}&\equiv& a_0 + \lambda \, a_1;\,\,\,\,\,\,A_{\eta}\equiv b_0 + \lambda \, b_1,\label{AB}\\
a_0& \equiv & - 2 \frac{ (\pa_{\xi} w)^3}{\pa_{\xi}^2 w} ;\,\,\,\,b_0 \equiv \zeta = \int^{\xi} d\xi' \Big[6 V^{(1)}  \frac{(\pa_{\xi'} w)^2}{\pa_{\xi'}^2 w}
- 2 V^{(2)}   \frac{(\pa_{\xi'} w)^4 }{(\pa_{\xi'}^2 w)^2}\Big];\label{ab00}\\
a_1 &\equiv &  \frac{1}{2} (\pa_{\xi} w)^2 ;\,\,b_1 \equiv - 2 - V  
\label{ab11}
\er
So, consider the identify $\pa_{\eta}(\frac{\pa_{\xi}\Phi }{\Phi}) - \pa_{\xi}(\frac{\pa_{\eta}\Phi}{\Phi})=0$, which taking into account the linear system  (\ref{lin1})-(\ref{lin2}) becomes
\br
\label{ABd}
\pa_{\eta} A_{\xi} - \pa_{\xi} A_{\eta} &=& 0.
\er
So, using (\ref{AB}) one can write
\br
\label{nl0} \pa_{\eta} a_0 - \pa_{\xi} b_0 &=& 0\\
\pa_{\eta} a_1 - \pa_{\xi} b_1 &=& 0  \label{em1}
\er  
In fact, these eqs. define two conservation laws. Then considering the expresions of $a_0$ and $b_0$ in (\ref{ab00}) the eq. (\ref{nl0}) defines a  non-local conservation law. However,  the eq. (\ref{nl0}) can be written conveniently as 
\br
\pa_{\eta} [\frac{1}{3} (\pa_{\xi} w)^3] + \pa_{\xi}[V \pa_{\xi} w] = V \pa^2_{\xi} w.
\er
This is just the eq. (\ref{qc}) for $N=3$, i.e.  the rational non-local conservation law (\ref{nl0}) hides a quasi-conservation law.  

Whereas, the second eq. (\ref{em1}) together with the expressions for $a_1$ and $b_1$ in (\ref{ab11}) provides the energy-momentum conservation law $\pa_{\eta} [\frac{1}{2} (\pa_
{\xi} w)^2] + \pa_{\xi} [ V-2 ] =0$, which has already been discussed in the pseudo-potential approach (\ref{n1}).

For the fields and potentials satisfying the symmetry (\ref{pxt}) one can get another linear system. So,  from  (\ref{lin1})-(\ref{ab11}), performing the transformation (\ref{pxt})-(\ref{pxt1}), one gets 
\br
\pa_{\xi} \chi &=& \widetilde{A}_{\xi} \chi; \label{lin1tx}\\
\pa_{\eta} \chi &=& \widetilde{A}_{\eta} \chi \label{lin2tx}\\
\widetilde{A}_{\xi}&\equiv& a_0 - \lambda \, a_1;\,\,\,\,\,\,\widetilde{A}_{\eta}\equiv b_0 - \lambda\,   b_1\label{ABtx}.
\er
The new potentials $\widetilde{A}_{\xi}$ and $\widetilde{A}_{\eta}$ can be obtained by making $\l \rightarrow -\l$ in the potentials of (\ref{AB}). The system (\ref{lin1tx})-(\ref{ABtx}) reproduces the same eq. of motion (\ref{sg1}) and the conservation laws (\ref{nl0})-(\ref{em1}).

In addition, taking into account the system of eqs. (\ref{ricc11}) and (\ref{psi2d}), as well as  the eq. (\ref{xiPsid}), of the dual Riccati-type representation, one can  introduce a new  pseudo-potential $\widetilde{\phi}$ through a transformation analogous to the one in (\ref{trans1}). Then, following similar steps as above, one can define a dual linear system of eqs. for the new quantity $\widetilde{\Phi}$ and associate to it the new pair of operators $\{ \widetilde{{\cal L}}_1,\,\widetilde{{\cal L}}_2 \}$. So, this construction will provide the dual linear representation of the DSG model.

\subsection{Non-local conservation laws}
 \label{ss:nonlo}

For non-linear equations, not necessarily integrable, which can be derived from compatibility conditions of an associated  
linear system possessing a spectral  parameter, a technique for obtaining explicit expressions of local and non-local  currents have been developed in the literature (see e.g. \cite{vladimirov}). In certain models the non-local conserved charges, as in the non-linear $\sigma-$model, imply absence of  particle production and the first non-trivial one alone fixes almost completely the on-shell dynamics of the model (see e.g. \cite{abdalla, luscher}). These charges may be constructed through an iterative procedure introduced by Br\'ezin, et.al. \cite{brezin}. We follow this method to construct a set of infinite number of non-local conservation laws for the system (\ref{lin1})-(\ref{lin2}). In fact, this system satisfies the properties: i) $(A_{\xi}, A_{\eta})$ is a ``pure gauge"; i.e. $A_{\mu}=\pa_{\mu}\Phi \Phi^{-1},\,\,\mu=\xi,\eta$; ii) $J_{\mu}=(A_{\xi}, A_{\eta})$ is a conserved current satisfying (\ref{ABd}). So, we can construct an infinite set of non-local 
conserved currents through an inductive procedure. Let us define the currents 
\br
J_{\mu}^{(n)} &=& \pa_{\mu} \chi^{(n)},\,\,\,\mu\equiv \xi, \eta;\,\,\,\,n=0,1,2,...\\
d \chi^{(1)} &=& A_{\xi} d\xi + A_{\eta} d\eta ,\,\,\\
  &\equiv& d{\cal I}_0(\xi,\eta)+ \l\,  d{\cal I}_1(\xi,\eta);\\
J_{\mu}^{(n+1)} &=& \pa_{\mu} \chi^{(n)}-A_{\mu} \chi^{(n)};\,\,\,\,\,\chi^{(0)}=1,
\er
where
\br
\label{expan11}
d{\cal I}_0(\xi,\eta)\equiv  a_0(\xi, \eta) d\xi +b_0(\xi, \eta) d\eta ,\,\,\,\,d{\cal I}_1(\xi,\eta) \equiv a_1(\xi, \eta) d\xi +b_1(\xi, \eta) d\eta.
\er
Then one can show by an inductive procedure that the  (non-local) currents $J_{\mu}^{(n)}$ are conserved
\br
\label{nlcl}
\pa_{\mu} J^{(n)\,\mu} =0,\,\,\,\,n=1,2,3,...
\er
The first non-trivial current conservation law $\pa_{\mu} J^{(1)\,\mu}=0$  reduces to the eq. (\ref{ABd}), and then provides the first two conservation laws (\ref{nl0})-(\ref{em1}). The next conservation law $\pa_{\mu} J^{(2)\,\mu}=0$ becomes
\br
\label{expan1}
\pa_{\eta}\Big[A_{\xi} - a_0 {\cal I}_0 -(a_0 {\cal I}_1 +a_1 {\cal I}_0)\l - a_1 {\cal I}_1 \l^2\Big]-\pa_{\xi}\Big[A_{\eta} -b_0 {\cal I}_0 -(b_0 {\cal I}_1 +b_1 {\cal I}_0)\l - b_1 {\cal I}_1 \l^2\Big]=0,
\er
where ${\cal I}_0$ and ${\cal I}_1$ are defined in (\ref{expan11}). From (\ref{expan1}), in addition to the conservation laws (\ref{ABd}) or  (\ref{nl0})-(\ref{em1}), one can get the new non-local conservations laws order by order in powers of $\lambda$
\br
\pa_{\eta} (a_0 {\cal I}_0) -\pa_{\xi} (b_0 {\cal I}_0)=0,\\
\pa_{\eta}(a_0 {\cal I}_1 +a_1 {\cal I}_0) -\pa_{\xi}(b_0 {\cal I}_1 +b_1 {\cal I}_0)=0,\\ 
\pa_{\eta}( a_1 {\cal I}_1)-\pa_{\xi}( b_1 {\cal I}_1)=0.
\er 
 
Notice that the linear systems (\ref{lin1})-(\ref{lin2}) and (\ref{lin1tx})-(\ref{lin2tx}) have been obtained by deforming the potential $V(w)$ away from sine-Gordon. The construction of analogous  linear systems may be relevant for the deformations of the well known integrable models related to the eq. of motion (\ref{sg1}), such as the Boullogh-Dodd model \cite{jhep6}. In addition, it would be interesting to uncover the classical Yangian as a Poisson-Hopf type algebra underlying those set of non-local currents and charges \cite{mackay} for the deformations of the integrable models such as the sine-Gordon, sinh-Gordon, Boullogh-Dodd and Liouville. We will postpone those important issues and some relevant applications for a future work.

\section{Conclusions and future prospects}
 
\label{sec:conclu}

In this paper, we have made the first steps toward deformations of the pseudo-potential approach to the sine-Gordon integrable model and applied, as an example, to the models of Bazeia et. al. \cite{bazeia}. We showed that  when the Riccati-type pseudo-potential equations are deformed, away from the sine-Gordon model, one can construct infinite towers of quasi-conservation laws associated to the deformed sine-Gordon models of type (\ref{sg1}). The first order set of conserved charges are related to the usual energy and momentum charges. In addition, a related linear system of equations allowed us to construct an infinite tower of non-local conservation laws. Moreover, by direct construction, we have obtained additional towers of quasi-conservation laws.  

Then, we have shown analytically that the second, third and fifth order set of exactly conserved charges becomes the first, second and fourth order time-derivatives of the energy (E) and momentum (P) charges, respectively. In this way, they are trivially conserved. It is also shown that the fourth order conservation law is a trivial identity. The redefined third and fifth order asymptotically conserved charges, which are embedded into the  relevant conservation laws, have been decomposed as a summation of $x-$integrals of certain densities, such that each density exhibits space-reflection symmetries for definite parity soliton configurations.

It has been verified, up to the fifth order and through numerical simulation, that there exist embedded into the relevant third and fifth order conservation laws, a pair of independent exactly conserved charges $q^{(3)}_{a\, \pm}$ (\ref{qq31})-(\ref{qq322}) and a pair of asymptotically conserved charges $q^{(5)}_{a\, \pm}$ (\ref{qq5p})-(\ref{qq5m2}), within numerical accuracy. The pair of exactly conserved charges $q^{(3)}_{a\, \pm}$ have been regarded as simply asymptotically conserved ones in the quasi-integrability approach \cite{jhep1, cnsns}. In general, in the pseudo-potential approach for the DSG models we have shown the absence of the so-called `anomalies' which are present in the quasi-conservation laws of \cite{jhep1}. We were not able  to trace the relation between our numerical results for conserved charges $q^{(3)}_{a\, \pm}$ and the corresponding analytical  construction of their associated exact conservation laws. 
             
We have checked through numerical simulations of soliton collisions (kink-kink, kink-antikink and breather configurations) the  conservation properties of the first two sets of higher order charges: the two third and two fifth order ones. We have used, as a particular example, the models of Bazeia et. al., which depend on a deformation parameter $q$ (such that  for $q=2$ it reduces to the ordinary sine-Gordon model) and have one
kink solutions (for any $q \in \IR$) and no other analytic solutions of these models (when $q \neq 2$) are known yet. We have studied these models numerically and computed their first six  nontrivial charges $E$, $P$, $q_{a\, \pm}^{(3)}$ and $q_{\pm}^{(5)}$ for various two-soliton and breather configurations.  Our numerical simulations allow us to argue that for general two-soliton configurations the charges   $q_{a\, \pm}^{(3)}$ are exactly conserved, within numerical accuracy; whereas the charges  $q_{a\, \pm}^{(5)}$ can be considered, in general, as asymptotically conserved ones. In addition, the charge  $q_{a\, -}^{(5)}$ becomes exactly conserved for soliton configurations possessing definite parities. Similarly, for definite parities  of  kink-kink and kink-antikink solutions, the lowest order anomalies of the new quasi-conservation laws presented in sec. 3 vanish.
 
Moreover, in sec. \ref{sec:newtowers} we have found new towers of quasi-conservation laws with true anomalies. We discussed some of their properties and simulated their relevant anomalies in sec. \ref{sec:secth}. It is remarkable that the anomalies $\mbox{a}_{-}$ and $\g_{-}$ vanish  for symmetric kink-antikink collision (see Fig. 4). Similarly, for anti-symmetric kink-kink collision (see Fig. 1) the anomalies $\mbox{a}_{+}$ and $\g_{-}$ vanish. These kind of anomalous charges also appear in the standard sine-Gordon model, and they are expected to appear in the other integrable systems and their quasi-integrable deformations \cite{pla1, new}.   

In the framework of the Riccati-type pseudo-potential approach we have constructed a pair of  linear system of equations,  (\ref{lin1})-(\ref{ab11}) and  (\ref{lin1tx})-(\ref{ABtx}), respectively,  whose relevant compatibility conditions furnish the DSG model (\ref{sg1}). The study of the properties of these linear systems, as well as their dual constructions, deserves a carefull consideration. In particular, the relation of their associated non-local currents with  the so-called classical Yangians \cite{mackay}.   

In view of our results above, one can ask if the quasi-integrable systems studied in the literature, such as the deformations of the non-linear Schr\"odinger, Bullough-Dodd, Toda, SUSY sine-Gordon and KdV systems \cite{jhep3, jhep4, jhep5, jhep6, toda, epl, arxiv1}, might possess more specific integrable structures, such as an infinite number of (non-local) conservation laws. So, they deserve careful considerations in the lines discussed above.  

Finally, following the work of Krasil'shchik and Vinogradov \cite{krasil} about non-local trends
in the geometry of differential equations, in which the partial differential equations (PDEs) have been regarded as infinite-dimensional manifolds, it has been introduced the so-called differential coverings, which have been used to study some of the PDEs properties including the constructions like Lax pairs and Backlund transformations (see e.g.  \cite{igonin}).  Moreover, it has been shown that all kinds of Lax pairs, zero-curvature representations and B\"acklund transformations in soliton theory are special types of coverings \cite{krasil}. In particular, an auto-B\"acklund transformation is associated to an automorphism of the covering. Then, it would be interesting to study the properties of the system (\ref{ricc1})-(\ref{ricc2}) and (\ref{ricc1r})-(\ref{ricc2s}) as some types  of differential coverings of the DSG model (\ref{sg1}).      
 
\section{Acknowledgments}

We would like to thank A. C. R. do Bonfim and A. R. Aguirre for discussions and JPRC acknowledges brazilian CAPES for financial support. HB thanks FC-UNI (Lima-Per\'u) and FC-UNASAM (Huaraz-Per\'u) for partial support and kind hospitality during the final stage of the work. 

\appendix
 
\section{The $u_n'$s of the first set of charges}
\label{fsca1}
The $u_n'$s can be determined recursively by substituting the expansion (\ref{expan}) into (\ref{ricc1}). Then the first quantities become
\br
\nonumber
u_1 &=& \frac{1}{2} \pa_{\xi} w\\\nonumber
u_2 &=& - \frac{1}{2} \pa_{\xi} u_1\\\nonumber
&=& - \frac{1}{2^2} \pa^2_{\xi} w\\\nonumber
u_3 &=& \frac{1}{2}\(  u_1^2 \pa_{\xi} w - \pa_{\xi} u_2 \)\\\nonumber
&=& \frac{1}{2^3} \( (\pa_{\xi} w)^3 + \pa_{\xi}^3 w \)\\\nonumber
u_4&=& \frac{1}{2} [ (u_1 u_ 2+u_2 u_1  )\pa_{\xi} w- \pa_{\xi} u_3 ] \\\label{terms}
&=&  - \frac{1}{2^3} (\pa_{\xi} w)^2 \pa_{\xi}^2 w -\frac{1}{2^4} \pa_{\xi}\( (\pa_{\xi}w)^3 + \pa_{\xi}^3 w\)\\\nonumber
u_5&=& \frac{1}{2} [ (u_2^2 + 2 u_1 u_3)\pa_{\xi} w - \pa_{\xi} u_4 ] \\\nonumber
&=&  \frac{11}{32}(\pa_{\xi}^2 w)^2 (\pa_{\xi} w) + \frac{7}{32}  \pa_{\xi}^3 w (\pa_{\xi} w)^2  +  \frac{1}{16}(\pa_{\xi} w)^5 + \frac{1}{32} \pa_{\xi}^5 w\\\nonumber
u_6 &=&\frac{1}{2} [ (2 u_2 u_3 + 2 u_1 u_4)\pa_{\xi} w - \pa_{\xi} u_5 ]\\
u_7 &=&\frac{1}{2} [ (u_3^2 + 2 u_1 u_5 + 2 u_2 u_4)\pa_{\xi} w - \pa_{\xi} u_5 ]\\
&&......\nonumber
\er

The above expressions can be written recursively for the  $u_n$'s      
\br
\label{comp1}
u_n &=& \frac{1}{2} [ ( \sum_{p+q=n-1} u_p u_q ) \pa_{\xi} w - \pa_{\xi} u_{n-1}  ],\,\,\,\,\,\,\,\,\, n=2,3,4,...\\
u_1 &=& \frac{1}{2} \pa_{\xi} w;\,\,\,\,\,\,\,\, u_0 = 0. \label{comp2}
\er
\section{The $r_n'$s and  $s_n'$s of the first set of charges}

\label{fsca11}

Substituting  the expansions (\ref{expan}) into the system of eqs.   (\ref{ricc1r})-(\ref{ricc2s}) the components $r_n$'s and $s_n$'s  can be written recursively as
\br
r_m &=& -\frac{1}{2} \( \pa_{\xi} r_{m-1} - s_{m-1} \pa_{\xi} w - \pa_{\xi} w \sum_{k=1}^{m-1} r_{m-k-1} u_k\),\,\,\,\,m=1,2,3,...; r_0 = \frac{1}{2} X, \\
\pa_{\xi} s_m &=& X u_{m+1} - r_{m} \pa_{\xi} w + \pa_{\xi} w \sum_{k=1}^{m} s_{m-k} u_{k} ,\,\,\,\,m=0,1,2,...
\er      

Next, making use of the expressions for the $u_n'$s presented in the appendix \ref{fsca1} and the recursion relations above we list the first components $\pa_{\xi}s_n $ for $n=0,1,2,3,4,5$
\br
\label{s0}
 \pa_{\xi} s_0 &=& 0,\\
 \pa_{\xi} s_1 &=& -\frac{1}{2^2} X \pa_{\xi}^2 w + \frac{1}{2^2} \pa_{\xi} w \pa_{\xi} X,\\
 \pa_{\xi} s_2 &=&   \frac{1}{2^3}  X \pa_{\xi}^3 w - \frac{1}{2^3}  \pa_{\xi} w \pa_{\xi}^2 X, \\
  \pa_{\xi} s_3 &=& - \frac{1}{2^4}  X \pa_{\xi}^4 w + \frac{1}{2^4}  \pa_{\xi} w \Big[3 (\pa_{\xi}w)^2   \pa_{\xi} X - 3 X \pa_{\xi} w  \pa_{\xi}^2 w + \pa^3_{\xi} X\Big],\\
   \pa_{\xi} s_4 &=&  \frac{1}{2^5} X \pa_{\xi}^5 w - \frac{1}{2^5}  \pa_{\xi} w \Big[ 5  (\pa_{\xi} w)^2  \pa_{\xi}^2 X -10 X (\pa^2_{\xi}w)^2   + 5 \pa_{\xi} w \( 2 \pa_{\xi} X \pa_{\xi}^2 w- X \pa_{\xi}^3 w\) + \pa_{\xi}^4 X\Big],\\
   \pa_{\xi} s_5 &=& - \frac{1}{2^6} X \pa_{\xi}^6 w + \frac{1}{2^6}  \pa_{\xi} w \Big[ 11  (\pa^2_{\xi} w)^2  \pa_{\xi} X -  36 X \pa^2_{\xi}w  \pa^3_{\xi}w + \pa^5_{\xi} X\Big] +\\
   && \frac{1}{2^6} (\pa_{\xi} w)^2 \Big[ 22  \pa^2_{\xi}w  \pa^2_{\xi}X +14  \pa_{\xi}X  \pa^3_{\xi}w -7 X  \pa^4_{\xi}w\Big] +\\
&&    \frac{1}{2^6} \Big[ 10 (\pa_{\xi}w)^5 \pa_{\xi} X -10 X (\pa_{\xi}w)^4 \pa^2_{\xi}w -11 X (\pa^2_{\xi}w)^3 + 7 (\pa_{\xi}w)^3 \pa^3_{\xi}X \Big],\\
   .....
\er 
\section{The $\widetilde{u}_n'$s of the second set of charges}
\label{fsca2}

Similarly, the $\widetilde{u}_n'$s can be determined recursively by substituting the expansion (\ref{expan2}) into (\ref{ricc11}). The first quantities become
\br
\nonumber
\widetilde{u}_1 &=& \frac{1}{2} \pa_{\eta} w\\\nonumber
\widetilde{u}_2 &=& - \frac{1}{2} \pa_{\eta} \widetilde{u}_1\\\nonumber
&=& - \frac{1}{2^2} \pa^2_{\eta} w\\\nonumber
\widetilde{u}_3 &=& \frac{1}{2}\(  \widetilde{u}_1^2 \pa_{\eta} w - \pa_{\eta} \widetilde{u}_2 \)\\\nonumber
&=& \frac{1}{2^3} \( (\pa_{\eta} w)^3 + \pa_{\eta}^3 w \)\\\nonumber
\widetilde{u}_4&=& \frac{1}{2} [ (\widetilde{u}_1 \widetilde{u}_ 2+\widetilde{u}_2 \widetilde{u}_1  )\pa_{\eta} w- \pa_{\eta} \widetilde{u}_3 ] \\
\label{terms2}
&=&  - \frac{1}{2^3} (\pa_{\eta} w)^2 \pa_{\eta}^2 w -\frac{1}{2^4} \pa_{\eta}\( (\pa_{\eta}w)^3 + \pa_{\eta}^3 w\)\\\nonumber
\widetilde{u}_5&=& \frac{1}{2} [ (\widetilde{u}_2^2 + 2 \widetilde{u}_1 \widetilde{u}_3)\pa_{\eta} w - \pa_{\eta} \widetilde{u}_4 ] \\\nonumber
&=& \frac{1}{2^5} \pa_{\eta}w (\pa_{\eta}^2 w )^2 + \frac{1}{2^4} \((\pa_{\eta} w)^3 + \pa_{\eta}^3 w\)(\pa_{\eta} w)^2 \\\nonumber
&& + \frac{1}{2^4} \pa_{\eta} \( (\pa_{\eta} w)^2 \pa_{\eta}^2 w + \frac{3}{2} (\pa_{\eta} w)^2  \pa_{\eta}^2 w + \frac{1}{2} \pa_{\eta}^4 w\)\\\nonumber
\widetilde{u}_6 &=&\frac{1}{2} [ (2 \widetilde{u}_2 \widetilde{u}_3 + 2 \widetilde{u}_1 \widetilde{u}_4)\pa_{\eta} w - \pa_{\eta} \widetilde{u}_5 ]\\
&&......\nonumber
\er

The recursion relation for the  $\widetilde{u}_n$'s becomes       
\br
\label{comp12}
\widetilde{u}_n &=& \frac{1}{2} [ ( \sum_{p+q=n-1} \widetilde{u}_p \widetilde{u}_q ) \pa_{\eta} w - \pa_{\eta} \widetilde{u}_{n-1}  ],\,\,\,\,\,\,\,\,\, n=2,3,4,...\\
\widetilde{u}_1 &=& \frac{1}{2} \pa_{\eta} w;\,\,\,\,\,\,\,\, \widetilde{u}_0 = 0. \label{comp22}
\er

\section{The $\widetilde{r}_n'$s and  $\widetilde{s}_n'$s of the second set of charges}

\label{aprs222}

Substituting  the expansions (\ref{expan2}) into the system of eqs.   (\ref{ricc1rd})-(\ref{ricc2sd}) the components $\widetilde{r}_n$'s and $\widetilde{s}_n$'s  can be written recursively as
\br
\widetilde{r}_m &=& -\frac{1}{2} \( \pa_{\eta} \widetilde{r}_{m-1} - \widetilde{s}_{m-1} \pa_{\eta} w - \pa_{\eta} w \sum_{k=1}^{m-1} \widetilde{r}_{m-k-1} \widetilde{u}_k\),\,\,\,\,m=1,2,3,...; \widetilde{r}_0 = \frac{1}{2} \widetilde{X}, \\
\pa_{\eta} \widetilde{s}_m &=& \widetilde{X} \widetilde{u}_{m+1} - \widetilde{r}_{m} \pa_{\eta} w + \pa_{\eta} w \sum_{k=1}^{m} \widetilde{s}_{m-k} \widetilde{u}_{k} ,\,\,\,\,m=0,1,2,...
\er      

Next, we list the first components $\pa_{\eta}\widetilde{s}_n,\, n=0,1,2,3,4,5$
\br
 \pa_{\eta} \widetilde{s}_0 &=& 0,\\
 \pa_{\eta} \widetilde{s}_1 &=& -\frac{1}{2^2} \widetilde{X} \pa_{\eta}^2 w + \frac{1}{2^2} \pa_{\eta} w \pa_{\eta} \widetilde{X},\\
 \pa_{\eta} \widetilde{s}_2 &=&   \frac{1}{2^3}  \widetilde{X} \pa_{\eta}^3 w - \frac{1}{2^3}  \pa_{\eta} w \pa_{\eta}^2 \widetilde{X}, \\
  \pa_{\eta} \widetilde{s}_3 &=& - \frac{1}{2^4}  \widetilde{X} \pa_{\eta}^4 w + \frac{1}{2^4}  \pa_{\eta} w \Big[3 (\pa_{\eta}w)^2   \pa_{\eta} \widetilde{X} - 3 \widetilde{X} \pa_{\eta} w  \pa_{\eta}^2 w + \pa^3_{\eta} \widetilde{X}\Big],\\
   \pa_{\eta} \widetilde{s}_4 &=&  \frac{1}{2^5} \widetilde{X} \pa_{\eta}^5 w - \frac{1}{2^5}  \pa_{\eta} w \Big[ 5  (\pa_{\eta} w)^2  \pa_{\eta}^2 \widetilde{X} -10 \widetilde{X} (\pa^2_{\eta}w)^2   + 5 \pa_{\eta} w \( 2 \pa_{\eta} \widetilde{X} \pa_{\eta}^2 w- \widetilde{X} \pa_{\eta}^3 w\) + \pa_{\eta}^4 \widetilde{X}\Big],\\
   \pa_{\eta} \widetilde{s}_5 &=& - \frac{1}{2^6} \widetilde{X} \pa_{\eta}^6 w + \frac{1}{2^6}  \pa_{\eta} w \Big[ 11  (\pa^2_{\eta} w)^2  \pa_{\eta} \widetilde{X} -  36 \widetilde{X} \pa^2_{\eta}w  \pa^3_{\eta}w + \pa^5_{\eta} \widetilde{X}\Big] +\\
   && \frac{1}{2^6} (\pa_{\eta} w)^2 \Big[ 22  \pa^2_{\eta}w  \pa^2_{\eta}\widetilde{X} +14  \pa_{\eta}\widetilde{X}  \pa^3_{\eta}w -7 \widetilde{X}  \pa^4_{\eta}w\Big] +\\
&&    \frac{1}{2^6} \Big[ 10 (\pa_{\eta}w)^5 \pa_{\eta} \widetilde{X} -10 \widetilde{X} (\pa_{\eta}w)^4 \pa^2_{\eta}w -11 \widetilde{X} (\pa^2_{\eta}w)^3 + 7 (\pa_{\eta}w)^3 \pa^3_{\eta}\widetilde{X} \Big],\\
   .....
\er 
  
\section{The $\Psi_n'$s of the first set of charges}
\label{aPsi1}
 
Next, we provide the first components of the pseudo-potential $\Psi$. Substituting the expansion (\ref{seriesPsi}) into the eq.  (\ref{xiPsi}) and taking into account of the components $u_n$ provided in the appendix \ref{fsca1} one can get
\br
\nonumber
\Psi_1 &=& -\frac{1}{4} Y^{(1)} (\pa_{\xi}w)^3\\ \nonumber
   &=& -\frac{1}{8} \pa_{\xi}\Big[(\pa_{\xi}w)^2 V^{(2)}\Big]-\frac{1}{8} \pa_{\eta}\Big[ (\pa_{\xi}^2w)^2 - (\pa_{\xi}w)^4\Big]\\\nonumber
\Psi_2 &=& -\frac{1}{2} \pa_{\xi} \Psi_1 +\frac{1}{2} \frac{\pa^2_{\xi}w}{ \pa_{\xi}w} \Psi_1 + \frac{1}{8} Y^{(1)} (\pa_{\xi}w)^2 \pa_{\xi}^2w\\ \nonumber
    &=& -\frac{1}{2} \pa_{\xi} \Psi_1\\\nonumber
    \Psi_3 &=&  -\frac{1}{2} \pa_{\xi} \Psi_2 + \frac{1}{2} \frac{\pa^2_{\xi}w}{ \pa_{\xi}w} \Psi_2 +
     u_1\pa_{\xi}w \Psi_1 -\frac{1}{16} Y^{(1)}\Big[ (\pa_{\xi}w)^5 + (\pa_{\xi}w)^2 \pa_{\xi}^3w  \Big].\\\nonumber
     &=& \pa_{\eta} R_3 + \pa_{\xi} S_3,\\
     \nonumber
    \Psi_4 &=&  -\frac{1}{2} \pa_{\xi} \Psi_3 + \frac{1}{2} \frac{\pa^2_{\xi}w}{ \pa_{\xi}w} \Psi_3 +
       u_1\pa_{\xi}w \Psi_2 + u_2 \pa_{\xi}w \Psi_1  -\frac{1}{2}(\pa_{\xi}w)^2\, u_4 \, Y^{(1)}.\\\nonumber
    \Psi_5 &=&  -\frac{1}{2} \pa_{\xi} \Psi_4 + \frac{1}{2} \frac{\pa^2_{\xi}w}{ \pa_{\xi}w} \Psi_4 +  u_1 \pa_{\xi}w \Psi_3 +u_2 \pa_{\xi}w \Psi_2 + u_3 \pa_{\xi}w \Psi_1  -\frac{1}{2} (\pa_{\xi}w)^2\, u_5 \, Y^{(1)}.\\\nonumber
\Psi_6 &=&  -\frac{1}{2} \pa_{\xi} \Psi_5 + \frac{1}{2} \frac{\pa^2_{\xi}w}{ \pa_{\xi}w} \Psi_5 +  u_1\pa_{\xi}w \Psi_4 + u_2 \pa_{\xi}w \Psi_3 + u_3 \pa_{\xi}w \Psi_2 +u_4 \pa_{\xi}w \Psi_1 -\frac{1}{2} (\pa_{\xi}w)^2\, u_6 \, Y^{(1)},\\\nonumber
&&......\er
where $R_3, S_3$ in the expression for $\Psi_3$ are defined  as 
\br
\label{nn52}
R_3 & \equiv & \frac{1}{32} (\pa_{\xi}^3 w)^2 + \frac{1}{16} (\pa_{\xi} w)^6 - \frac{5}{16} (\pa_{\xi}^2 w \pa_{\xi} w)^2,\\
S_3 & \equiv & -\frac{1}{16} \pa_{\xi} [(\pa_{\xi} w)^3  Y^{(1)}] + \frac{1}{16} \pa_{\xi}^2 w (\pa_{\xi} w)^2 Y^{(1)} -\frac{1}{8}  \pa_{\xi}^3 w\pa_{\xi} w Y +\frac{1}{16}(\pa_{\xi}^2 w)^2 Y -\frac{1}{2} \pa_{\xi}^3 w \pa_{\xi} w + \frac{1}{4} (\pa_{\xi}^2 w)^2+ \nonumber \\
&& \frac{1}{4}  \pa_{\xi}^3 w \pa_{\xi} w V- \frac{1}{4} \pa_{\xi}^2 w (\pa_{\xi} w )^2 V^{(1)} - \frac{1}{8} (\pa_{\xi}^2 w)^2 V + \frac{1}{16} \pa_{\xi}^3 w \pa_{\xi} w V^{(2)}-\frac{3}{32} (\pa_{\xi} w)^4 V^{(2)}
\label{nn53}.\\
Y &\equiv &  \frac{1}{2}V^{(2)}+ 2 V - 4; \,\,\,\,\,\, Y^{(1)}\equiv \frac{d}{dw} Y. \label{y11}
\er
The recursion relation for the components $\Psi_n$ becomes
\br
\label{Psin}
\Psi_n &=&  -\frac{1}{2} \pa_{\xi} \Psi_{n-1} + \frac{1}{2} \frac{\pa^2_{\xi}w}{ \pa_{\xi}w} \Psi_{n-1} +  \pa_{\xi}w \sum_{k=1}^{n-2} u_k \Psi_{n-k-1} -\frac{1}{2} (\pa_{\xi}w)^2\, u_n \, Y'.
\er

\section{The $\widetilde{\Psi}_n'$s of the second set of charges}
\label{aPsi11}

Finally, we provide the first components of the pseudo-potential $\widetilde{\Psi}$. Substituting the expansion (\ref{seriesPsid}) into the eq.  (\ref{xiPsid}) and taking into account of the components $\widetilde{u}_n$ provided in the appendix \ref{fsca2} one can get
\br
\nonumber
\widetilde{\Psi}_1 &=& -\frac{1}{4} Y^{(1)} (\pa_{\eta}w)^3\\ \nonumber
   &=& -\frac{1}{8} \pa_{\eta}\Big[(\pa_{\eta}w)^2 V^{(2)}\Big]-\frac{1}{8} \pa_{\eta}\Big[ (\pa_{\eta}^2w)^2 - (\pa_{\eta}w)^4\Big]\\\nonumber
\widetilde{\Psi}_2 &=& -\frac{1}{2} \pa_{\eta} \widetilde{\Psi}_1 +\frac{1}{2} \frac{\pa^2_{\eta}w}{ \pa_{\eta}w} \widetilde{\Psi}_1 + \frac{1}{8} Y^{(1)} (\pa_{\eta}w)^2 \pa_{\eta}^2w\\ \nonumber
    &=& -\frac{1}{2} \pa_{\eta} \widetilde{\Psi}_1\\\nonumber
    \widetilde{\Psi}_3 &=&  -\frac{1}{2} \pa_{\eta} \widetilde{\Psi}_2 + \frac{1}{2} \frac{\pa^2_{\eta}w}{ \pa_{\eta}w} \widetilde{\Psi}_2 +
     \widetilde{u}_1\pa_{\eta}w \widetilde{\Psi}_1 -\frac{1}{16} Y^{(1)}\Big[ (\pa_{\eta}w)^5 + (\pa_{\eta}w)^2 \pa_{\eta}^3w  \Big].\\\nonumber
     &=& \pa_{\xi} \widetilde{R}_3 + \pa_{\eta} \widetilde{S}_3,\\
     \nonumber
   \widetilde{\Psi}_4 &=&  -\frac{1}{2} \pa_{\eta} \widetilde{\Psi}_3 + \frac{1}{2} \frac{\pa^2_{\eta}w}{ \pa_{\eta}w} \widetilde{\Psi}_3 +
       \widetilde{u}_1\pa_{\eta}w \widetilde{\Psi}_2 + \widetilde{u}_2 \pa_{\eta}w \widetilde{\Psi}_1  -\frac{1}{2}(\pa_{\eta}w)^2\, \widetilde{u}_4 \, Y^{(1)}.\\\nonumber
   \widetilde{\Psi}_5 &=&  -\frac{1}{2} \pa_{\eta} \widetilde{\Psi}_4 + \frac{1}{2} \frac{\pa^2_{\eta}w}{ \pa_{\eta}w} \widetilde{\Psi}_4 +  \widetilde{u}_1 \pa_{\eta}w \widetilde{\Psi}_3 +\widetilde{u}_2 \pa_{\eta}w \widetilde{\Psi}_2 + \widetilde{u}_3 \pa_{\eta}w \widetilde{\Psi}_1  -\frac{1}{2} (\pa_{\eta}w)^2\, \widetilde{u}_5 \, Y^{(1)}.\\\nonumber
\widetilde{\Psi}_6 &=&  -\frac{1}{2} \pa_{\eta} \widetilde{\Psi}_5 + \frac{1}{2} \frac{\pa^2_{\eta}w}{ \pa_{\eta}w} \widetilde{\Psi}_5 +  \widetilde{u}_1\pa_{\eta}w \widetilde{\Psi}_4 + \widetilde{u}_2 \pa_{\eta}w \widetilde{\Psi}_3 + \widetilde{u}_3 \pa_{\eta}w \widetilde{\Psi}_2 +\widetilde{u}_4 \pa_{\eta}w \widetilde{\Psi}_1 -\frac{1}{2} (\pa_{\eta}w)^2\, \widetilde{u}_6 \, Y^{(1)}.\\\nonumber
&&......\er 
where $\widetilde{R}_3, \widetilde{S}_3$ in the expression for $\widetilde{\Psi}_3$ are defined as
\br
\label{nn52d}
\widetilde{R}_3 & \equiv & \frac{1}{32} (\pa_{\eta}^3 w)^2 + \frac{1}{16} (\pa_{\eta} w)^6 - \frac{5}{16} (\pa_{\eta}^2 w \pa_{\eta} w)^2,\\
\widetilde{S}_3 & \equiv & -\frac{1}{16} \pa_{\eta} [(\pa_{\eta} w)^3  Y^{(1)}] + \frac{1}{16} \pa_{\eta}^2 w (\pa_{\eta} w)^2 Y^{(1)} -\frac{1}{8}  \pa_{\eta}^3 w\pa_{\eta} w Y +\frac{1}{16}(\pa_{\eta}^2 w)^2 Y -\frac{1}{2} \pa_{\eta}^3 w \pa_{\eta} w + \frac{1}{4} (\pa_{\eta}^2 w)^2+ \nonumber \\
&& \frac{1}{4}  \pa_{\eta}^3 w \pa_{\eta} w V- \frac{1}{4} \pa_{\eta}^2 w (\pa_{\eta} w )^2 V^{(1)} - \frac{1}{8} (\pa_{\eta}^2 w)^2 V + \frac{1}{16} \pa_{\eta}^3 w \pa_{\eta} w V^{(2)}-\frac{3}{32} (\pa_{\eta} w)^4 V^{(2)}
\label{nn53d}. 
\er
The recursion relation for the components $\widetilde{\Psi}_n$ becomes
\br
\label{Psin2}
\widetilde{\Psi}_n &=&  -\frac{1}{2} \pa_{\eta} \widetilde{\Psi}_{n-1} + \frac{1}{2} \frac{\pa^2_{\eta}w}{ \pa_{\eta}w} \widetilde{\Psi}_{n-1} +  \pa_{\eta}w \sum_{k=1}^{n-2} \widetilde{u}_k \widetilde{\Psi}_{n-k-1} -\frac{1}{2} (\pa_{\eta}w)^2\, \widetilde{u}_n \, Y'.
\er

\end{document}